\documentclass[11pt]{article}

\usepackage[reqno]{amsmath}
\usepackage{epsfig}
\usepackage{array}
\usepackage{float}

\begin{document}


\def\ltap{\ \raisebox{-.4ex}{\rlap{$\sim$}} \raisebox{.4ex}{$<$}\ }
\def\gtap{\ \raisebox{-.4ex}{\rlap{$\sim$}} \raisebox{.4ex}{$>$}\ }
\newcommand{\deltaatm}{\mbox{$\Delta  m^2_{\mathrm{atm}} \ $}}
\newcommand{\deltasol}{\mbox{$ \Delta  m^2_{\odot} \ $}}
\newcommand{\utre}{\mbox{$|U_{\mathrm{e} 3}|$}}
\newcommand{\betabeta}{\mbox{$(\beta \beta)_{0 \nu}  $}}
\newcommand{\mefff}{\mbox{$ < \! m  \! > $}}
\newcommand{\meff}{\mbox{$\left|  < \! m  \! > \right| \ $}}
\newcommand{\hbeta}{$\mbox{}^3 {\rm H}$ $\beta$-decay \ }
\newcommand{\eV}{\mbox{$ \  \mathrm{eV} \ $}}
\newcommand{\deltatre}{\mbox{$ \ \Delta m^2_{32} \ $}}
\newcommand{\deltadue}{\mbox{$ \ \Delta m^2_{21} \ $}}
\newcommand{\ueuno}{\mbox{$ \ |U_{\mathrm{e} 1}|^2 \ $}}
\newcommand{\uedue}{\mbox{$ \ |U_{\mathrm{e} 2}|^2 \ $}}
\newcommand{\uetre}{\mbox{$ \ |U_{\mathrm{e} 3}|^2  \ $}}


\hyphenation{par-ti-cu-lar}
\hyphenation{ex-pe-ri-men-tal}
\hyphenation{dif-fe-rent}
\hyphenation{bet-we-en}
\hyphenation{mo-du-lus}
\hyphenation{}
\rightline{Ref. SISSA 1/2001/EP}
\rightline{January 2001}
\vskip 0.6cm
\begin{center}
{\bf
Majorana Neutrinos, Neutrino Mass Spectrum, CP-Violation and \\ 
Neutrinoless Double $\beta$-Decay: I. The Three-Neutrino Mixing Case\\
} 

\vspace{0.3cm} 
S. M. Bilenky$^{(a,b)}$,~S. Pascoli$^{(b,c)}$~and~
S. T. Petcov$^{(b,c)}$ 
\footnote{Also at: Institute of Nuclear Research and
Nuclear Energy, Bulgarian Academy of Sciences, 1784 Sofia, Bulgaria}

\vspace{0.3cm}
{\em $^{(a)}$ Joint Institute for Nuclear Research, Dubna, Russia\\
}
\vspace{0.2cm}   
{\em $^{(b)}$ Scuola Internazionale Superiore di Studi Avanzati, 
I-34014 Trieste, Italy\\
}
\vspace{0.2cm}   
{\em $^{(c)}$ Istituto Nazionale di Fisica Nucleare, 
Sezione di Trieste, I-34014 Trieste, Italy\\
}
\end{center}

\vskip 0.4cm
\begin{abstract}
Assuming three-neutrino mixing 
and massive Majorana neutrinos,
we study the implications of the neutrino 
oscillation solutions of the solar 
and atmospheric neutrino problems
and of the results of the CHOOZ experiment
for the predictions of the
effective Majorana mass in
neutrinoless double beta
(\betabeta-)decay, \meff. 
The general case of 
CP-nonconservation 
is investigated.
The predicted values of  \meff, 
which determines the magnitude of the
\betabeta-decay rate,
depend strongly on the type of the
neutrino mass spectrum, on the solution of 
the solar neutrino problem, as well as 
on the values of the two
Majorana CP-violating phases,
present in the lepton mixing matrix.
We find that i) $\meff \ltap 0.02~$eV 
for a hierarchical neutrino mass spectrum,
ii) $\meff \ltap 0.09~$eV
if the spectrum is of the inverted hierarchy type, 
and iii) $\meff \leq m$ in the case of 
three quasi-degenerate neutrinos,
$m > 0$ being the common neutrino mass scale
which is limited by the bounds 
from the $^3$H 
$\beta-$decay experiments,
$m < 2.5~$eV.
The indicated maximal values of \meff 
are reached in the cases i), ii) and iii) 
respectively, for the large 
mixing angle (LMA) MSW solution,
the small mixing angle (SMA) MSW, and
for all current solutions, of the 
solar neutrino problem.
If CP-invariance holds, 
\meff is very sensitive
to the values of the relative 
CP-parities of the massive 
Majorana neutrinos.
The cases of neutrino mass 
spectra which interpolate
between the hierarchical 
or inverted hierarchy type
and the quasi-degenerate 
one are also studied.
The observation of the
\betabeta-decay with a rate
corresponding to 
$\meff \gtap 0.02~$eV 
can provide unique information 
on the neutrino mass spectrum.
Combined with information on the lightest
neutrino mass or the type of neutrino mass spectrum, 
it can give also information 
on the CP-violation in the lepton sector,
and if CP-invariance holds - on the 
relative CP-parities
of the massive Majorana neutrinos.

\end{abstract}
\vfill
\newpage
\section{Introduction} \indent 

   The observation of a significant
up-down asymmetry 
in the rate of (multi-GeV) 
$\mu-$like events
produced by the atmospheric 
$\nu_{\mu}$ and $\bar{\nu}_{\mu}$ in the Super-Kamiokande
experiment \cite{SKatm98},
brought to a new level the 
investigation of the neutrino mass 
and neutrino mixing problem: for the first time 
model-independent 
experimental evidence for neutrino
oscillations was obtained. 
Important evidences 
for existence of neutrino mixing were            
found  in the experiments
with solar neutrinos as well: in all five
experiments Homestake, Kamiokande, SAGE, GALLEX 
and Super-Kamiokande, 
which have provided data on the solar neutrino flux so far
\cite{Cl98,Fu96,Ab99,Ha99,SK99,SKYSuz00}
(see also \cite{GNO,SNO}),
considerably smaller signals than expected 
\cite{BP00} were observed.
These results are compatible with 
a depletion of the solar $\nu_{e}$ flux
on the way to the Earth,
i.e., with a disappearance of the solar $\nu_{e}$.
Indications for $\bar{\nu}_{\mu}\leftrightarrow \bar{\nu}_{e}$
oscillations were reported by
the accelerator LSND experiment \cite{LSND}.

   The existing evidences for nonzero neutrino 
masses and neutrino mixing will be thoroughly tested
in the next generation of neutrino oscillation experiments.
The Super-Kamiokande results on the oscillations of the 
atmospheric neutrinos will be checked  
in the accelerator long baseline experiments K2K \cite{K2K},
MINOS \cite{MINOS} and in the CERN-Gran-Sasso (CNGS) 
experiment \cite{CERNGS}; K2K is taking data while
MINOS and CNGS experiments are under preparation.
The solar neutrino experiments SNO \cite{SNO}, which 
began operation approximately an year 
and a half ago, and 
BOREXINO \cite{BOREX}, and the reactor 
antineutrino experiment
KamLAND \cite{KamL},
will provide new crucial information 
on the oscillations of solar neutrinos.
Both BOREXINO and KamLAND detectors are 
under construction
and are expected to be operative in 2002. The 
LSND results on 
$\bar{\nu}_{\mu}\leftrightarrow \bar{\nu}_{e}$
oscillations will be tested in the 
accelerator experiment
MiniBooNE \cite{MiniB} which is under preparation.

    The high intensity neutrino beams from
a neutrino factory, with known and well-controlled 
spectra and fluxes, 
will allow to study the neutrino 
oscillation phenomena in considerable detail \cite{NuFact1}. 
These studies can provide, in particular,
a very precise information on the values 
of neutrino mass squared differences
and of the elements of the 
neutrino (lepton) mixing matrix.
The possibility to built a neutrino factory is 
extensively investigated at present \cite{NuFact2}.

  There is no doubt that 
the indicated experimental
studies will allow to make a big step forward
in the understanding of the 
patterns of neutrino mass squared differences and 
of the neutrino mixing. However, no information
about the absolute values of the neutrino masses
and about the physical nature of the neutrinos
with definite mass can 
be obtained in these experiments.

   The problem of the nature of massive 
neutrinos (see, e.g., \cite{BiPe87}) -
are they Dirac particles possessing 
distinctive antiparticles,
or Majorana fermions, i.e., truly neutral 
particles identical with their antiparticles, 
is one of the fundamental problems in the studies
of neutrino mixing.
In the minimal version of the Standard
Theory the individual lepton charges 
$L_e$, $L_{\mu}$ and $L_{\tau}$ are conserved and 
neutrinos are massless. Massive Dirac neutrinos 
and neutrino mixing arise in gauge theories
in which the lepton charges 
$L_e$, $L_{\mu}$ and $L_{\tau}$ are 
not conserved, but a specific combination
of the latter, which could be the total 
lepton charge $L = L_e + L_{\mu} + L_{\tau}$,
or, e.g., the charge \cite{SPPD82}
$L' = L_e - L_{\mu} + L_{\tau}$,
is conserved. 
Thus, the existence of massive Dirac neutrinos is
associated in  gauge theories with massive neutrinos
with the presence of a conserved lepton charge.
Massive Majorana neutrinos arise 
if no lepton charge
is conserved by the 
electroweak interactions.
Thus, the question of the nature of 
massive neutrinos is directly related 
to the question of the 
fundamental symmetries of the 
elementary particle interactions.

   The relative smallness of the neutrino masses,
following from the data, can be naturally 
explained by the violation 
of the total lepton charge conservation 
at a scale which is 
much larger than the electroweak
symmetry breaking scale
\footnote{Let us note that this explanation is by 
no means unique.} \cite{SeeSaw}.
In this case the neutrinos with definite 
mass are Majorana particles. Thus, establishing 
the Majorana nature
of the massive neutrinos
could imply the  
existence of a new fundamental scale
in physics.

    Experiments studying the oscillations of
neutrinos cannot answer the question
regarding the nature of the massive neutrinos
\cite{BHP80,LPST87}~\footnote{If the massive neutrinos 
are Majorana particles,
e.g., the {\it transitions} $\nu_l \rightarrow \bar{\nu}_{l}$,
$l=e,\mu,\tau$ are effectively possible \cite{BKRS}.
However, the amplitudes of these transitions
are proportional to the ratio 
$m_j/E$, $m_j$ and $E$ being the neutrino mass and energy.
For the values of the neutrino masses 
suggested by the existing 
$^3$H $\beta-$decay data and the data on the 
oscillations of the solar and
atmospheric neutrinos (see Section 2),
$m_j \ltap $ few eV, the transitions
$\nu_l \rightarrow \bar{\nu}_{l}$
are unobservable at present.}.
Neutrino oscillations, in particular,
are not sensitive
to the Majorana CP-violating phases
\cite{BHP80,Kobz80}
which enter into the expression for the
lepton mixing matrix in the case 
of massive Majorana neutrinos
and which are absent if the 
massive neutrinos are Dirac particles.
Thus, the neutrino oscillation 
experiments cannot provide also information
on CP-violation caused by the Majorana
CP-violating phases, and in the case of 
CP-invariance - on the relative CP-parities of 
the massive Majorana neutrinos.

    The Majorana nature of the massive neutrinos
can manifest itself in the existence of processes
in which the total lepton charge $L$
is not conserved and changes by two units,
$\Delta L = 2$.
The process most sensitive to the 
existence of massive Majorana neutrinos 
(coupled to the electron)
is the neutrinoless double $\beta$ 
($\betabeta-$) decay of
certain even-even nuclei (see, e.g., \cite{bb0nu1,BiPe87}):

\begin{equation}
(A,Z) \rightarrow (A,Z+2) + e^{-} + e^{-}.
\end{equation}

\noindent 
If the $\betabeta-$ decay is generated 
{\it only by the 
left-handed (LH) charged current weak interaction
through the
exchange of virtual 
massive Majorana neutrinos}, 
the probability amplitude of this process 
is proportional 
in the case of Majorana 
neutrinos having masses not exceeding few MeV 
to the so-called ``effective Majorana mass parameter''
\begin{equation}
\mefff \equiv  \sum_{j=1} U_{\mathrm{e} j}^2~ m_j,
\label{meff}
\end{equation}

\noindent where $m_j$ is the mass of the 
Majorana neutrino $\nu_j$  
and $U_{ej}$ is the element of neutrino (lepton)
mixing matrix.

   A large number of experiments are searching  for 
$\betabeta-$decay of different 
nuclei at present (for a rather complete list 
see, e.g., \cite{bb0nu2}). 
No indications that
this process takes place were found so far.
A rather stringent constraint on the 
value of the effective Majorana mass parameter 
was obtained in the $^{76}$Ge Heidelberg-Moscow 
experiment \cite{76Ge00}: 
\begin{equation}
\meff < 0.35 \ \mathrm{eV},~~~90\%~{\rm C.L.}.
\label{76Ge0}
\end{equation}
\noindent Taking into account a factor of 3 
uncertainty associated with the calculation of the relevant
nuclear matrix element (see, e.g., \cite{bb0nu1,bb0nu2})
we get
\begin{equation}
\meff < (0.35 \div 1.05) \ \mathrm{eV},~~~90\%~{\rm C.L.}.
\label{76Ge00}
\end{equation}

\noindent The IGEX collaboration  has obtained \cite{IGEX00}:
\begin{equation}
\meff < (0.33 \div 1.35) \ \mathrm{eV},~~~90\%~{\rm C.L.}
\end{equation}

  Considerably higher sensitivity to the value of 
$\meff$ is planned to be 
reached in several $\betabeta-$decay experiments
of a new generation. 
The NEMO3 experiment \cite{NEMO3}, 
scheduled to start in 2001, 
will search for $\betabeta-$decay of $^{100}$Mo
and $^{82}$Se. In its first stage,
after 5 years of data taking, 
this experiment will reach 
a sensitivity to values of $\meff \cong 0.1~$eV. 
A similar sensitivity  
is planned to be reached with 
the cryogenic detector CUORE
\cite{CUORE} which will search
for the $\betabeta-$decay of $^{130}$Te. 
An order of magnitude better sensitivity,
i.e., to $\meff \cong 10^{-2}~$eV,
is planned to be achieved 
in the GENIUS experiment \cite{GENIUS}  
utilizing one ton of enriched
$^{76}$Ge. Finally, there is a very 
interesting proposal \cite{EXO} to study
the $\betabeta-$decay of $^{136}$Xe in a 
background-free experiment with detection of 
the two electrons and
the $^{+}$Ba atom in the final state. 
The estimated sensitivity of 
this experiment is 
$\meff \cong (1 - 5)\times 10^{-2}~$eV. 

   Having in mind the future 
progress in the
experimental searches for
$\betabeta-$decay,
we derive in the present article 
the constraints on the effective Majorana
mass that can be obtained 
from the analyzes of 
the latest neutrino oscillation
data. More specifically, 
assuming three-neutrino mixing 
and massive Majorana neutrinos,
we study in detail 
the implications of the neutrino 
oscillation fits of the solar 
and atmospheric neutrino data
and of the results of the 
reactor long baseline CHOOZ and Palo Verde 
experiments,
for the predictions of the
effective Majorana mass parameter 
\meff, which controls 
the \betabeta-decay rate. 
We consider the possible types of neutrino
mass spectrum: hierarchical, with two
quasi-degenerate in mass Majorana neutrinos
(the third having a different mass),
and with three quasi-degenerate Majorana neutrinos.
In the case of two quasi-degenerate
massive neutrinos, the neutrino mass 
spectrum can be of three
varieties - with inverted hierarchy, 
with partial hierarchy and with partial inverted
hierarchy, and they are also considered 
in this paper. 

   The indicated approach for
obtaining the values of \meff 
which are compatible
with solar, atmospheric and 
accelerator neutrino oscillation data
was first used in ref. \cite{SPAS94} and in
ref. \cite{BBGK9596}.
In ref. \cite{BGKP96} it was noticed 
that in the case of neutrino mass spectrum with
inverted hierarchy and the large mixing angle (LMA)
MSW solution of the solar neutrino problem,
the measurement of \meff can give information
about the CP-violation in the lepton sector,
induced by the Majorana CP-violating phases
\footnote{Let us note that 
the most recent solar neutrino 
data from the Super-Kamiokande experiment
favors the LMA MSW solution 
of the solar neutrino problem.}.
Various aspects of CP-violation in the lepton sector, 
generated by the Majorana CP-violating phases,
were studied, e.g., in 
\cite{Branco86,NievPal87,ODonSark95,Branco00}.  
The approach indicated above was further exploited, e.g., 
in refs. \cite{BG-bb,Vissani-bb,Minakata-Yasuda-bb,Barger99}.
In the articles \cite{BGGKP99,SPWIN99,BGWIN99}, 
in which the solar and atmospheric neutrino data
available in 1998 - 1999 were used, 
it was concluded, in particular, that
the observation of the \betabeta-decay with a rate 
within the sensitivity of the next generation of 
\betabeta-decay experiments
can give information about 
the neutrino mass spectrum.  
Further analyzes along the indicated 
lines were performed more recently in
\cite{Giunti99,FKNT00,Pols00,KPS00,Rodej00}.

 In the present study we investigate  
the general case of 
CP-nonconservation 
and we use the results of the
analyzes of the latest solar and 
atmospheric neutrino data.
For each of the  five types of 
possible neutrino mass spectrum indicated above we 
give detailed predictions for \meff
for the three solutions of the solar neutrino problem,
favored by the current solar neutrino data:
the LMA MSW, the small mixing angle (SMA) MSW,
the LOW - quasi-vacuum oscillation (LOW-QVO) one.
In each case we identify the ``just-CP-violation'' region
of values of \meff: a value of \meff in this
region would unambiguously signal the existence of
CP-violation in the lepton sector, caused by 
Majorana CP-violating phases.
Analyzing the case of CP-conservation, we 
derive predictions for \meff corresponding 
to all possible values of the
relative CP-parities of the three massive Majorana
neutrinos for the five 
different types of neutrino mass spectrum.
The possibility of cancellation 
between the different terms contributing to
\meff is investigated. We identify the 
cases when such a cancellation
is impossible and there exist
non-trivial lower bounds 
on \meff and we give these bounds. 
We find, in particular, that  
the observation of the
\betabeta-decay with a rate
corresponding to 
$\meff \gtap 0.02~$eV 
can provide unique information 
on the neutrino mass spectrum.
Combined with data
on the mass of the lightest neutrino
or/and on the type of neutrino mass spectrum, 
it can give also information 
on the CP-violation in the lepton sector,
and if CP-invariance holds - on the 
relative CP-parities
of the massive Majorana neutrinos.

  In the present article we consider
only the mixing of the three
flavour neutrinos, involving
three massive Majorana neutrinos.
We assume that the \betabeta-decay is 
induced by the (V-A) charged current weak
interaction via the
exchange of the massive Majorana neutrinos.
Results for the case of four-neutrino 
mixing will be presented
in a separate publication \cite{BPPbb4nu}.


\section{\textbf{Neutrino Masses and Mixing 
from the Existing Data}}
\label{data}

\indent We summarize in the present Section
the data on the neutrino masses, 
neutrino mass squared differences and on the
neutrino (lepton) mixing 
which will be used in our further analyzes.

     Strong evidences for neutrino 
oscillations have been obtained
in the experiments   
with atmospheric and solar neutrinos.
The Super-Kamiokande atmospheric neutrino data 
\cite{SKatm98,SKatm00}
is best interpreted in terms of (dominant)
$\nu_\mu \rightarrow \nu_\tau$
and $\bar{\nu}_\mu \rightarrow \bar{\nu}_\tau$
oscillations \cite{SKatm00}. 
Assuming two-neutrino
mixing, one obtains a description of the data
at $90~(99)\%$ C.L.
for the following values of  
the relevant two-neutrino 
oscillation parameters -
the mass-squared difference 
$\deltaatm$ and the mixing angle 
$\theta_{\mathrm{atm}}$ \cite{SKatm00}: 
\begin{gather}
2.0~(1.5) \times 10^{-3} \mathrm{eV}^2  \leq 
|\deltaatm | \leq 6.5~(8.0) \times 10^{-3} \mathrm{eV}^2, \\
\label{deltamatmo}
0.87~(0.75) \leq \sin^2 (2 \theta_{\mathrm{atm}}) \leq 1. 
\end{gather} 

\noindent The sign of $\deltaatm$ is undetermined 
by the data. The best fit value of 
$|\deltaatm |$ found in \cite{SKatm00} 
is: $|\deltaatm |_{\mathrm{BF}} = 
3.2 \times 10^{-3} \mathrm{eV}^2$.

 Global neutrino oscillation analyzes 
of the most recent solar neutrino data 
\cite{Cl98,Ab99,Ha99,SKYSuz00,GNO}
were performed, e.g., in 
refs. \cite{SKYSuz00,Fogli00,PK00,Gonza3nu}.
The new high precision Super-Kamiokande 
results \cite{SKYSuz00} on
the spectrum of the recoil electrons 
and on the day-night (D-N) effect 
were included in the analyzes.
In refs. \cite{SKYSuz00,Fogli00,PK00}
the solar neutrino data were analyzed 
in terms of the hypothesis of
two-neutrino oscillations of solar neutrinos,
characterized by the two parameters 
$\deltasol > 0$ and $\theta_{\odot}$ (or $\tan^2 \theta_{\odot}$). 
In ref. \cite{Gonza3nu} a three-neutrino
oscillation analysis was performed (see further).
The regions of the large mixing angle (LMA) MSW, 
small mixing angle (SMA)
MSW (see, e.g., \cite{SP97,BGG99}), of the LOW and the 
quasi-vacuum oscillation (QVO)
solutions of the solar neutrino problem
(see, e.g., \cite{Fogli00}), allowed by the 
data at a given C.L. 
were determined. The studies performed in 
refs. \cite{SKYSuz00,Fogli00,PK00} showed, 
in particular, that 
i) a relatively large part of the 
previously allowed region of the SMA MSW 
$\nu_e \rightarrow \nu_{\mu (\tau)}$
transition solution is ruled out by the
Super-Kamiokande data on the
recoil--e$^{-}$ spectrum and the D-N effect, 
ii) the LMA MSW $\nu_e \rightarrow \nu_{\mu (\tau)}$
transition solution provides a better
description of the data than the SMA MSW solution, and
that iii) the purely vacuum 
oscillation (VO) solution due to the
$\nu_e \leftrightarrow \nu_{\mu (\tau)}$
oscillations (see, e.g., \cite{Po67,Vosc}),
as well as the VO and MSW solutions with solar $\nu_e$ 
transitions into a sterile neutrino 
(see e.g., \cite{KLPSterile96}), $\nu_s$,
are strongly disfavored (if not ruled out) 
by the data. The Super-Kamiokande 
collaboration, using a method of analysis
which is different from those
employed in refs. \cite{Fogli00,PK00},
rules out the SMA MSW $\nu_e \rightarrow \nu_{\mu (\tau)}$
transition solution at 95\% C.L.
In Tables \ref{tabella1} and \ref{tabella2} 
we give the results for the
different solution regions derived 
in the two-neutrino 
$\nu_e \rightarrow \nu_{\mu (\tau)}$
transition/oscillation analyzes in 
refs. \cite{SKYSuz00,Fogli00}. The Super-Kamiokande
collaboration \cite{SKYSuz00} did not produce 99\% C.L.
results, so we quote and use only their 
95\% C.L. allowed regions. 

\begin{table}[h]
\caption{Values of \deltasol obtained in the analyzes of
the solar neutrino data in refs. \cite{SKYSuz00,Fogli00,Gonza3nu}
and used in the present study (see text for details).
The best fit values (B.F.V.) in the different 
solution regions are also given.}
\setlength{\extrarowheight}{4pt}
\begin{center}
\begin{tabular}{|c|c|c|c|}  \hline
\setlength{\tabcolsep}{1pt} 
\label{tabella1}
\mbox{} &           & $\deltasol [\mathrm{eV}^2]$            & B.F.V. $[ \mathrm{eV}^2]$   \\ \hline \hline
Ref. \cite{SKYSuz00}  & LMA & $ 2.5 \times 10^{-5} \div  1.3 \times 10^{-4}$    &$ 4.0 \times 10^{-5}$  \\
$(95 \% \, \mathrm{ C.L.} )$         & LOW-QVO & $ 5.6 \times 10^{-8}  \div    2.0 \times 10^{-7}$   &  \\ \hline
Ref. \cite{Fogli00}    & LMA       & $1.6 \  ( 1.6 ) \times 10^{-5} \div  1.0 \ (1.0) \times 10^{-4}$   & $4.8 \times 10^{-5}$  \\
$(90 \ (99 \%) \, \mathrm{ C.L.} )$ & SMA       & $(4.0 \times 10^{-6} \hspace{0.1cm}  \div \hspace{0.1cm} 1.0 \times 10^{-5})$   & $ 8.0 \times 10^{-6}$ \\
\mbox{} & LOW-QVO & $ 5.5 \ (0.4) \times 10^{-9} \div  1.4 \ (2.4) \times 10^{-8}$   & $1.0 \times 10^{-9}$ \\ \hline
Ref. \cite{Gonza3nu}   & LMA       & $  1.6  \ (1.5) \times 10^{-5} \div  2.0 \ (2.0) \times 10^{-4}$   & $3.5 \times 10^{-5}$  \\
$(90 \ (99 \%) \, \mathrm{ C.L.})$ & SMA       & $ 4.0 \ (3.0) \times 10^{-6}  \div  9.0 \ (10.0)  \times 10^{-6} $   & \\
 & LOW-QVO & $ 8.0 \ (5.0) \times 10^{-10} \div  3.0 \ (4.0) \times 10^{-7} $   &  \\ 
 \hline
\end{tabular}
\end{center}
\end{table}

\begin{table}
\setlength{\extrarowheight}{4pt}
\caption{Values of $\theta_\odot$ obtained in the analyzes of
the solar neutrino data in refs. \cite{SKYSuz00,Fogli00,Gonza3nu}
and used in the present study.}
\begin{center}
\begin{tabular}{|c|c|c|c|} \hline
\setlength{\tabcolsep}{1pt}
\label{tabella2}
\mbox{} &      &   $\tan^2 \theta_{\odot}$           
          & B.F.V. \\ \hline \hline
Ref. \cite{SKYSuz00}	        & LMA &
$0.21 \div 0.67$ & 0.38 \\
$(95 \%\,  \mathrm{ C.L.} )$           & LOW-QVO &
$0.3  \div 0.5$ &  \\ \hline
Ref. \cite{Fogli00}   & LMA & $  0.20 \ (0.19) \div 0.6  \ (1.0)$ & 0.35 \\
$(90 \ (99 \%) \, \mathrm{ C.L.} )$  & SMA & $(1.2 \times 10^{-4} \hspace{1mm}  \div  \hspace{1mm}  1.8 \times 10^{-3})$ & $6.0 \times 10^{-4}$ \\
\mbox{}        & LOW-QVO & $ 0.57 \ (0.42) \div 0.90 \ (3.0)$ & 0.70 \\ \hline
Ref. \cite{Gonza3nu}         & LMA & $  0.18  \ (0.16) \div 1.0  \ (4.0)$ & 0.35 \\
$(90 \ (99 \%) \, \mathrm{ C.L.} )$  & SMA & $ 2.0  \ (1.0)  \times 10^{-4} \div  2.0 \  (2.5) \times 10^{-3}$ &  \\
       & LOW-QVO & $ 0.4 \ (0.2) \div 3.0 \ (4.0)$ & \\ \hline
\end{tabular}
\end{center}
\end{table}

  Very important constraints on the oscillations of 
electron (anti-)neutrinos were obtained 
in the CHOOZ and Palo Verde disappearance 
experiments with reactor $\bar{\nu}_e$ \cite{CHOOZ,PaloV}.
This experiment was sensitive to values of
$\Delta m^2 \gtap 10^{-3}~{\rm eV^2}$,
which includes 
the corresponding atmospheric neutrino region,
eq. (5). No disappearance of the reactor
$\bar{\nu}_e$ was observed. Performing a 
two-neutrino oscillation
analysis, the following rather stringent
upper bound on the value of the
corresponding mixing angle, $\theta$,
was obtained by the CHOOZ collaboration
\footnote{The possibility of large 
$\sin^2  \theta > 0.9$
which is admitted by the CHOOZ data alone
is incompatible with the neutrino oscillation
interpretation of the solar neutrino deficit
(see, e.g., \cite{BBGK9596,BGKP96,BGG99,SPWIN99}).
Let us note also that the CHOOZ limit 
does not depend on the sign of 
$\Delta m^2$: we have assumed that 
$\Delta m^2 > 0$ for convenience.
}
at 95\% C.L. for 
$\Delta m^2 \geq 1.5 \times 10^{-3} \mathrm{eV}^2$:
\begin{equation}
\sin^2  \theta < 0.09.
\label{chooz}
\end{equation}

\noindent The precise upper limit in eq. (\ref{chooz}) 
is $\Delta m^2$-dependent. More concretely, 
it is a decreasing function of 
$\Delta m^2$ as 
$\Delta m^2$ increases up to 
$\Delta m^2 \simeq 6 \times 10^{-3}~\mathrm{eV}^2$
with a minimum value 
$\sin^2  \theta \simeq 1 \times 10^{-2}$. 
The upper limit becomes an
increasing function of 
$\Delta m^2$ when
the latter increases further up to   
$\Delta m^2 \simeq 8 \times 10^{-3}~\mathrm{eV}^2$,
where $\sin^2  \theta < 2 \times 10^{-2}$.
Let us note that this dependence
is accounted for, whenever necessary,
in our analysis. The sensitivity 
to the value of the  parameter
$\sin^2  \theta$ is expected
to be considerably 
improved by the MINOS 
experiment \cite{MINOS} in which 
the following upper limit can be reached:
\begin{equation}
 \sin^2 \theta < 5 \times 10^{-3}.
\end{equation}

 As we have already mentioned, 
a comprehensive 3-flavour 
neutrino oscillation analysis of the 
most recent solar neutrino, atmospheric neutrino 
and  CHOOZ data has recently been 
performed in ref. \cite{Gonza3nu}.
The analysis was done under the assumption
$\deltasol \ll |\deltaatm |$, suggested by the
results of the two-neutrino oscillation
studies \cite{SKYSuz00,Fogli00,PK00}.
As it follows from 
eq. (5), the indicated inequality holds for  
\begin{equation}
\deltasol \ltap 2.0\times 10^{-4} ~\mathrm{eV}^2,
\label{3nudmsol}
\end{equation}

\noindent i.e., for the SMA MSW and LOW-QVO
solutions and in most of 
the LMA MSW solution region
found in \cite{SKYSuz00,Fogli00,PK00}
(see Table 1). 
Under the condition
$\deltasol \ll |\deltaatm |$,
the solar neutrino data 
constrain the parameters 
\deltasol, $\theta_\odot$ and 
$\theta$ (see, e.g., 
\cite{SPWIN99,BGG99} and the references 
quoted therein).
It was found in ref. \cite{Gonza3nu}
that the LMA MSW, SMA MSW and
LOW-QVO solutions of the solar neutrino problem
are allowed and the 
corresponding solution regions were
determined.
The best fit point was found to lie in 
the LMA MSW solution 
region. Although the latter extends 
to values of 
$\deltasol \sim (7.0 - 8.0)\times 10^{-4}~ \mathrm{eV}^2$,
for \deltasol exceeding the upper 
bound in eq. (\ref{3nudmsol})
the reliability (accuracy) of the results 
thus derived  
is questionable. For this reason 
in our further analysis
we will use as a maximal value of 
\deltasol for the LMA MSW solution 
the value given in (\ref{3nudmsol}), 
which is reflected in Table 1 (see below).
The atmospheric neutrino 
data analysis 
restricts in the case of 
$\deltasol \ll |\deltaatm |$
the mass-squared difference $|\deltaatm |$ and 
the angles $\theta_{\mathrm{atm}}$ and 
$\theta$ (see, e.g., \cite{SPWIN99,BGG99}). 
The parameter \deltaatm is found in \cite{Gonza3nu}
to lie at 90 (99)\%~C.L. in the following interval:
\begin{equation}
1.4 \ (1.1) \times 10^{-3}~{\rm eV^2} \leq 
|\deltaatm | 
\leq 6.1 \ (7.3) \times 10^{-3}~{\rm eV^2}
\label{deltaatmgarcia}
\end{equation}
with best fit value given by 
$|\Delta m^2_{\mathrm{atm}}|_{\mathrm{BF}} = 
3.1 \times 10^{-3}~{\rm eV^2}$.
Note that the upper and the lower limits in 
eq. (\ref{deltaatmgarcia}) depend on the 
value of $\sin^2 \theta$. More specifically, 
the upper bound is a decreasing function, while 
the lower one is an increasing function 
of $\sin^2 \theta$ and the two bounds reach a common 
value at $|\Delta m^2_{\mathrm{atm}} | = 
2.0 \times 10^{-3} \eV^2$ for $\sin^2 \theta = 0.04  
\ (0.08)$ at 90 (99)\%~C.L.
Combining all the bounds it is possible 
to constrain further the value of 
$\theta$. The result at $90~(99)\%$ C.L. reads \cite{Gonza3nu}:
\begin{equation}
\sin^2 \theta < 0.05 \ (0.08).
\label{Ue3glob}
\end{equation}
\noindent 
The best fit value of  
$\sin^2 \theta$ was 
found to be \cite{Gonza3nu}:
$\sin^2 \theta_{\mathrm{BF}} = 0.005$. 
The allowed regions for the different parameters 
can be read from the plots reported in \cite{Gonza3nu}
and are given in Tables 1 and 2. 
We point out that, in general, the various 
bounds are correlated  and 
such interdependencies, whenever relevant, have
been taken into account in our analysis.

  In the next few years 
the constraints on the
values of \deltasol, $\theta_\odot$,
\deltaatm, $\theta_{\mathrm{atm}}$
and $\theta$ will be 
improved due to the increase of the 
statistics of the currently running experiments
(e.g., SAGE, GNO, Super-Kamiokande, SNO, K2K)
and the upgrade of some of them,
as well as due to the data 
from the new experiments 
BOREXINO, KamLand, MINOS and CNGS.
Hopefully, this will lead, for instance,
to the identification of a unique solution 
of the solar neutrino problem
instead of the three solutions 
allowed by the current data. 

 Let us note that if the three-flavour-neutrino 
mixing takes place,
it would be possible to determine the sign of 
\deltaatm, e.g., by studying the transitions
$\nu_{\mu}\rightarrow \nu_e$ ($\nu_{e}\rightarrow \nu_{\mu}$)
and $\bar{\nu}_{\mu}\rightarrow \bar{\nu}_e$ 
($\bar{\nu}_{e}\rightarrow \bar{\nu}_{\mu}$)
in the Earth under the condition that the 
Earth matter effects are non-negligible
\cite{signdeltaatm}. A negative 
\deltaatm  would correspond to 
neutrino mass spectrum with
inverted hierarchy or to three quasi-degenerate
neutrinos. In the inverted mass hierarchy case we
adapt the notations as to have 
$\deltaatm >0$. This leads to 
a different correspondence between
$\theta_{\odot}$,
$\theta_{\mathrm{atm}}$ and 
$\theta$ and the elements of the 
neutrino mixing matrix
in comparison with the case of a hierarchical
neutrino mass spectrum.
For the quasi-degenerate 
neutrino mass spectrum
which is considered in Section 6,
the formulae corresponding 
to the case of $\deltaatm < 0$ 
can be obtained from those 
given in Section 6 and 
corresponding to $\deltaatm > 0$,
by replacing \deltaatm with $|\deltaatm |$.
In what regards the 
\betabeta-decay, under the condition
$\deltasol \ll |\deltaatm |$
the results in the two cases coincide.

   The Troitzk~\cite{MoscowH3} and Mainz~\cite{Mainz} 
\hbeta experiments, studying the electron spectrum, 
provide information on the electron (anti-)neutrino mass  
$m_{\nu_e}$. The data contain features which 
require further investigation
(e.g., a peak in the end-point region
which varies with time \cite{MoscowH3}). 
The upper bounds given 
by the authors (at 95\%~C.L.) read:
\begin{eqnarray}
m_{\nu_e}  <  2.5 \eV~~\cite{MoscowH3},\ \ 
m_{\nu_e} <  2.9  \eV~~~\cite{Mainz}.
\label{H3beta}
\end{eqnarray}

\noindent There are prospects to 
increase the sensitivity  
of the \hbeta experiments and 
probe the region of
values of $m_{\nu_e}$
down to $m_{\nu_e} \sim (0.3 - 0.4)$ eV
\cite{KATRIN}. 

 Cosmological and astrophysical data provide
information on the sum of the neutrino masses.
The current upper bound reads (see, e.g., \cite{Cosmo}
and the references quoted therein):
\begin{eqnarray}
\sum_{j} m_{j}  \ltap  5.5 \eV.
\end{eqnarray}

\noindent The future experiments MAP and PLANCK 
can be sensitive to \cite{MAP}
\begin{eqnarray}
\sum_{j} m_{j}  \cong  0.4 \eV.
\end{eqnarray}

 In the next Sections we show that the data
from the new generation of $\betabeta-$ decay 
experiments, which will be sensitive to values of 
$\meff \gtap (0.01 - 0.10)~$eV, can provide 
unique information on the neutrino mass spectrum 
as well as on the CP-violation in the lepton sector
in the case of massive Majorana neutrinos.


\section{\textbf{The Formalism of Neutrino Mixing and \betabeta-Decay}}

\indent The explanation of the atmospheric and 
solar neutrino data in terms of 
neutrino oscillations
requires the existence of 
mixing of the three left-handed flavour neutrino 
fields, $\nu_{lL}$, $l  = e,\mu,\tau$,
in the weak charged lepton current: 
\begin{equation}
\nu_{l \mathrm{L}}  = \sum_{j=1}^{3} U_{l j} \nu_{j \mathrm{L}},
\label{3numix}
\end{equation}

\noindent where $\nu_{j \mathrm{L}}$ is the 
left-handed field of the 
neutrino $\nu_j$ having a mass $m_j$
and $U$ is a $3 \times 3$ unitary mixing matrix - 
the lepton mixing matrix. We will assume that 
the neutrinos $\nu_j$ are Majorana particles
whose fields satisfy the Majorana condition:
\begin{equation}
C(\bar{\nu}_{j})^{T} = \nu_{j},~j=1,2,3,
\label{Majcond}
\end{equation}
\noindent where $C$ is the charge conjugation matrix.
We will also assume (without loss of generality) 
that $m_1 < m_2 < m_3$.
One of the standard parametrizations of the matrix $U$ reads:

\vspace{0.4cm} \hspace{-2cm}
\begin{equation}
U \hspace{3mm} = \hspace{3mm} 
                 \begin{pmatrix}
                 c_{12} c_{13} & s_{12} c_{13} & s_{13}  \\
                 -s_{12} c_{23} - c_{12} s_{23} s_{13} e^{ i \delta} &
                 -c_{12} c_{23} - s_{12} s_{23} s_{13} e^{ i \delta} &
                  s_{23} c_{13}e^{i \delta} \\
                 s_{12} s_{23} - c_{12} c_{23} s_{13} e^{ i \delta} &
                 -c_{12} s_{23} - s_{12} c_{23} s_{13} e^{ i \delta} &
                  c_{23} c_{13}e^{i \delta} \\
                 \end{pmatrix}
                                       \mbox{diag} 
(1, e^{i \frac{\alpha_{21}}{2}}, e^{i\frac{\alpha_{31}}{2}}) 
\label{umatrix}
\end{equation}

\vspace{0.4cm}
\noindent where $c_{ij} \equiv \cos \theta_{ij}$, 
$s_{ij} \equiv \sin \theta_{ij}$, 
$0 \leq \theta_{ij} \leq \pi/2$, 
$\delta$ is the so-called 
Dirac CP-violating phase and
$\alpha_{21}$ and  $\alpha_{31}$ are 
the Majorana CP-violating phases 
\cite{BHP80} (see also, e.g, \cite{BiPe87}).

  We are interested in 
the effective Majorana mass parameter
\begin{equation}
\meff \equiv \left|~ m_1 U_{\mathrm{e} 1}^2 + 
m_2 U_{\mathrm{e} 2}^2 + m_3 U_{\mathrm{e} 3}^2~\right|,
\label{meffective}
\end{equation}

\noindent whose value is determined by 
the values of the neutrino masses $m_j$
and by the elements of the first row of the
lepton mixing matrix, $U_{ej}$. The latter  satisfy
the unitarity condition:
\begin{equation}
\sum_{j=1}^{3} ~|U_{\mathrm{e} j}|^2 = 1.
\label{Uejunit}
\end{equation}
\noindent We have 
\begin{equation}
U_{\mathrm{e} j} = |U_{\mathrm{e} j}|~e^{i{\alpha_j}\over 2},
\label{Uejphase}
\end{equation}
\noindent where $\alpha_j$, $j=1,2,3$, are three
phases. Only the phase differences
$(\alpha_j - \alpha_k) \equiv \alpha_{jk}$ ($j > k$)
can play a physical role. 
The mixing matrix in eq. (\ref{umatrix})
is written taking this into account.

\indent In the  general case, one can have 
\begin{equation}
C(\bar{\nu}_{j})^{T} = (\xi_j^{\ast})^2~\nu_{j},~j=1,2,3,
\label{Majcond2}
\end{equation}
\noindent where 
$\xi_j$, $j=1,2,3$, are three
phases. Only two combinations 
of the six phases 
$\alpha_j$ and $\xi_j$
represent physical Majorana
CP-violating phases.
The effective Majorana mass parameter
now has the form:
\begin{equation}
\meff \equiv \left|~ m_1 U_{\mathrm{e} 1}^2 \xi_1^2
+ m_2 U_{\mathrm{e} 2}^2 \xi_2^2 + 
m_3 U_{\mathrm{e} 3}^2 \xi_3^2~\right|.
\label{meffectivexi}
\end{equation}
\noindent Obviously, the two physical Majorana
CP-violating phases on which \meff depends 
are $\alpha_{21} \equiv  
{\rm arg}(U_{\mathrm{e} 2}^2 \xi_2^2)
-{\rm arg}(U_{\mathrm{e} 1}^2 \xi_1^2)$ and
$\alpha_{31} \equiv
{\rm arg}(U_{\mathrm{e} 3}^2 \xi_3^2)
-{\rm arg}(U_{\mathrm{e} 1}^2 \xi_1^2)$.
Actually, all CP-violation effects
associated with the Majorana nature of
the massive neutrinos are generated 
by $\alpha_{21} \neq k\pi$ and
$\alpha_{31} \neq k'\pi$, $k,k'=0,1,2,...$. 
Indeed,  under a rephasing of the charged lepton, l(x), and the neutrino,
$\nu_j(x)$, fields in the weak charged lepton
current, $l(x) \rightarrow e^{i \eta_l}l(x)$ \ and \ 
$\nu_j(x) \rightarrow e^{i \beta_j} \nu_j(x)$, 
the elements of the lepton mixing matrix and 
the phase factors in the Majorana condition
for the Majorana neutrino fields change as follows:
\begin{align}
U_{\mathrm{l} j} & \rightarrow 
U_{\mathrm{l} j} e^{ -i ( \eta_l - \beta_j)},~l=e,\mu,\tau,~j=1,2,3, \\
\xi_j  & \rightarrow \xi_j  e^{ -i \beta_j}.
\end{align}
%
As was shown in \cite{NievPal87}, in the lepton sector
of the theory under discussion 
with mixing of three massive Majorana neutrinos 
there exist three rephasing invariants.
The first is the standard
Dirac one, $J$, present in the case of mixing of 
three massive Dirac 
neutrinos \cite{CeciliaJ,NievPal87} (see also \cite{KP88Earth}): 
\begin{equation}
J = \mathit{Im} \ \ ( U_{\mu 2} U_{e 3} U_{\mu 3}^{\ast} U_{e2}^{\ast}).
\end{equation}
\noindent The existence of the other two, $S_1$ and $S_2$, 
is related to the Majorana nature of 
the massive neutrinos $\nu_j$ \cite{NievPal87}: 
\begin{align}
S_1  & \equiv  \mathit{Im} \ \ (  U_{e1} U_{e3}^{\ast} 
\xi_3^{\ast} \xi_1), \\
S_2  & \equiv  \mathit{Im} \ \ (  U_{e2} U_{e3}^{\ast} \xi_3^{\ast} \xi_2). 
\label{Majrephinv}
\end{align}
\noindent A geometrical representation 
of CP-violation in the lepton sector 
in terms of unitarity triangles 
in the case of three-neutrino mixing 
and presence of Majorana CP-violating
phases in the lepton mixing matrix 
is given in \cite{Branco00}.
The two  Majorana CP-violating phases 
$\alpha_{21}$ and $\alpha_{31}$
are determined by the two independent 
rephasing invariants, $S_{1,2}$. We have:
\begin{align}
\cos \alpha_{31} & =  1 - 2 \frac{S_1^2}{ |U_{\mathrm{e} 1}|^2 
|U_{\mathrm{e} 3}|^2}, \\
 \cos(\alpha_{31} -  \alpha_{21}) & =
\cos(\alpha_{3} - \alpha_{2}) =    
1 - 2 \frac{S_2^2}{ |U_{\mathrm{e} 2}|^2 |U_{\mathrm{e} 3}|^2}, \\
\intertext{and}
\cos \alpha_{21} & =  \cos(\alpha_{31} -  \alpha_{21}) \cos \alpha_{31} +  
\sin(\alpha_{31} -  \alpha_{21}) \sin \alpha_{31}.
\end{align}
%
\noindent One can express \meff in terms of 
rephasing-invariant quantities as:
%
\begin{equation}
\begin{split}
\meff^2   = &  m_1^2 |U_{\mathrm{e} 1}|^4 +  
m_2^2 |U_{\mathrm{e} 2}|^4 +  m_3^2 |U_{\mathrm{e} 3}|^4 \\
           &  + 2 m_1 m_2 |U_{\mathrm{e} 1}|^2 |U_{\mathrm{e} 2}|^2 
\cos \alpha_{21} + 
2 m_1 m_3 |U_{\mathrm{e} 1}|^2 |U_{\mathrm{e} 3}|^2 \cos \alpha_{31}\\
          & +  2 m_2 m_3 |U_{\mathrm{e} 2}|^2 |U_{\mathrm{e} 3}|^2 
\cos (\alpha_{31} -  \alpha_{21}).
\end{split}
\end{equation}
\noindent Note that \meff depends only on $S_{1,2}$ 
and does not depend on the Dirac rephasing invariant $J$.
This is a consequence of the specific choice of 
the rephasing invariants $S_{1,2}$, eq. (\ref{Majrephinv}),  
which is not unique \cite{NievPal87}. 
With this choice
the amplitude of the
$K^{+} \rightarrow \ \pi^{-} + \mu^{+} + \mu^{+}$ decay,
for instance, which, as like the \betabeta-decay, is generated 
by the exchange of the three virtual massive Majorana neutrinos
in the scheme under discussion,
depends, as can be shown, on all three rephasing 
invariants $S_{1}$, $S_{2}$ and $J$.

  If CP-invariance holds in the lepton sector 
we have, in particular, $S_1,S_2 = 0$,
or $\mathit{Re} \ (U_{e1} U_{e3}^{\ast} 
\xi_3^{\ast} \xi_1) = 0,\mathit{Re} 
(U_{e2} U_{e3}^{\ast} \xi_3^{\ast} \xi_2) = 0$.
In this case the unitarity triangles 
reduce to lines oriented either along the horizontal 
or the vertical axis on the plane \cite{Branco00}.
In terms of constraints on the phases
$\alpha_{21}$ and $\alpha_{31}$ this implies 
$\alpha_{21} = k \pi$,
$\alpha_{31} = k^\prime \pi$ with $k, 
k^\prime= 0,1,2...$.

  In all our subsequent analyzes we will
set for convenience (and without loss of generality) 
$\xi_j = 1$, j=1,2,3, i.e., we will assume that 
the fields of the Majorana neutrinos $\nu_j$ satisfy
the Majorana conditions (\ref{Majcond}).
In this case
$\alpha_{21} = \alpha_{2} - \alpha_{1}$ and 
$\alpha_{31} = \alpha_{3} - \alpha_{1}$,
where $\alpha_{j}$ are determined by eq. 
(\ref{Uejphase}).
The CP-invariance constraint on the elements
of the lepton mixing matrix of interest reads 
\footnote{This constraint is obtained 
from the requirement of CP-invariance
of the charged current weak interaction 
Lagrangian ${\cal L}^{cc}$, 
by choosing the arbitrary phase factors 
in the CP-transformation laws of the electron 
and the $W$-boson fields equal to 1.}  
\cite{BNP84,Kayser84} (see also \cite{BiPe87}):
\begin{equation}
U^{\ast}_{\mathrm{e} j} = \eta^{CP}_j~U_{\mathrm{e} j},
\label{UejCPinv}
\end{equation}
\noindent where $\eta^{CP}_j = i \phi_j = \pm i$ is 
the CP-parity of the Majorana neutrino $\nu_j$
with mass $m_j >0$.
In this case \meff is given by:
\begin{equation}
\meff \equiv 
\left| \sum_{j=1}^{3}  \eta^{CP}_j |U_{\mathrm{e} j}|^2 m_j \right| =
      \left| \sum_{j=1}^{3}  \phi_j |U_{\mathrm{e} j}|^2 m_j \right|.
\label{meffCP}
\end{equation}

  For establishing a direct relation with the
mixing angles constrained by the solar neutrino data
and the data from the CHOOZ experiment, 
it proves convenient to express 
(e.g., in the case of neutrino mass hierarchy)
$|U_{\mathrm{e} 1}|$ and $|U_{\mathrm{e} 2}|$ in the form:
\begin{equation}
|U_{\mathrm{e} 1}| = \cos \varphi \sqrt{1 - |U_{\mathrm{e} 3}|^2},
~~|U_{\mathrm{e} 2}| = \sin \varphi \sqrt{1 - |U_{\mathrm{e} 3}|^2},
\label{Ue12}
\end{equation}
\noindent where $\varphi$ is an angle and we have used the relation 
$|U_{\mathrm{e} 1}|^2 + |U_{\mathrm{e} 2}|^2 = 1 - |U_{\mathrm{e} 3}|^2$.
In certain cases (e.g., inverted neutrino mass hierarchy)
the analogous relations,
$|U_{\mathrm{e} 2}| = \cos \tilde{\varphi} \sqrt{1 - |U_{\mathrm{e} 1}|^2}$,
$|U_{\mathrm{e} 3}| = \sin \tilde{\varphi} \sqrt{1 - |U_{\mathrm{e} 1}|^2}$,
are more useful. Depending on the type of the
neutrino mass spectrum, we will have either
$\varphi = \theta_{\odot}$, $|U_{\mathrm{e} 3}|^2 = \sin^2\theta$, or 
$\tilde{\varphi} = \theta_{\odot}$, $|U_{\mathrm{e} 1}|^2 = \sin^2\theta$.

 The neutrino oscillation 
experiments, as is well-known, provide information
on $\Delta m^2_{jk} = m^2_{j} - m^2_{k}~ (j > k)$.
In the case of 3-neutrino mixing (\ref{3numix})
there are two independent $\Delta m^2$ parameters:
$\Delta m^2_{31}$, for instance, can be expressed as
$\Delta m^2_{31} = \Delta m^2_{32} +  \Delta m^2_{21}$.
Correspondingly, 
as an independent set of three neutrino mass
parameters we can choose $m_1$,
$\sqrt{\Delta m^2_{21}}$ and $\sqrt{\Delta m^2_{32}}$.
We have:
\begin{equation} 
m_{2} = \sqrt{m^2_{1} +  \Delta m^2_{21}}~,
\label{m2}
\end{equation}
\begin{equation} 
m_{3} = \sqrt{m^2_{1} + \Delta m^2_{21} +  \Delta m^2_{32}}~.
\label{m3}
\end{equation}
 The mass-squared difference 
inferred from the neutrino oscillation interpretation of the 
atmospheric neutrino data,
\deltaatm, is equal to $\Delta m^2_{31}$,
\begin{equation} 
\deltaatm = \Delta m^2_{31} = \Delta m^2_{21} +  \Delta m^2_{32}~,
\label{dmatm0}
\end{equation}

\noindent while for  the one deduced from 
the solar neutrino data, \deltasol, we 
have two possibilities:
\begin{equation}
\deltasol \equiv \deltatre  \qquad \mbox{or} 
\qquad \deltasol \equiv \deltadue.
\label{dmsol0}
\end{equation}

\noindent  Depending on the relative magnitudes of
 $m_1$, $\sqrt{\deltadue}$ and $\sqrt{\deltatre}$,  
one recovers the different possible 
types of neutrino mass spectrum:

\begin{enumerate}
\item
if $ m_1 \ll \sqrt{\deltadue} \ll \sqrt{\deltatre}$, 
one has  $m_1 \ll m_2 \ll m_3$, i.e., 
hierarchical neutrino mass spectrum;
\item
 $m_1 \ll \sqrt{\deltatre} \ll \sqrt{\deltadue}$ implies 
$m_1  \ll  m_2  \simeq   m_3$, i.e.,
neutrino mass spectrum with inverted hierarchy;
\item
for $\sqrt{\deltadue},\sqrt{\deltatre} \ll m_1$, 
we have $m_1  \simeq  m_2  \simeq   m_3$, i.e.,
quasi-degenerate neutrino mass spectrum;
\item
if $\sqrt{\deltadue} \ll \sqrt{\deltatre} \sim \boldsymbol{O}( m_1)$, 
one finds $m_1 \simeq  m_2 < m_3$ , 
i.e., spectrum with ``partial mass hierarchy''. 
This pattern of neutrino masses interpolates between 
the hierarchical one and  
the quasi-degenerate one.
\item
for $\sqrt{\deltatre} \ll \sqrt{\deltadue} \sim \boldsymbol{O}( m_1)$, 
we have $m_1 < m_2 \simeq m_3$ , 
i.e., spectrum with ``partial inverted mass hierarchy''. 
This pattern interpolates between the  
inverted mass hierarchy spectrum and the quasi-degenerate one.
\end{enumerate}

  For each of the 
possible patterns of neutrino masses 
indicated above, we will study in detail 
the implications of the data on neutrino
oscillations, obtained in the experiments
with solar and atmospheric neutrinos and in the  
CHOOZ experiment, for the searches for
\betabeta-decay.


\section{\textbf{Hierarchical Neutrino Mass Spectrum }}
\label{masshierarchy1}

\indent The hierarchical neutrino mass spectrum is
characterized by the 
following pattern of the neutrino masses
$m_j$:
\begin{equation}
m_1  \ll  m_2  \ll   m_3.
\label{hierarchy}
\end{equation}
\noindent This type of neutrino mass spectrum
is predicted by the standard versions of the 
see-saw mechanism of 
neutrino mass generation \cite{SeeSaw}.
The pattern corresponds 
to the inequalities
\footnote{The case of neutrino mass spectrum with partial hierarchy, 
$m_1  \simeq  m_2 < m_3$ will be treated in Section 7.}
\begin{equation}
m_1  \ll \sqrt{\Delta m^2_{21}} \ll \sqrt{\Delta m^2_{32}}~.
\label{hierarchy2}
\end{equation}
\noindent Using  (\ref{hierarchy}) and (\ref{hierarchy2})
it is possible to make the identification 
(see, e.g., \cite{SPWIN99,BGG99}):
\vspace{3mm}
\begin{gather}
 \Delta  m^2_{\odot}  \equiv \Delta m^2_{21},~~~~
\Delta  m^2_{\mathrm{atm}} \equiv \Delta m^2_{32}, \nonumber \\ 
\ueuno = \cos^2 \theta_\odot (1 - \uetre),  \nonumber \\
\uedue = \sin^2 \theta_\odot  (1 - \uetre), \nonumber\\
\uetre \equiv \sin^2\theta  < 0.09 ~~~~~({\rm CHOOZ}).
\label{obs1}
\end{gather}

\vspace{2mm}
\noindent We will suppose that \deltaatm 
lies in the interval (5) or 
(\ref{deltaatmgarcia}), 
\deltasol and $\theta_{\odot}$ 
take values in the regions given in Tables 1 and 2,
and that \uetre satisfies the CHOOZ upper bound. 
Equations~(\ref{hierarchy}) and (\ref{obs1}) further imply:
\begin{equation}
m_2  \simeq  \sqrt{\deltasol},~~~~~
m_3 \simeq  \sqrt{\deltaatm}~. 
\label{m2m3}
\end{equation}
\noindent Using eqs. (\ref{meffective}),
(\ref{Uejphase}),
(\ref{obs1}) and (\ref{m2m3}) 
we can express the effective Majorana mass 
parameter in terms of the quantities,
whose values are determined in the
solar and atmospheric neutrino experiments, 
and of the phase $(\alpha_3 - \alpha_2)$:

\begin{equation}
\meff \simeq \left| \sqrt{\deltasol} (1 - |U_{\mathrm{e}3}|^2) 
\sin^2 \theta_\odot + 
\sqrt{\deltaatm} |U_{\mathrm{e}3}|^2 e^{i(\alpha_3 - \alpha_2)} \right|
\label{eqmasshierarchy01}
\end{equation}
\noindent where the phases $\alpha_{2,3}$ 
are determined by eq. (\ref{Uejphase})
and we have neglected the 
contribution of the term  $\sim m_1$.
Note that although in this case
one of three massive Majorana neutrinos effectively
``decouples'' and does not give a contribution to 
\meff, the value of \meff
still depends on the Majorana CP-violating 
phase $\alpha_{32} = \alpha_{3} - \alpha_{2}$.
This reflects the fact that in contrast
to the case of massive Dirac neutrinos (or quarks),
{\it CP-violation can take place
in the mixing of only two massive 
Majorana neutrinos} \cite{BHP80}.

   Due to the presence of the first term in 
eq.~(\ref{eqmasshierarchy01}), one obtains 
different predictions for \meff
for the LMA, SMA and LOW-QVO solutions 
of the solar neutrino problem,
as the three solutions require different ranges 
of values of $\sin^2\theta_{\odot}$ and \deltasol.

 Consider first the case of CP-conservation.
There exist two possibilities.
\vskip 0.3truecm

\noindent {\bf Case A.} The neutrinos $\nu_2$ and $\nu_3$ can
have the same CP parities, i.e., $\phi_2=\phi_3$.
In this case (with the phase conventions we are 
using, see Section 3) 
$\alpha_{32} \equiv 
\alpha_{3} - \alpha_{2} = 0$
(or $\alpha_{21} = \alpha_{31} = 0, \pm \pi$),
 and the effective Majorana mass \meff is given by:
\begin{equation}
\meff \simeq \sqrt{\deltasol} \sin^2 \theta_\odot (1 - |U_{\mathrm{e} 3}|^2)  
+ \sqrt{\deltaatm} |U_{\mathrm{e} 3}|^2~.
\end{equation}

\noindent Using the results on \deltaatm, \deltasol,
$\sin^2 \theta_\odot$ and $|U_{\mathrm{e} 3}|^2$
obtained in ref.~\cite{Gonza3nu} at $90\ (99) \%$ C.L.
(quoted in Section 2 and in Tables 1 and 2),
we find that in the cases of the LMA, SMA and
LOW-QVO solutions of the solar neutrino problem
\meff is bounded to lie in the following 
intervals:
%
\begin{align}
 6.1 \ (5.3) \times 10^{-4} \ \mathrm{eV} \quad  &  \leq &  \meff \quad &  \leq &   7.4 \  
(14) \times 10^{-3}  
\ \mathrm{eV}~, & \quad \quad  \quad \text{LMA}; \\
2.0 \ (0.8) \times 10^{-7} \ \mathrm{eV} \quad &  \leq &  \meff \quad &  \leq &  1.9  \  (3.6) \times 10^{-3} \ \mathrm{eV}~, & \quad \quad  \quad\text{SMA}; \\
6.7 \  (0.5)\times 10^{-5}\  \mathrm{eV} \quad &  \leq &  \meff \quad  & \leq &  3.5 \  (3.6) \times 10^{-3}  \ \mathrm{eV}~, & \quad \quad \quad  \text{LOW-QVO}. 
\label{3nuh++}
\end{align}
\noindent For the values of the parameters
providing the best fit of the solar 
and atmospheric neutrino data \cite{Gonza3nu} one finds:
$\meff_{\mathrm{BF}} = 1.8 \times 10^{-3}~ \mathrm{eV}$. 
Note, that in the case of the LMA solution, \meff 
can reach 
$\meff \cong (1 - 2) \times 10^{-2}~\mathrm{eV}$.
If the LMA solution admits values of 
$\deltasol \sim (7 - 8) \times 10^{-4}~\mathrm{eV^2}$,
as  suggested in \cite{Gonza3nu},
\meff can be as large as $\sim 3 \times 10^{-2}~\mathrm{eV}$. 
This is the maximal value $\meff$ can have in the case of 
hierarchical neutrino mass spectrum. 
Values of $\meff \gtap 10^{-2}~\mathrm{eV}$
can be tested in the next generation of 
\betabeta-decay experiments, as we have discussed in the 
Introduction.

    The maximal allowed values of \meff,
${\rm max}~(\meff$), in the case of the SMA and LOW-QVO 
solutions are by a factor of $\sim (2 - 4)$ smaller than 
that for the LMA solution. This is related 
to the fact that for  
the SMA and LOW-QVO 
solutions 
\begin{equation}
{\rm max}~(\meff) =
{\rm max}~(\sqrt{\deltaatm} |U_{\mathrm{e}3}|^2),
\label{hmaxmeffSMA}
\end{equation}
\noindent while for the LMA solution 
$\sqrt{\deltasol} \sin^2\theta_{\odot}$ can be as large as 
$\sim 1.1 \times 10^{-2}~\mathrm{eV}$
and both terms in eq. (\ref{eqmasshierarchy01}) contribute to
${\rm max}~(\meff)$.

   The minimal allowed values of \meff,
${\rm min}~(\meff)$, given in eq. (46) - (\ref{3nuh++}),
are exceedingly small. Let us note nevertheless
that for the given solution of the solar neutrino problem 
they correspond to 
$m_1$ having a value much smaller than 
the ${\rm min}~(\meff)$ in (46) - (\ref{3nuh++}),
i.e., to $m_1 \ll 7.0 \times 10^{-4}~\mathrm{eV}$
for the LMA solution, etc. If, however,
$m_1 \ll \sqrt{\deltasol}$, but 
$m_1 |U_{\mathrm{e} 1}|^2 = 
m_1(1 - |U_{\mathrm{e} 3}|^2)\cos^2\theta_{\odot}$ 
is of the order of, or larger than,
some of the ${\rm min}~(\meff)$ given 
in (46) -  (\ref{3nuh++}), the corresponding 
lower bounds will be modified.
This can happen in the case of the SMA MSW solution,
while ${\rm min}~(\meff)$ for the
LMA and LOW-QVO solutions are practically
stable with respect to ``finite'' $m_1$ 
corrections. Indeed, for the 
LMA (LOW-QVO) solution,
$m_1 \ll \sqrt{\deltasol}$ implies
$m_1 \ltap  4.0~(0.025) \times 10^{-4}~\mathrm{eV}$
and thus
$m_1|U_{\mathrm{e} 1}|^2 \ltap 2.0~(0.013) 
\times 10^{-4}~\mathrm{eV}$, 
which can change the relevant 
values of ${\rm min}~(\meff)$ in 
eq. (46) (eq. (\ref{3nuh++}))
by not more than $\sim 30\%$.
In contrast, in the case of the MSW solution
for which $\cos^2\theta_{\odot} \cong 1$,
the inequality 
$m_1 \ll \sqrt{\deltasol}$ is satisfied for
$m_1 \ltap  1.8 \times 10^{-4}~\mathrm{eV}$.
Correspondingly,
$m_1(1 - |U_{\mathrm{e} 3}|^2)\cos^2\theta_{\odot} \ltap 
1.8 \times 10^{-4}~\mathrm{eV}$
and ${\rm min}~(\meff)$ can range from 
$\sim 1.8 \times 10^{-4}~\mathrm{eV}$
to $0~\mathrm{eV}$. Let us note also that 
in the case of the SMA solution,
a value of $\meff \gtap 8.0 \times 10^{-4}~\mathrm{eV}$ 
would imply (for a hierarchical neutrino mass spectrum)
that  
\begin{equation}
\hspace{-5.5cm} {\rm SMA},~\meff \gtap 
8.0 \times 10^{-4}~\mathrm{eV}: \hspace{0.5cm}
\meff \simeq \sqrt{\deltaatm} |U_{\mathrm{e}3}|^2.
\label{hmaxmeffSMA2}
\end{equation}

  Using the results of the two-neutrino oscillation analyzes
of refs. \cite{SKYSuz00,Fogli00} (see Tables 1 and 2),
the results on \deltaatm from ref. \cite{SKatm00}, 
eq. (\ref{deltamatmo}),
and the upper limit on $|U_{\mathrm{e}3}|^2$, 
obtained in the CHOOZ experiment \cite{CHOOZ},
we get the following intervals of 
possible values of \meff :
%
\begin{align}
 0.8 \times 10^{-3} \ \mathrm{eV} \quad &  
\leq & \meff \quad & \leq &  7.6 \times 10^{-3}  
\ \mathrm{eV}~, & \quad \quad  \quad  \text{LMA}~~\cite{SKYSuz00,SKatm00,CHOOZ}; \nonumber\\
5.0 \times 10^{-5}\  \mathrm{eV} \quad &  \leq &  \meff \quad &  \leq & 3.6 \times 10^{-3}  \ \mathrm{eV}~, & \quad  \quad  \quad  \text{LOW-QVO}~~\cite{SKYSuz00,SKatm00,CHOOZ}, 
\label{SKh++}
\end{align}
\noindent and 
%
\begin{align}
 6.7 \ (5.4) \times 10^{-4} \ \mathrm{eV} \quad  &  \leq &  \meff  \quad & \leq &  7.4 \  
(8.5) \times 10^{-3}  
\ \mathrm{eV}~, & \quad \quad  \quad  \text{LMA}~~\cite{SKatm00,Fogli00,CHOOZ}; \nonumber\\
(2.4 \times 10^{-7} \ \mathrm{eV} \quad &  \leq &  \meff \quad  &  \leq &  4.0
 \times 10^{-3} \ \mathrm{eV})~, & \quad \quad  \quad  \text{SMA}~~
\cite{SKatm00,Fogli00,CHOOZ};  \nonumber\\
2.7 \  (0.6)\times 10^{-5}\  \mathrm{eV} \quad &  \leq & \meff \quad  &  \leq &  3.5 \  (4.0) \times 10^{-3}  \ \mathrm{eV}~, & \quad \quad  \quad   \text{LOW-QVO}~~
\cite{SKatm00,Fogli00,CHOOZ}, 
\label{Foglih++}
\end{align}

\noindent where the numbers (the numbers in brackets) 
correspond to the 90\% (99\%)~C.L. solutions 
of the $\nu_\odot$-problem in \cite{Fogli00}.

  The values of  ${\rm max}~(\meff)$ 
for the LMA solution derived 
in the two-neutrino oscillation 
analyzes \cite{SKYSuz00,Fogli00}
do not exceed $8.5 \times 10^{-3}~\mathrm{eV}$
and are smaller than the corresponding 
one for the LMA solution
obtained in the three-neutrino oscillation
analysis \cite{Gonza3nu}.
This difference is due to the difference in the
maximal allowed values of 
$\sqrt{\deltasol} (1 - |U_{\mathrm{e}3}|^2) 
\sin^2 \theta_\odot$ in the two cases. 
The minimal $\meff $ 
for the SMA solution in eq. 
(\ref{Foglih++})
is again unstable with respect to
``finite'' $m_1$ corrections 
and actually can have any value from 
$\sim  10^{-4}~\mathrm{eV}$
to $0~\mathrm{eV}$.

\vskip 0.3truecm
\noindent {\bf Case B.} $\nu_2$ and $\nu_3$  
have opposite CP parities, i.e., $\phi_2 = - \phi_3$. 
We have $\alpha_{32} = \pm \pi$ (or 
$\alpha_{21} = \alpha_{31} + \pm \pi  = 0, \pm \pi$), 
and the effective Majorana mass parameter is given by:
\begin{equation}
\meff = \left| \sqrt{\deltasol} 
\sin^2 \theta_\odot (1 - |U_{\mathrm{e} 3}|^2) - 
\sqrt{\deltaatm} |U_{\mathrm{e} 3}|^2 \right|.
\label{hmeffoppcp}
\end{equation}
\noindent In this case there exists the possibility
of cancellation \cite{LW81} between the two terms in 
eq. (\ref{hmeffoppcp}). Correspondingly, one can have 
$\meff \simeq 0~$eV for all 
solutions of the solar neutrino problem.
 
 The 90 (99)\%~C.L. results 
of the three-neutrino oscillation analysis 
of ref.~\cite{Gonza3nu} imply the
following ranges of allowed values of \meff:
\begin{align}
0 \quad \leq  \quad \meff \leq &  \  6.4 \  (10) \times 10^{-3} \ \mathrm{eV}~, 
& \quad  \text{LMA}; \nonumber  \\
0  \quad \leq  \quad \meff \leq &  \  1.8 \  (3.1) \times 10^{-3}  \ \mathrm{eV}~, 
& \quad  \text{LOW-QVO}; \nonumber  \\
0 \quad \leq  \quad \meff \leq &  \  1.9 \  (3.5) \times 10^{-3}  \ \mathrm{eV}~, 
& \quad \text{SMA}. 
\end{align}
\noindent To the best fit values of the parameters 
found in \cite{Gonza3nu} (the best fit
\deltasol and $\sin^2\theta_{\odot}$
lie in the LMA solution region), 
there corresponds  $\meff_{\mathrm{BF}} = 
1.2 \times 10^{-3} \mathrm{eV}$.
 
 The results of the  two-neutrino oscillation analyzes 
of refs. \cite{SKYSuz00} and \cite{Fogli00}, and 
of \cite{SKatm00,CHOOZ}
lead to the following possible values of
\meff:
\begin{align}
0 \leq  \meff \leq &  \ 4.5\times 10^{-3} \ \mathrm{eV}~, 
& \quad  \text{LMA}~~\cite{SKYSuz00,SKatm00,CHOOZ}; \nonumber  \\
0 \leq  \meff \leq &  \ 3.5 \times 10^{-3}  \ \mathrm{eV}~, 
& \quad  \text{LOW-QVO}~~\cite{SKYSuz00,SKatm00,CHOOZ},
\end{align}
\noindent and
\begin{align}
0 \leq  \meff \leq  &  \  3.7 \  (5.0) \times 10^{-3} \ \mathrm{eV}~, 
& \quad  \text{LMA}~~\cite{SKatm00,Fogli00,CHOOZ}; \nonumber \\
(0 \leq  \meff \leq &  \  4.0 \times 10^{-3}  \ \mathrm{eV})~, 
& \quad  \text{SMA}~~\cite{SKatm00,Fogli00,CHOOZ};\nonumber \\
0 \leq  \meff \leq &  \  3.5 \  (4.0) \times 10^{-3}  \ \mathrm{eV}~, 
& \quad \text{LOW-QVO}~~\cite{SKatm00,Fogli00,CHOOZ}. 
\end{align}
 We see that the 99\% C.L. results of the three-neutrino
oscillation analysis \cite{Gonza3nu} allow values of 
${\rm max}~(\meff) \simeq 10^{-2}~\mathrm{eV}$
for the LMA solution.
Using the results of the two-neutrino oscillation analyzes
\cite{SKYSuz00,SKatm00,Fogli00,CHOOZ}
one gets ${\rm max}~(\meff) \ltap 5.0\times 10^{-3}~\mathrm{eV}$.

 The possibility of partial or complete 
cancellation between the two terms in eq. (\ref{hmeffoppcp})
for \meff is extremely important in view of 
the future searches for the \betabeta-decay.
Barring finite $m_1$ corrections which can be relevant
in the case of the SMA solution 
of the solar neutrino problem,
the cancellation can take place if
$|U_{\mathrm{e}3 }|^2$
has a value given by:
\begin{equation}
\hspace{-5cm} \meff = 0: \hspace{1.7cm} ~~~~|U_{\mathrm{e}3 }|^2_{0} =  
\  \frac{\sqrt{\deltasol}\sin^2 \theta_\odot}{ \sqrt{\deltasol}
\sin^2 \theta_\odot + \sqrt{\deltaatm}}~~.
\label{meff0mh}
\end{equation}

\noindent For the ranges of allowed values of
\deltasol, $\sin^2\theta_{\odot}$ for the three
solutions of the $\nu_{\odot}-$problem, and of
\deltaatm, found in \cite{Gonza3nu},
the corresponding values of 
$|U_{\mathrm{e}3 }|^2_{0}$
for which the cancellation can
occur belong to the intervals:
%
\begin{align}
  8.0 \ (8.0) \times 10^{-3} \quad   & \leq  &  |U_{\mathrm{e}3} |^2_{0} \quad  &
\leq  & 1.3  \  (3.7) \times 10^{-1}  & \quad  \quad  \quad  \quad  \quad  \quad \quad  \quad  \quad \text{LMA}, \nonumber\\
  1.0 \ (0.3) \times 10^{-5} \quad  & \leq  & |U_{\mathrm{e}3} |^2_{0} \quad   &
 \leq  & 7.0 \  (20) \times 10^{-5}  & \quad \quad  \quad  \quad  \quad  \quad \quad  \quad  \quad \text{SMA}, \nonumber \\
  8.0  \ (0.7) \times 10^{-4} \quad  & \leq   & |U_{\mathrm{e}3} |^2_{0} \quad  &   
\leq  & 0.5 \  (13) \times 10^{-2}  & \quad  \quad  \quad \quad  \quad  \quad  \quad  \quad  \quad \text{LOW-QVO}.
\end{align}
\noindent
These regions for $|U_{\mathrm{e}3 }|^2_{0}$ 
overlap with the experimentally 
allowed one and, as we noticed earlier, 
there can be a complete 
cancellation between the different contributions 
in \meff for all the solutions of the solar neutrino problem. 
Note, however, that the parameters 
entering into the expression (\ref{meff0mh})
for $|U_{\mathrm{e}3 }|^2_{0}$ 
are not related, in general, 
 and therefore the
complete cancellation of the terms in
the expression for \meff seems to require
a fine tuning or the existence
of a specific symmetry 
(see, e.g., \cite{LW81,SPZee82,BiPe87}). 
Note also that the cancellation cannot occur for 
the best fit 
value of $|U_{\mathrm{e}3}|^2 = 0.005$ (LMA)
found in \cite{Gonza3nu}.
One arrives to analogous conclusions
performing this analysis exploiting the 
two-neutrino oscillation 
results of refs. \cite{SKYSuz00,SKatm00,
Fogli00,CHOOZ}.

   The MINOS experiment \cite{MINOS} currently under preparation
will be able to search for $\nu_{\mu} \rightarrow \nu_{e}$ 
transitions and can probe values of 
$|U_{\mathrm{e}3}|^2 \geq 5 \times 10^{-3}$.
These data can provide information on the possibility
of cancellation of the two terms in 
\meff only for the case of the LMA solution of 
the solar neutrino problem. 
If it is found that 
$|U_{\mathrm{e}3}|^2 \gtap 8 \times 10^{-3}$,
the two terms in eq. (\ref{hmeffoppcp})
for \meff can cancel and one can have
$\meff \cong 0~$eV. If, on the other hand,
$|U_{\mathrm{e}3}|^2 \ltap 8 \times 10^{-3}$,
there cannot be a complete cancellation.
However, the dominant contribution to \meff,
eq. (\ref{hmeffoppcp}), 
would be given by the term 
$\sqrt{\deltasol}(1 - 
|U_{\mathrm{e}3}|^2) \sin^2 \theta_\odot $
and in this case \meff would be relatively small:
\begin{equation}
\hspace{-4.5cm} |U_{\mathrm{e}3}|^2 \ltap 8 \times 10^{-3}:\hspace{7mm} ~~~~~~~~
\meff \sim (\mathrm{few} \times 10^{-4} 
\div \mathrm{few} \times 10^{-3}) \ \mathrm{eV}.
\end{equation}

  Our results are summarized in 
Fig.~\ref{figure:gmassh01}, Fig.~\ref{figure:gmassh02} 
and Fig.~\ref{figure:gmassh03} in which we show
the allowed 
ranges of \meff as a function of $|U_{\mathrm{e}3}|^2$ 
for the LMA, LOW-QVO and SMA solutions, respectively, 
found in ref.~\cite{Gonza3nu}. 
The figures are obtained for
any possible values of the CP-violating phase 
$(\alpha_3 - \alpha_2)$.
The regions of allowed values of \meff
in the cases of CP-conservation ($\phi_2 = \pm \phi_3$, or
$\alpha_3 - \alpha_2 = 0,\pi$)
cover completely that in the case of 
CP-violation ($(\alpha_3 - \alpha_2) \neq k\pi,~ k=0,1,...$). 
This is mainly due to the uncertainties in the value 
of \deltasol. With the improvement of the precision  
on \deltasol it will be possible to 
restrict the allowed 
ranges of values of \meff in the indicated cases.
The latter can lead to a separation between the
allowed regions of \meff, corresponding to the 
cases of CP-conservation and of CP-violation.
Let us note also that the cancellation
leading to $\meff \cong 0~$eV  can 
be present in the case of CP-violation as well.

  Equation (\ref{eqmasshierarchy01}) permits to express 
the cosine of the  CP-violating phase
($\alpha_3 - \alpha_2$) in terms of measurable quantities:  
\begin{equation}
\cos(\alpha_3 - \alpha_2) = \frac{ \meff^2 - \, \deltasol \,  
\sin^4 \! \theta_\odot (1 - |U_{\mathrm{e}3}|^2)^2  - \deltaatm \, 
(|U_{\mathrm{e}3}|^2)^2}{2 \sqrt{\deltasol} \sqrt{\deltaatm} \, 
\sin^2 \! \theta_\odot |U_{\mathrm{e}3}|^2 (1-|U_{\mathrm{e}3}|^2)}.
\end{equation}
\noindent
Thus, if \meff and $|U_{\mathrm{e}3}|^2$ 
will be found to be nonzero
and their values (together
with the values of \deltasol, \deltaatm and
$\sin^2\theta_{\odot}$) will be determined
experimentally with sufficient precision,
one can get direct information
about the CP-violation
in the lepton sector, caused by Majorana
CP-violating phases.
In Fig.~\ref{figure:gmassh04} we show 
the allowed range of $\cos(\alpha_2 - \alpha_3)$ 
as a function of \meff for different values of 
$|U_{\mathrm{e}3}|^2$, using the best fit values 
for \deltaatm, \deltasol and $\theta_\odot$ 
from \cite{Gonza3nu}.

Let us note that even if it is found
that $\alpha_3 - \alpha_2 = 0, \pm \pi$, 
this will not necessarily mean there is
no CP-violation due to the Majorana CP-violating phases
in the lepton sector 
because the second of the two physical 
CP-violating phases is not constrained 
and can be a source of CP-violation
in other $\Delta L = 2$ processes.

   It follows from the analysis presented in this Section 
that in the case of 3-neutrino mixing and 
a hierarchical neutrino mass spectrum,
the values of \meff compatible with 
the neutrino oscillation/transition interpretation of the
solar and atmospheric neutrino data and with the CHOOZ limit
are smaller than $5\times 10^{-3}~$eV for the SMA and 
LOW-QVO solutions of the solar neutrino problem.
For the LMA solution one can have 
${\rm max}(\meff) \cong (1 - 2)\times 10^{-2}~$eV.
If the exchange of three 
relatively light Majorana neutrinos
{\it is the only mechanism generating 
the \betabeta-decay},
an observation of the latter with 
a lifetime corresponding to
$\meff \gtap 10^{-2}~$eV
would practically rule out a 
hierarchical neutrino mass spectrum for
the SMA and LOW-QVO solutions;
an experimentally measured value of 
$\meff \gtap 2\times 10^{-2}~$eV
would strongly disfavor 
the hierarchical neutrino mass spectrum
for the LMA solution as well.
In such a situation one would be led to conclude that
at least two of the three massive Majorana neutrinos
are quasi-degenerate in mass.


\section{\textbf{Inverted Mass Hierarchy Spectrum}}

\indent The inverted mass hierarchy spectrum is characterized 
by the following relation between the neutrino masses $m_j$:
\begin{equation}
m_1  \ll  m_2  \simeq   m_3.
\label{invmhspec}
\end{equation}
\vskip -0.1truecm
The identification with the neutrino oscillation
parameters probed in the solar and atmospheric neutrino
experiments and in CHOOZ reads 
(see, e.g., \cite{BGKP96,SPWIN99,BGG99}):
%
\begin{gather}
\Delta  m^2_{\odot} \equiv \Delta m^2_{32}, \nonumber \\
\Delta  m^2_{\mathrm{atm}} \equiv \Delta m^2_{31} \simeq \Delta m^2_{21}, 
\nonumber\\
\ueuno = \sin^2\theta < 0.09~~~~({\rm CHOOZ}),\nonumber\\
\uedue = \cos^2 \theta_\odot (1- \ueuno),\nonumber \\
\uetre = \sin^2 \theta_\odot (1- \ueuno) . 
\label{invparam}
\end{gather}
\vspace{1mm}
\noindent We also have:
\begin{equation}
m_2 \simeq m_3 \simeq \sqrt{\Delta  m^2_{\mathrm{atm}}}.
\label{invm23}
\end{equation}
\noindent The inverted mass hierarchy spectrum can also
be defined by the inequalities
\begin{equation}
m_1 \ll~(<)~\sqrt{\Delta  m^2_{32}} \ll \sqrt{\Delta  m^2_{21}}.
\label{invm1}
\end{equation}

 The term $m_1 |U^2_{\mathrm{e} 1}|$ in the expression for \meff,
eq. (\ref{meffective}), can be neglected, 
since $m_1 \ll m_{2,3}$ and $|U^2_{\mathrm{e} 1}| \ll 1$.
This approximation would be valid 
as long as the sum of the two other
terms in eq. (\ref{meffective}) exceeds 
$\sim 5.0\times 10^{-3}~\mathrm{eV}$.  
Under the assumption that $m_1 |U^2_{\mathrm{e} 1}|$
gives a negligible contribution to \meff we have 
\cite{BGKP96,BGGKP99}:
\begin{equation}
\meff \simeq \left| |U_{\mathrm{e} 2}|^2 m_2 + 
|U_{\mathrm{e} 3}|^2 m_3 e^{i(\alpha_3 - \alpha_2)} \right| 
\label{meffinvmh1}
\end{equation}
\begin{equation}
= m_{2,3} \sqrt{1 - 4 |U_{\mathrm{e} 2}|^2 | U_{\mathrm{e} 3}|^2 
\sin^2\left( \frac{\alpha_3 - \alpha_2}{2} \right)} 
\label{meffinvmh2}
\end{equation}
\begin{equation}
= \sqrt{\deltaatm} (1 - \ueuno) \sqrt{1 - \sin^22\theta_{\odot} 
\sin^2\left( \frac{\alpha_3 - \alpha_2}{2} \right)},
\label{meffinvmh3}
\end{equation}
where we have used (\ref{invparam}) and (\ref{invm23}).
We see again that even though one of the
three massive Majorana neutrinos 
``decouples'', the value of \meff
depends on the Majorana CP-violating phase
$(\alpha_3 - \alpha_2)$ \cite{BHP80}.
In contrast, there cannot be CP-violation effects 
in the mixing of {\it only two} 
massive Dirac neutrinos. 

  It follows, e.g.,  from eq. (\ref{meffinvmh3}) that 
\meff satisfies the inequalities \cite{BGKP96,BGGKP99}:
\begin{equation}
\sqrt{\deltaatm}(1 - \ueuno) | \cos 2 \theta_\odot|
\leq \meff \leq \sqrt{\deltaatm}(1 - \ueuno).
\label{meffinvmhbs}
\end{equation}
\noindent The upper and the lower limits
in (\ref{meffinvmhbs})
correspond respectively to the 
CP-conserving cases $\phi_2=\phi_3$ 
($\alpha_3 - \alpha_2 = 0$, or
$\alpha_{21} = \alpha_{31} = 0, \pm \pi$) 
and $ \phi_2= - \phi_3$ 
($\alpha_3 - \alpha_2 = \pm \pi$,
or $\alpha_{21} = \alpha_{31} + \pi = 0, \pm  \pi$).

   If CP-invariance holds in the leptonic sector, the 
parameters which determine \meff are
i) $\deltaatm$, on the 
possible values of which 
there's a general 
agreement, ii) $(1 - \ueuno) > 0.91$
and iii) $|\cos 2 \theta_\odot|$, if $\phi_2= - \phi_3$, 
whose allowed range varies with the analysis
(see Table 2). Consider the two cases of CP-conservation.

\vskip 0.3truecm
\noindent {\bf Case A.} If $\nu_2$ and $\nu_3$ 
have the same CP parities, $\phi_2=\phi_3$ 
($\alpha_{32} = 0$, or
$\alpha_{21}=\alpha_{31}=0, \pm \pi$), the effective 
Majorana mass \meff does not depend 
on $|\cos 2\theta_\odot|$,
\begin{equation}
\meff = \sqrt{\deltaatm}(1 - \ueuno). 
\label{meffinvmh++}
\end{equation}
Exploiting the 90\% (99)\%~C.L. results of
ref.~\cite{Gonza3nu} we find that \meff can take the values 
\begin{equation*}
 3.7  \ (3.3) \times 10^{-2} \eV \leq \meff 
\leq 6.8 \  (8.1) \times 10^{-2} \  \mathrm{eV}.
\label{mmm1}
\end{equation*}
The ``best fit'' value is found to be 
\begin{equation*}
\meff_{\mathrm{BF}} = 5.6 \times 10^{-2} \  \mathrm{eV}. 
\label{meffinvmhbf3nu}
\end{equation*}
\noindent For the results 
obtained in 
ref.~\cite{SKatm00} and 
ref.~\cite{CHOOZ}
 we get:
%
\begin{equation*}
 4.4 \ (3.6) \times 10^{-2} \eV \leq \meff 
\leq 8.1 \  ( 8.9) \times 10^{-2}  \eV.
\end{equation*}

\noindent All indicated  values of \meff can  
be probed  in the next generation of \betabeta-decay experiments
like GENIUS, EXO, etc.

\vskip 0.3truecm
\noindent {\bf Case B.} For $\nu_2$ and $\nu_3$  having opposite 
CP parities, $\phi_2 = - \phi_3$
($\alpha_{32} = \pm \pi$, or
$\alpha_{21} = \alpha_{31} + \pm \pi  = 0,  \pm \pi$), 
we have 
\begin{equation*}
\meff = \sqrt{\deltaatm} (1 - \ueuno)|\cos 2 \theta_\odot|.
\label{meffinvmh+-}
\end{equation*}

\noindent The ranges of values \meff can have in this case 
are reported in Table~\ref{table3}.
The ``best fit'' value, according to the results 
in ref.~\cite{Gonza3nu},
is  $\meff_{\mathrm{BF}} = 
2.1 \times 10^{-2} \mathrm{eV}$ 
(with the best fit \deltasol and $\sin^2\theta_{\odot}$
lying in the LMA solution region). 
Note that for the  SMA solution
one has $\cos2\theta_{\odot} \simeq 1$
and \meff has the same value in the two cases
$\phi_2 = -  \phi_3$ and $\phi_2 = \phi_3$.

\begin{table}[H] 
\setlength{\extrarowheight}{4pt}
\caption{Values \meff compatible with the 
neutrino oscillation/transition
interpretation of the solar and atmospheric neutrino data
and with the CHOOZ limit in the case of
neutrino mass spectrum with inverted hierarchy, 
eq. (\ref{invmhspec}),
CP-conservation and $\phi_2 = - \phi_3$ (see text for details).
}
\label{table3}
\begin{center}
\begin{tabular}{|c|c|c|} \hline 
Data from &     & \meff [eV]  \\ \hline \hline
Ref. \cite{SKYSuz00,SKatm00,CHOOZ}     & LMA & $ 7.3 \times 10^{-3}  \div 5.7 \times 10^{-2}   $ \\ 
$(95 \%\,  \mathrm{ C.L.} )$      & LOW-QVO  & $  0.9 \times 10^{-2}  \div 4.5 \times 10^{-2} $  \\ \hline
  Ref. \cite{SKatm00,Fogli00,CHOOZ}   & LMA & $  1.1\times 10^{-2}~(0.0) \div 5.3   \  (5.9) \times 10^{-2} $ \\ 
 $(90 \% \ (99 \%) \, \mathrm{ C.L.} )$    & SMA & $ (3.6 \times 10^{-2} \div  8.9 \times 10^{-2}) $  \\ 
\mbox{}     & LOW-QVO  & $   2.2\times 10^{-3}~(0.0) \div 2.1  \  (3.7)  \times 10^{-2}$  \\ \hline
Ref. \cite{Gonza3nu}     & LMA & $ 0  \ ( 0)  \div 4.8  \  (5.8) \times 10^{-2} $ \\
 $(90 \% \ (99 \%) \, \mathrm{ C.L.} )$    & SMA & $  3.7 \ (3.3) \times 10^{-2}  \div 6.8  \    ( 8.1) \times 10^{-2} $  \\ 
\mbox{}     & LOW-QVO  & $   0 \ (0)  \div 3.4  \  (5.4)  \times 10^{-2}$  \\ \hline
\end{tabular}
\end{center}
\end{table}

  The results derived in the present Section are
illustrated and summarized
in Figs. \ref{figuramassinv01},~\ref{figuramassinv02}, 
\ref{figuramassinv05} and \ref{figuramassinv06}.
In Figs.~\ref{figuramassinv01} and \ref{figuramassinv02} 
we show the allowed regions of \meff 
for the LMA and LOW-QVO solutions, respectively,
as a function of $\sqrt{\deltaatm}$. It is interesting to
note that 
there is a region, marked by dark-grey color, 
which can be spanned {\it only in the presence of CP-violation},
i.e., this is a ``just-CP-violation'' region. 
Thus, for the LMA and LOW-QVO 
solutions of the solar neutrino problem
and neutrino mass spectrum of
the type (\ref{invmhspec}), an
experimentally measured value of 
\meff lying in the indicated region
will signal the existence of CP-violation in the 
leptonic sector, caused by Majorana 
CP-violating phases. A similar region is present
and the above conclusion remains valid if 
we perform the analysis using the results of ref.~\cite{Fogli00} 
and  ref.~\cite{SKYSuz00} for the LMA and LOW-QVO solutions, 
as Figs.~\ref{figuramassinv05} and \ref{figuramassinv06} 
demonstrate. 

  It would be practically impossible to 
obtain information on the CP-violation in the leptonic
sector due to the Majorana CP-violating phases 
by measuring \meff if the SMA solution of 
the solar neutrino problem
turns out to be the valid one.

 For all solutions of the solar neutrino problem
we have $\meff \ltap (0.08 - 0.09)~$eV for the 
inverted mass hierarchy neutrino mass spectrum.
In large allowed regions of the corresponding
parameter space we have $\meff \gtap (0.01 - 0.02)~$eV,
which can be tested in the future \betabeta-decay 
experiments like GENIUS, EXO, etc.

  As the dependence of \meff on $\cos 2 \theta_\odot$ 
is rather strong, in Fig.~\ref{figure:massinv07} 
we plot \meff versus  $\cos 2 \theta_\odot$.
The region of values of \meff,
corresponding to the 
``just-CP-violation'', is marked by the dark-grey color. 
The uncertainty 
in the allowed ranges of \meff 
for the two different 
values of the relative CP-parity of
$\nu_2$ and $\nu_3$ 
is due mainly to the uncertainty in the 
value of \deltaatm. If the  value of
\deltaatm is better constrained, 
and \meff is found to lie in the 
regions corresponding to the LMA 
or LOW-QVO solution,  
it would be possible 
to establish in the case of the 
mass spectrum (\ref{invmhspec}) if CP is violated
and to get information on 
the value of one of the two CP-violating phases
$(\alpha_3 - \alpha_2)$. 

 Equation (\ref{meffinvmh3}) permits 
to relate the value of  
$\sin^2 (\alpha_3 - \alpha_2)/2$ to the experimentally 
measured quantities
\meff, \deltaatm and $\sin^22\theta_{\odot}$: 
\begin{equation}
\sin^2  \frac{\alpha_3 - \alpha_2}{2}  = 
\left( 1 - \frac{\meff^2}{\deltaatm (1 - \ueuno )^2} \right) 
\frac{1}{\sin^2 2 \theta_\odot}.
\end{equation}
A more precise 
determination of \deltaatm and  $\theta_\odot$ 
and a sufficiently accurate measurement of \meff 
could allow to get information about the value of 
$(\alpha_3 - \alpha_2)$,
provided the neutrino mass spectrum is of
the inverted hierarchy type (\ref{invmhspec}).  
This is illustrated in Fig.~\ref{figuramassinv03} 
for the cases of the LMA and LOW-QVO solutions.
We would like to point out also
that, as in the case of a hierarchical 
neutrino mass spectrum,
even if one finds using the 
\betabeta-decay data that the phase
$(\alpha_3 - \alpha_2) = 0, \pm \pi$
and does not violate CP-invariance,
the second phase,
$(\alpha_2 - \alpha_1) \equiv \alpha_{21}$,
is not constrained and can be a source of 
CP-violation in other $\Delta L = 2$
processes.

  The results derived in the present Section show that
in the case of massive Majorana neutrinos and 
neutrino mass spectrum of the
inverted hierarchy type, eq. (\ref{invmhspec}),
the description of the solar and atmospheric neutrino data
in terms of neutrino oscillations/transitions and  
the CHOOZ limit imply an upper limit on the 
effective Majorana mass parameter:
$\meff \ltap (8 - 9)\times 10^{-2}~$eV.
A measured value of $\meff \gtap (2 - 3)\times 10^{-1}~$eV
would strongly disfavor, under the general 
assumptions of the present study
(3-neutrino mixing, \betabeta-decay generated 
only by the charged (V-A) current weak interaction 
via the exchange of the three Majorana neutrinos),
the possibility of inverted neutrino mass hierarchy.


\section{\textbf{The Case of Three Quasi-Degenerate Neutrinos}}

\indent The neutrinos $\nu_{1,2,3}$ are quasi-degenerate in mass if 
\begin{equation}
m_1 \simeq  m_2  \simeq   m_3 \equiv m,
\label{mdegspec}
\end{equation}
\noindent  and 
\begin{equation}
m \gg \sqrt{\deltaatm}.
\label{mbig}
\end{equation}
\noindent When (\ref{mdegspec}) holds but 
$m \sim \mathrm{O}(\sqrt{\deltaatm})$,
we have a partial hierarchy or partial inverted hierarchy
between the neutrino masses. These two possibilities 
will be considered in the next Section.
The possible implications of quasi-degenerate
neutrino mass spectrum for \meff 
were discussed recently 
also in \cite{Rodej00}, but from a 
somewhat different point of view.

  As for the hierarchical neutrino mass spectrum we have:
\begin{gather}
\Delta  m^2_{\odot} \equiv \Delta m^2_{21},~~~
\Delta  m^2_{\mathrm{atm}} \equiv \Delta m^2_{32}, \nonumber \\
\ueuno = \cos^2 \theta_\odot (1- \uetre),\nonumber \\
\uedue = \sin^2    \theta_\odot (1- \uetre), \nonumber\\
\uetre = \sin^2\theta < 0.09~~~~({\rm CHOOZ}).
\label{degparam}
\end{gather}

\vspace{1mm}
\noindent
The parameters \deltaatm, \deltasol and  $\theta_{\odot}$ 
take values in the regions quoted in Section~\ref{data}.
Equation (\ref{degparam}) allows to express 
eqs. (\ref{mdegspec}) and (\ref{mbig}) in the 
compact form
\begin{equation}
\sqrt{\Delta m^2_{21}} \ll ~(<)~\sqrt{\Delta m^2_{32}}\ll m_1.
\label{mdegspec2}
\end{equation}

\indent As can be shown (see, e.g., \cite{BGWIN99,BGG99}),
the mass scale $m$ effectively coincides with 
the electron (anti-)neutrino mass $m_{\nu_e}$ 
measured in the $^{3}$H $\beta$-decay experiments:
\begin{equation}
m = m_{\nu_e}.   
\end{equation}
\noindent Thus, the experimental 
upper bounds in eq. (\ref{H3beta}) 
lead to $m < 2.5 \eV$.

 The quasi-degenerate neutrino mass 
spectrum under discussion is actually realized 
for values of the neutrino mass $m$, which is
measured in the $^3$H $\beta-$decay experiments,
$m = m_{\nu_e} \gtap (0.2 - 0.3)$ eV (see Section 7 for 
a more detailed discussion).
Recently,  Karlsruhe, Mainz and Troizk
collaborations proposed a new
$^3$H $\beta-$decay experiment 
KATRIN \cite{KATRIN}, which is planned 
to have a record sensitivity of 0.35 eV
to the neutrino mass $m_{\nu_e}$.
Clearly, the realization of this project
could be crucial for the test of the
possibility of three quasi-degenerate 
neutrinos. 

   It follows from eqs.~(\ref{mdegspec}) - (\ref{mbig})
 and eq.~(\ref{meffective})  
that \meff can be approximated by
\footnote{Under the assumption that the terms $\sim |U_{\mathrm{e} 3}|^2$
give a negligible contribution in \meff, this case was considered 
briefly recently in \cite{TomW01}.}:
\begin{equation}
\meff \simeq m \left| \cos^2 \theta_\odot (1- \uetre)e^{i \alpha_1} +  
\sin^2    \theta_\odot (1- \uetre)   
 e^{i \alpha_2} +|U_{\mathrm{e} 3}|^2 e^{i \alpha_3} \right|.
\label{meffmdeg}
\end{equation}

\indent If CP is conserved, we have the following possibilities,
depending on the relative CP-parities of the neutrinos $\nu_j$. 

\vskip 0.2truecm
{\bf Case A.} For $\phi_1 = \phi_2 = \pm \phi_3$ 
(i.e., $\alpha_{21} =0,~ \alpha_{31}=0, \pm \pi$), 
we get a very simple expression for \meff: 
\begin{equation}
\meff \simeq m \left(1 - |U_{\mathrm{e}3}|^2 \pm |U_{\mathrm{e}3}|^2 \right),
\end{equation}

\noindent i.e., \meff does not depend on \deltasol, $\theta_{\odot}$
and \deltaatm. Note that if $\phi_1 = \phi_2 = \phi_3$, 
$\meff = m$.
 Using the CHOOZ limit \cite{CHOOZ} on $|U_{\mathrm{e}3}|^2$
one finds the range
of allowed values of \meff:
\begin{equation}
0.8 \ m \leq \meff \leq 1.0 \ m.
\label{mdeg++}
\end{equation}

\noindent The above result and 
the bound on \meff obtained in the Heidelberg-Moscow
experiment \cite{76Ge00}, eq. (\ref{76Ge00}), imply 
\begin{equation}
 m < (0.4 \div  1.2) \eV,
\end{equation}
\noindent which is compatible with the 
$^{3}$H beta-decay limit on  $m$.

\vskip 0.2truecm
{\bf Case B.} If  $\phi_1 = - \phi_2 = \pm  \phi_3$ (i.e., 
 $\alpha_{21} = \pm \pi, \ \alpha_{31} = 0, \pm \pi$), 
\meff is given by:
\begin{equation}
\meff \simeq m \left| \cos 2 \theta_{\odot} 
(1 - |U_{\mathrm{e}3}|^2) \pm |U_{\mathrm{e}3}|^2  \right|.
\end{equation}

\noindent Now \meff depends $\cos 2 \theta_{\odot}$ 
and thus varies with the solution of the solar neutrino problem.
For the LMA and LOW-QVO solutions, the expression for 
\meff in the case of $\cos 2 \theta_\odot > 0$
and $\phi_1 = - \phi_2 =   \phi_3$ coincides with that
for $\cos 2 \theta_\odot < 0$ and 
$\phi_1 = - \phi_2 = - \phi_3$.
The allowed intervals of values of $\meff/m$ 
for the three solutions of the solar neutrino problem
derived in \cite{SKYSuz00,Fogli00,Gonza3nu} 
are reported in Table~\ref{table4}.
\begin{table}[H]
\caption{Values of $\meff/m$ for the 
quasi-degenerate neutrino mass spectrum, 
eq. (\ref{mdegspec}),
CP-conservation and $\phi_1 = - \phi_2 = 
\pm \phi_3$ (see text for details).}
\label{table4}
\begin{center}
\setlength{\extrarowheight}{4pt}
\begin{tabular}{|c|c|c|c|}
 \hline
Data from   &                 &    \multicolumn{2}{c|}{$\meff / m$} \\ \hline
\mbox{}     &                 &  $\phi_1 = - \phi_2 = \phi_3$ &  $\phi_1 = - \phi_2 = -  \phi_3$ \\ \hline \hline
Ref. \cite{SKYSuz00}	        &   LMA           & $0.21 \div 0.68$  & $ 0.10 \div 0.65 $ \\
$(95 \%\,  \mathrm{ C.L.} )$    & LOW-QVO         & $0.3 \div 0.58$  & $0.18 \div  0.54$  \\ \hline  
Ref. \cite{Fogli00}        &   LMA           & $ 0.25 \ (0)  \div 0.68 \ (0.70)$  & $  0.12 \ (0) \div 0.65 \ (0.67)$ \\
  $(90\% \ (99 \%) \, \mathrm{ C.L.} )$   & SMA             & $   (1.0)$  & $  (0.8  \div  1.0)$  \\
\mbox{}     & LOW-QVO         & $  0.05 \ (0)  \div 0.33 \ (0.49)$  & $ 0 \ (0) \div 0.26 \ (0.42)$  \\ \hline
Ref. \cite{Gonza3nu}   &  LMA & $  0 \ (0) \div 0.70 \ (0.74) $    & $ 0 \  (0) \div 0.68 \ (0.72)$ \\ 
 $(90\% \ (99 \%) \, \mathrm{ C.L.} )$    &   SMA           & $  1.0 \ (1.0) $    & $ 0.9 \ (0.8) \div 1.0 \ (1.0)$\\
\mbox{}     &   LOW-QVO       & $  0 \ (0) \div 0.50 \ (0.69)  $   & $  0 \ (0) \div 0.48 \ (0.67)$ \\ \hline
\end{tabular}
\end{center}
\end{table}

  Several remarks are in order.

\begin{enumerate} 
\item In the case of the LMA solution, 
the $99\%$ C.L. results of the analysis in 
ref.~\cite{Gonza3nu} and \cite{Fogli00} 
do not exclude the possibility of a cancellation
between the different contributions in \meff, so that
\begin{equation*}
0 \leq \meff \ltap 0.7~m. 
\end{equation*}
\noindent  The results of ref.~\cite{SKYSuz00} 
for the same solution
does not allow a complete 
cancellations to occur and one has
\begin{equation*}
0.2 \times 10^{-2}~m  \leq \meff \leq 0.7~m. 
\end{equation*}
%
\item The results of all three analyzes for the SMA solution 
agree. The case under discussion cannot be distinguished 
from the $\phi_1 = \phi_2 = \pm \phi_3$ one
since  $\cos 2 \theta_\odot  \simeq 1 $.
\item In the LOW-QVO solution case, the results 
of the analyzes \cite{Fogli00,Gonza3nu}
do not exclude the possibility of cancellations 
between the different contributions in \meff, as well as
of having $\cos 2 \theta_\odot < 0$, so that:
\begin{equation*}
0 \leq \meff \leq 0.5 \div 0.7~m. 
\end{equation*}
In contrast, according to the results of the analysis \cite{SKYSuz00}, 
$\cos 2 \theta_\odot > 0 $ and one finds a relatively 
large lower bound on \meff.

\end{enumerate}

 The regions of allowed values of \meff, corresponding to the 
LMA and the LOW-QVO solutions of ref.~\cite{Gonza3nu},
are shown as a function of 
$m$ in Figs.~\ref{figure:massdeg01} and ~\ref{figure:massdeg02},
respectively. One can have $\meff \gtap 0.1~$eV for values of
$m \geq 0.1~$eV. Taking into account the 
$^{3}$H $\beta-$decay upper bound 
$m < 2.5~$eV, we find that 
${\rm max}(\meff) \simeq 1.7~$eV. This implies that 
\meff can have the maximal value allowed by
the limit (\ref{76Ge00}): ${\rm max}(\meff) < (0.35 - 1.05)~$eV.

\begin{figure}[t]
\begin{center}
\epsfig{file=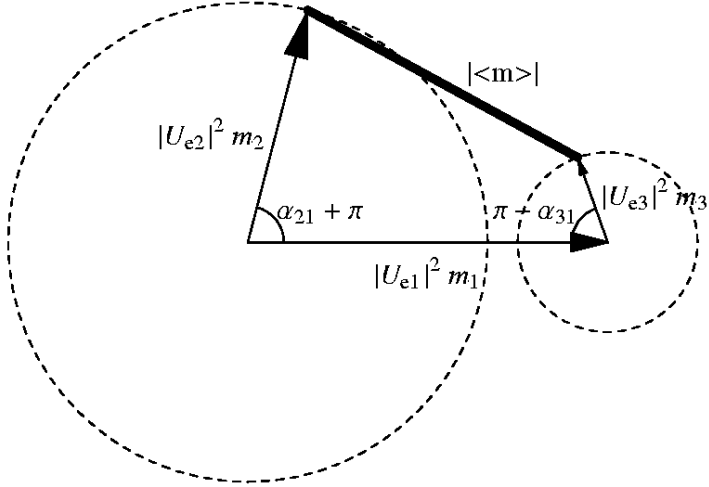, height=6.5cm
}
\end{center}
\caption{The effective Majorana mass 
$\meff$ represented as the absolute 
value of the vector
sum of the three contributions in eqs. (\ref{meffective})
and (\ref{Uejphase}), 
each of which is  
expressed as a vector in the complex plane
(see text for details).
The CP-violating phases $\alpha_{21}$ and $\alpha_{31}$ 
shown in the figure can vary between 0 and $2 \pi$.  
} 
\label{figure:massdeg00}
\end{figure}

\begin{figure}
\begin{center}
\epsfig{file=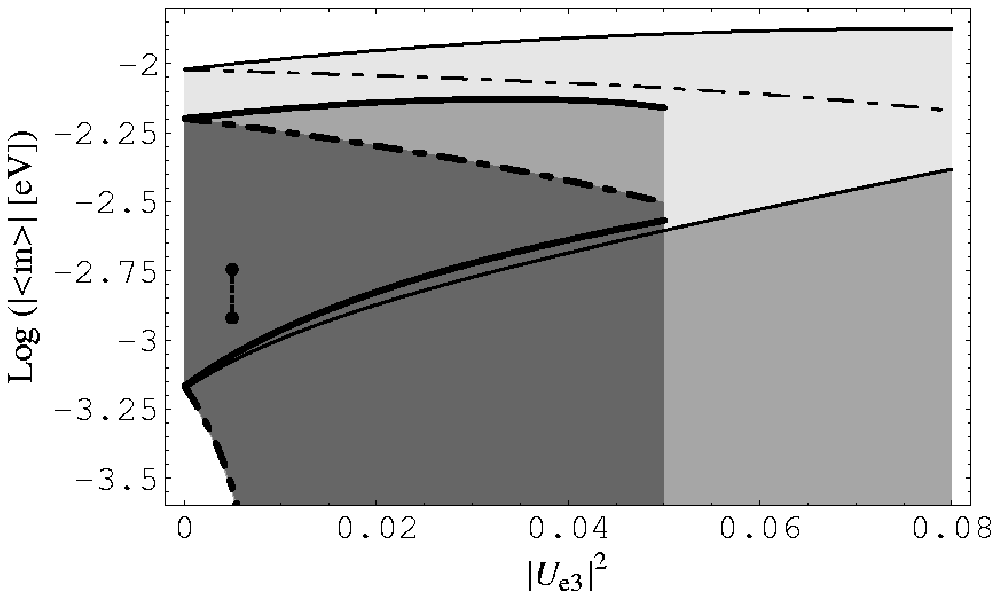, height=6.5cm
}
\end{center}
\caption[gmassh01]{The effective Majorana mass 
\meff, allowed by the data from the solar 
and atmospheric neutrino
and CHOOZ experiments, as a function 
of $|U_{\mathrm{e} 3}|^2$ 
in the case of 
hierarchical neutrino mass spectrum, eq. (\ref{hierarchy}). 
The values of \meff  are obtained 
for \deltasol, $\sin^2 \theta_\odot$ from the 
LMA MSW solution region,
and \deltaatm and $|U_{\mathrm{e} 3}|^2$, 
derived in \cite{Gonza3nu} at  
$90\%$ C.L. (medium grey and dark grey regions at 
$|U_{\mathrm{e} 3}|^2 <0.05$)
and at $99\%$ C.L. (light, medium   
and dark grey regions at $|U_{\mathrm{e} 3}|^2 <0.08$). 
The $90\%$ ($99\%$) C.L. allowed regions 
located \textit{i)} between the two thick 
(thin) solid lines and \textit{ii)} between 
the  thick (thin) dashed lines and the horizontal 
axis, correspond to the two cases of CP conservation: 
\textit{i)} 
$\phi_2 = \phi_3$, and \textit{ii)} 
$\phi_2 = - \phi_3$, 
$i\phi_{2,3}$ being the CP-parities of 
$\nu_{2,3}$.
The values of \meff, calculated for the best 
fit values of the input parameters, are 
indicated by thick dots 
(CP-conservation) 
and a  dotted line (CP-violation).
In the case of CP-violation all the regions 
marked with different grey colour scales are allowed.
}
\label{figure:gmassh01}
\end{figure}

\begin{figure}[p]
\begin{center}
\epsfig{file=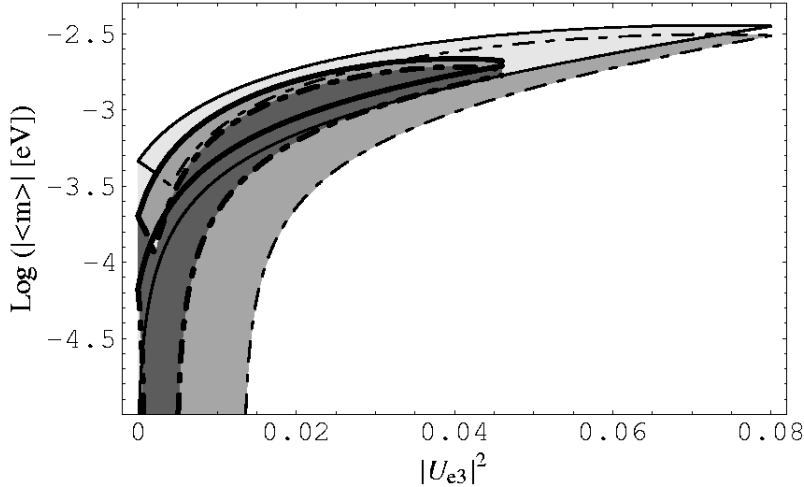, height=6.5cm
}
\end{center}
\caption[gmassh02]{The same as Fig.~\ref{figure:gmassh01} 
but  for \deltasol and  $\sin^2 \theta_\odot$ from the
$90\%$ ($99\%$) C.L. region of the LOW-QVO 
solution of the $\nu_\odot$-problem \cite{Gonza3nu}. 
The  regions limited by the two thick (thin) solid 
and dashed lines correspond to the two 
cases of CP-conservation: \textit{i)}  
$\phi_2 = \phi_3$, (regions within the solid lines) 
and \textit{ii)} $\phi_2 = - \phi_3$ (regions 
within the dashed lines). In the case of CP-violation, 
$(\alpha_3 - \alpha_2) \neq k \pi$, $k = 0,1,...$, 
all the regions marked by different grey color scales are allowed.
}
\label{figure:gmassh02}
\end{figure}

\begin{figure}[p]
\begin{center}
\epsfig{file=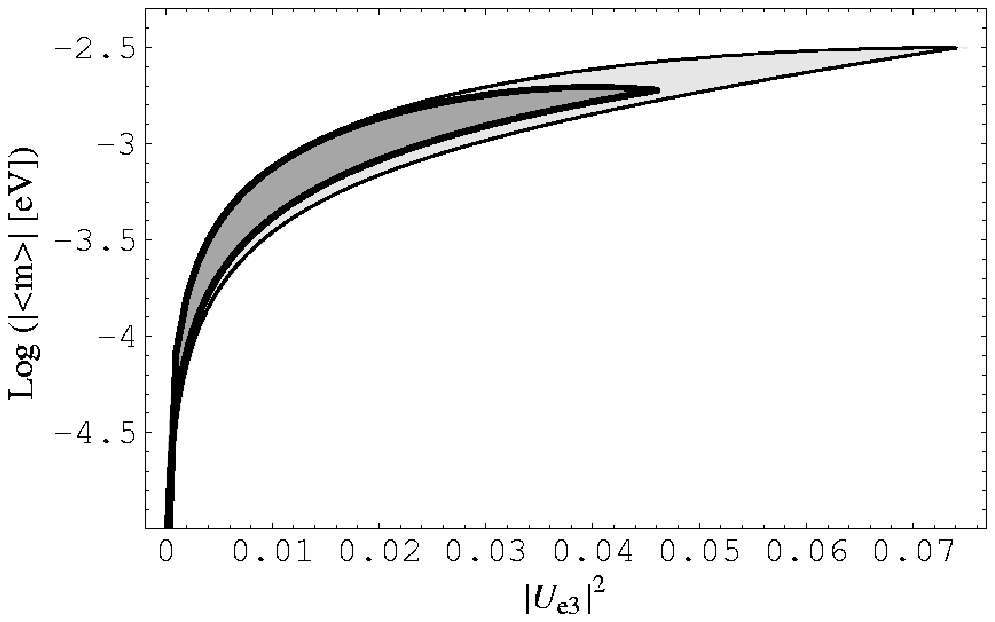, height=6.5cm
}
\end{center}
\caption[gmassh03]{The same as Fig.~\ref{figure:gmassh01} 
but  for \deltasol and  $\sin^2 \theta_\odot$ from 
the region of the SMA solution of the 
$\nu_\odot$-problem \cite{Gonza3nu} obtained 
at $90\%$ C.L. (region within the thick solid lines) 
and at $99\%$ C.L. (region within the thin solid lines) .
The results shown are derived assuming 
$m_1 \ll 10^{-4} \eV$. For the range of values 
of \meff in the figure, one has in this case 
$\meff \sim \sqrt{\deltaatm} |U_{\mathrm{e}  3}|^2$ 
and thus \meff does not depend on the 
CP-violating phase $(\alpha_3 - \alpha_2)$. 
}
\label{figure:gmassh03}
\end{figure}

\begin{figure}[p]
\begin{center}
\epsfig{file=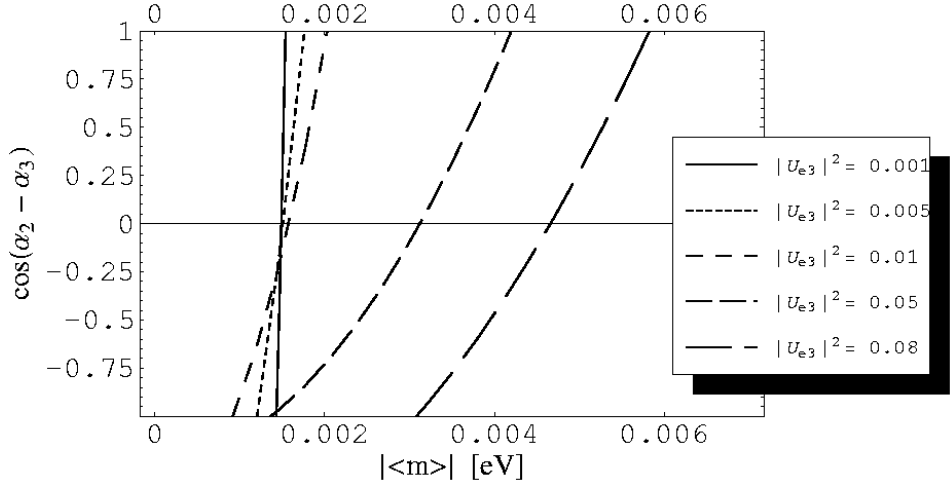, height=6.5cm
}
\end{center}
\caption[gmassh04]{The CP-violation
factor $\cos(\alpha_2 - \alpha_3)$, eq. (\ref{hierarchy})
(hierarchical neutrino mass spectrum), 
as a function of \meff for different values of $|U_{\mathrm{e}  3}|^2$ and for 
the best fit values of the parameters 
\deltasol, $\sin^2 \theta_\odot$ and \deltaatm, 
found in the analysis \cite{Gonza3nu}.
Values of $\cos(\alpha_2 - \alpha_3) = 0,\pm 1$,
correspond to CP-invariance.
}
\vspace{3mm}
\label{figure:gmassh04}
\end{figure}

\begin{figure}[p]
\begin{center}
\epsfig{file=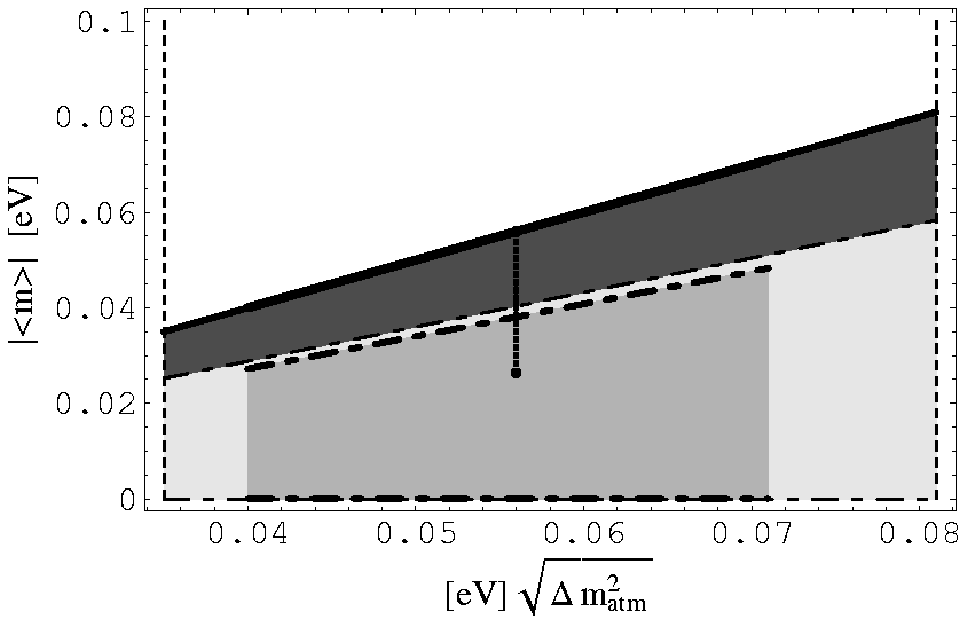, height=6.5cm
} 
\end{center} 
\caption[gmassinv01]{The effective Majorana mass 
\meff as a function of $\sqrt{\deltaatm}$ 
for the neutrino mass spectrum of the inverted hierarchy
type, eq. (\ref{invmhspec}). 
The allowed regions (in grey) 
correspond to the LMA solution 
of ref.~\cite{Gonza3nu}.  
In the case of CP-invariance and for the
$90\%~ (99\%)$ C.L. results for the 
solution region, \meff can have values 
{\it i}) for $\phi_2 = \phi_3$ - in the  medium grey  
(light grey and medium grey) upper  region, limited by the
doubly thick (thick and doubly thick) solid lines and 
{\it ii}) for $\phi_2 = - \phi_3$ - in the medium grey  
(light grey and medium grey) 
region limited
by the doubly thick (thick and doubly thick) 
dash-dotted lines .  If CP is not conserved, \meff can lie in any of the 
regions marked by different grey scales     
The dark-grey region 
corresponds to ``just-CP-violation'':
\meff can have value in this region 
{\it only if the CP-parity is not conserved}. 
The values of \meff corresponding 
to the best fit values of the 
input parameters, found in \cite{Gonza3nu}, 
are denoted by dots 
in the CP-conserving cases and by a 
dotted line in the CP-violating one.
}  
\label{figuramassinv01}
\end{figure}

\begin{figure}[p]
\begin{center}
\epsfig{file=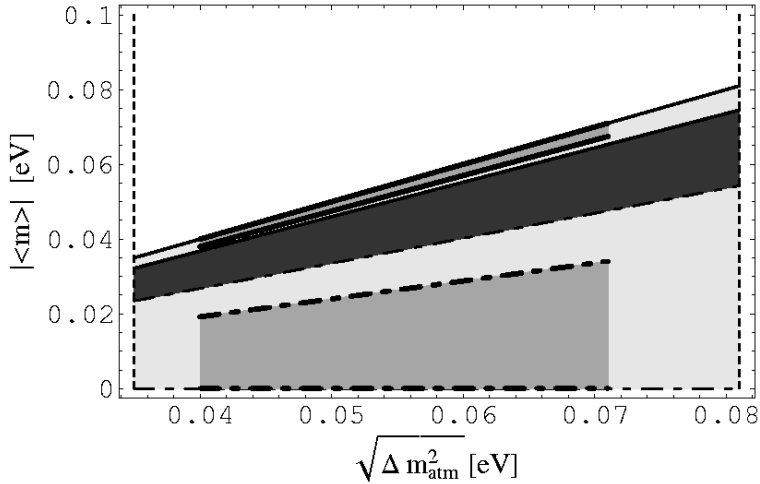, height=6.5cm
}
\end{center}
\caption[gmassinv02] {The same as in Fig. (\ref{figuramassinv01})
for the $90\%$ ($99\%$) C.L. LOW-QVO solution
of ref. \cite{Gonza3nu}. The medium
grey (light-grey and medium 
grey) region, bounded by the 
the thick (thick and doubly thick) solid lines, and
medium grey  
(light grey and medium grey) lower region,
bounded by the thick (thin) dash-dotted lines,
correspond to the two cases of CP-invariance, 
$\phi_2 =  \phi_3$  and $\phi_2 = -  \phi_3$,
respectively. If CP is not conserved, 
\meff can lie in any of the 
regions marked by different grey scales.      
The ``just-CP-violation'' region is shown in dark-grey color:
\meff can have value in this region 
{\it only if the CP-symmetry is violated}.
}  
\label{figuramassinv02}
\end{figure}

\begin{figure}[p]
\begin{center}
\epsfig{file=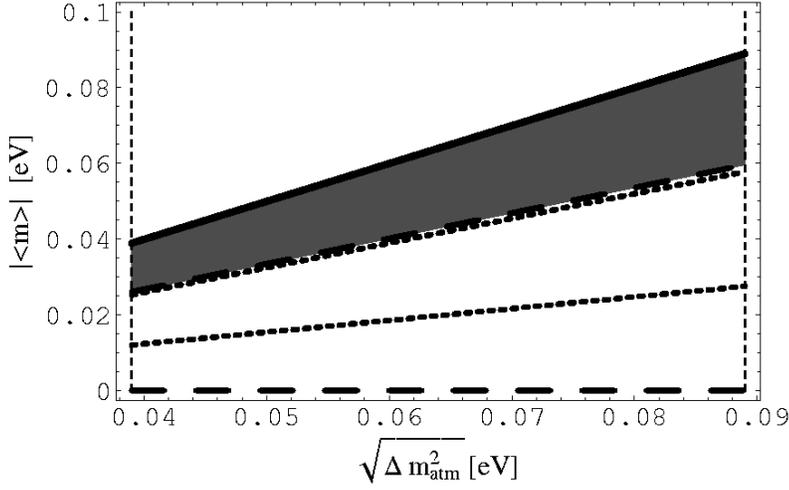, height=6.5cm 
}
\end{center}
\caption[gmassinv03]{\meff as a function of $\sqrt{ \deltaatm}$ 
for the inverted mass hierarchy spectrum and LMA solution
of the solar neutrino problem. The figure is obtained
using the results of ref.~\cite{SKYSuz00} (region between the
thick solid line and the lower dotted line) and of 
ref.~\cite{Fogli00} (at $99\%$ C.L. - region between the 
thick solid line and the lower long-dashed line). 
If CP-invariance does not hold,
\meff spans all the allowed regions
indicated (in brackets) above. 
The values of \meff in the two CP-conserving cases
lie i) on the doubly thick solid line if $\phi_2 = \phi_3$,
and ii) for $\phi_2 = - \phi_3$ -
between the two dotted lines (dashed lines)
for the LMA solution in ref. \cite{SKYSuz00} 
(ref. \cite{Fogli00}). 
The common ``just-CP-violation'' 
region for both analyzes is marked in
dark-grey : {\it a value of \meff 
in this region would signal CP-violation}.
} 
\label{figuramassinv05}
\end{figure}

\begin{figure}[p]
\begin{center}
\epsfig{file=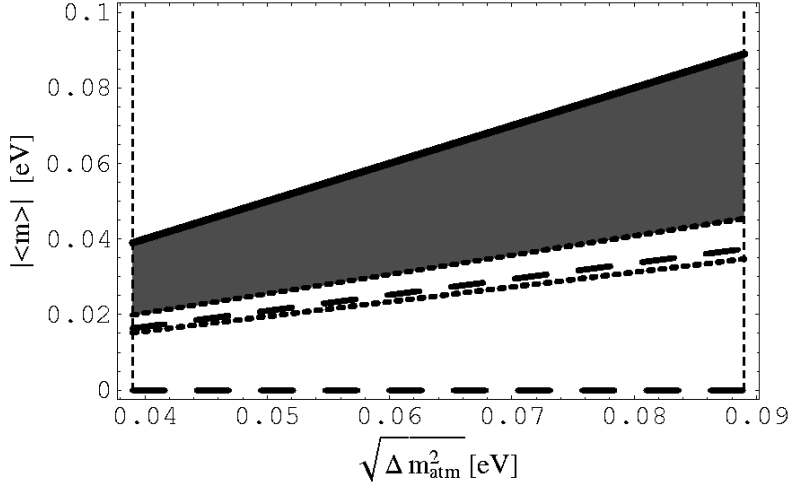, height=6.5cm
}
\end{center} 
\caption[gmassinv04]{The same as in Fig. 8 for the 
LOW-QVO solution found in ref.~\cite{SKYSuz00}
and in ref.~\cite{Fogli00} (at $99\%$ C.L.).
The ``just-CP-violation'' region, in particular,
is marked by dark-grey color. 
} 
\label{figuramassinv06}
\end{figure}

\begin{figure}[p]
\begin{center}
\epsfig{file=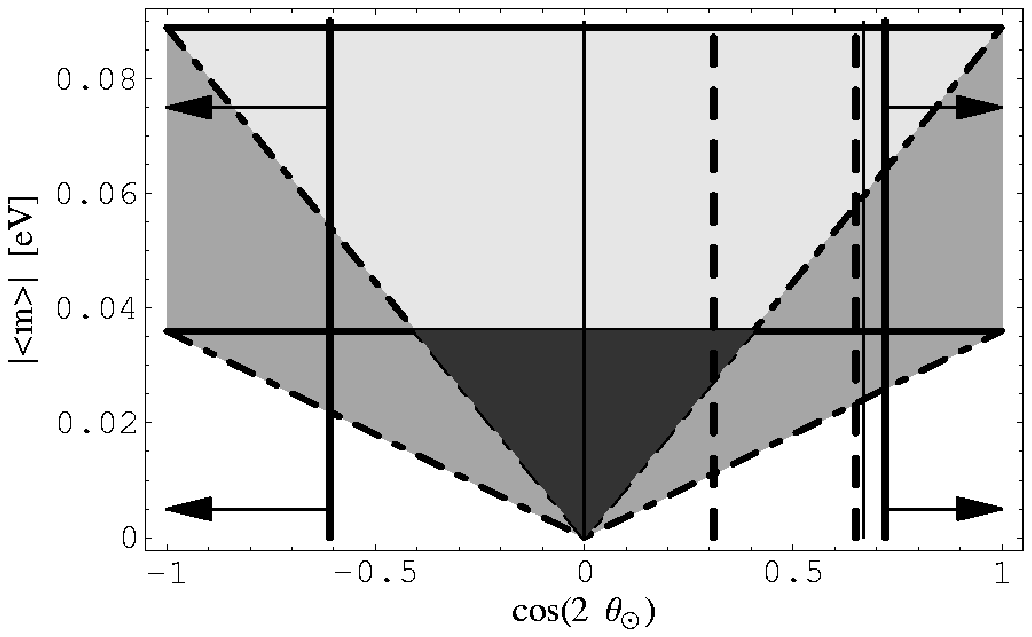, height=8cm 
}
\end{center}
\caption[gmassinv07]{The dependence of 
\meff on $\cos 2 \theta_\odot$ 
for the neutrino mass spectrum
with inverted hierarchy, 
eq. (\ref{invmhspec}), 
and the LMA solution of the 
solar neutrino problem.
The region between the two thick horizontal 
solid lines (in light  grey and medium grey colors)
and the two triangular regions between the thick 
dash-dotted lines (in medium grey color),
correspond to the two CP-
conserving cases, $\phi_2 = \phi_3$ 
and $\phi_2 = - \phi_3$, respectively.
The ``just-CP-violation'' region is denoted 
by dark-grey color. 
The regions between each of the three pairs 
of vertical lines of a given type
- solid, doubly thick solid and doubly thick dashed,
correspond to the intervals of values of 
$\cos 2 \theta_\odot$ 
for the LMA solution derived (at 99\% C.L.) 
in ref.~\cite{SKYSuz00}
(region between the two doubly thick dashed lines),
ref.~\cite{Fogli00} (region between the solid lines)
and ref.~\cite{Gonza3nu}. 
} 
\label{figure:massinv07}
\end{figure}

\begin{figure}[p]
\begin{center}
\epsfig{file=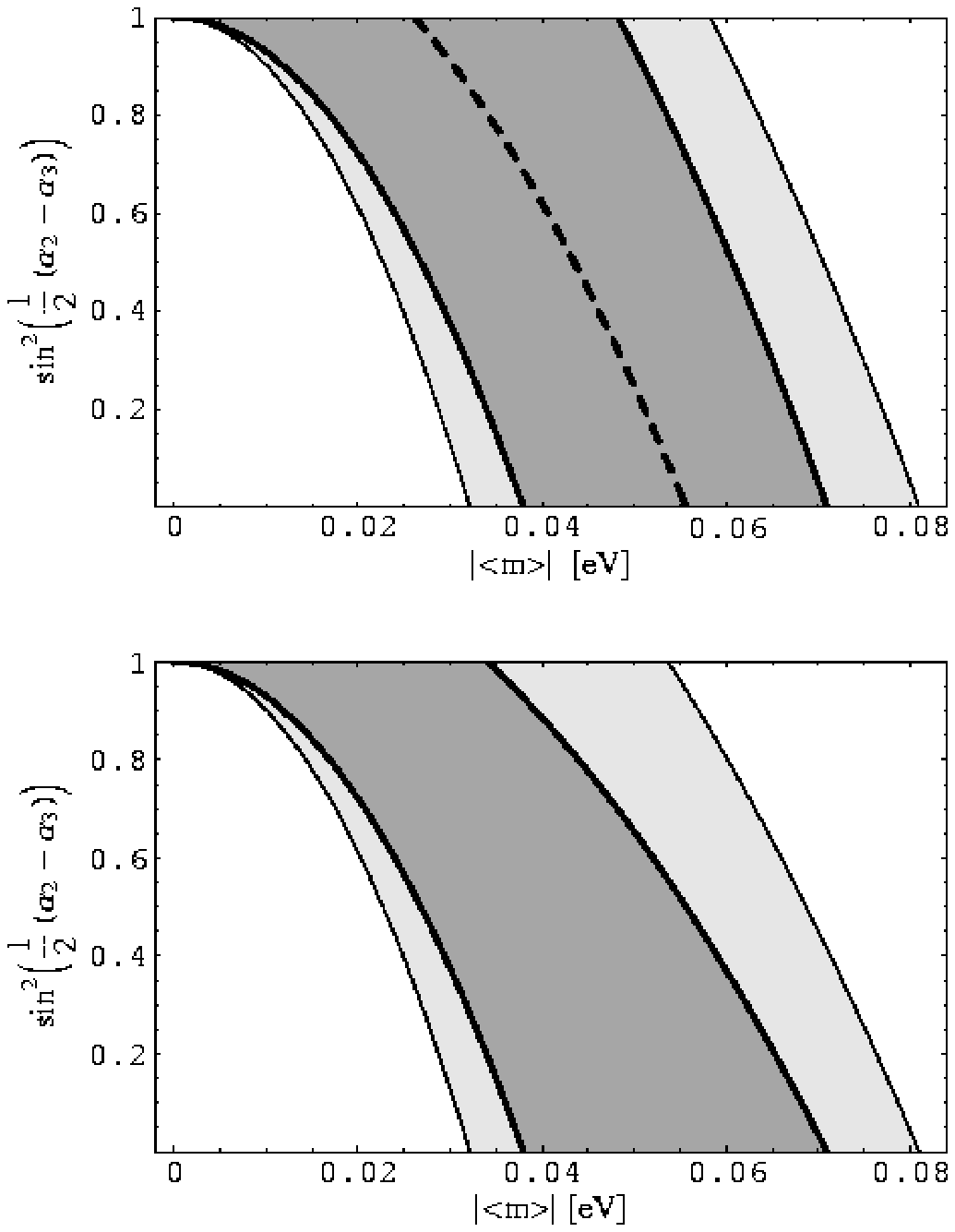, height=14cm
}
\end{center}
\caption[gmassinv06]{The CP-violation factor 
$\sin^2 (\alpha_2 - \alpha_3)/2$ 
as a function of \meff
in the case of inverted mass hierarchy spectrum, 
eq. (\ref{invmhspec}), and the LMA  (upper panel) and 
LOW-QVO (lower panel) solutions of ref.~\cite{Gonza3nu},
obtained at $90\%$ C.L. (medium grey region 
with thick contours) and $99\%$ C.L.
(light grey  and  medium grey region limited by 
ordinary solid lines). 
The values of \meff, corresponding to the best fit 
values of the input parameters, found in \cite{Gonza3nu},
are indicated by the dashed line.
A value of $\sin^2 (\alpha_2 - \alpha_3)/2 \neq 0,1$, 
would signal CP-violation.
} 
\label{figuramassinv03}
\end{figure}

\begin{figure}[p]
\begin{center}
\epsfig{file=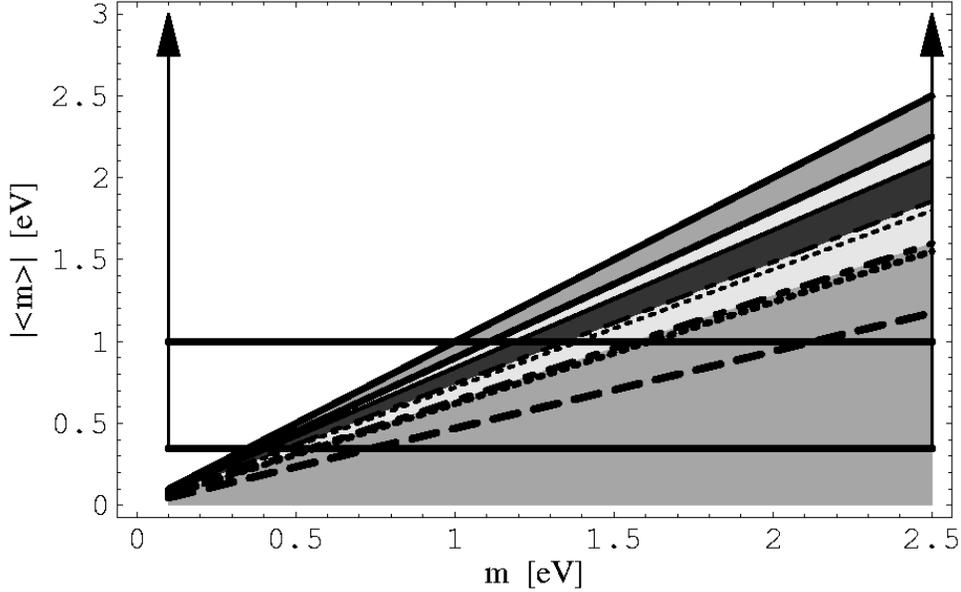, height=8cm
}
\end{center}
\caption[gmassdeg01]{ \meff as a function of the neutrino mass
$m$ for the quasi-degenerate neutrino mass spectrum, 
eqs. (\ref{mdegspec}) - (\ref{mbig}), 
and the LMA solution of the $\nu_{\odot}-$problem
of ref. \cite{Gonza3nu} at 90\%~C.L. (99\% C.L.). 
The regions allowed in the cases of 
CP-conservation are marked by 
{\it i}) $\phi_1= \phi_2 = \phi_3$ - the
thick solid (non-horizontal) line $\meff = m$,
{\it ii}) $\phi_1= \phi_2 = - \phi_3$ - 
medium grey color  triangular region between the two doubly thick solid lines (light grey and medium grey color  triangular region between the doubly thick and thick solid lines,  
{\it iii}) $\phi_1= - \phi_2 = \pm \phi_3$ (two cases) - 
medium grey color 
triangular regions between the
doubly thick dashed-dotted line and the horizontal axes 
  and between the  doubly thick
dotted line and the horizontal axes (light grey and medium grey region between the thin dashed-dotted line and the horizontal axes 
  and between the thin
dotted line and the horizontal axes). 
The ``just-CP-violation'' region is denoted by 
dark-grey color. The doubly thick solid line 
corresponding to $\meff = m$ and the doubly thick long dashed line 
indicate the ``best fit lines'' of values of 
\meff for the i)-ii) and  
the iii) cases, respectively. The two horizontal 
(doubly thick) lines show the upper limits 
\cite{76Ge00}, quoted in eq. (3).
} 
\label{figure:massdeg01}
\end{figure}

\begin{figure}[p]
\begin{center}
\epsfig{file=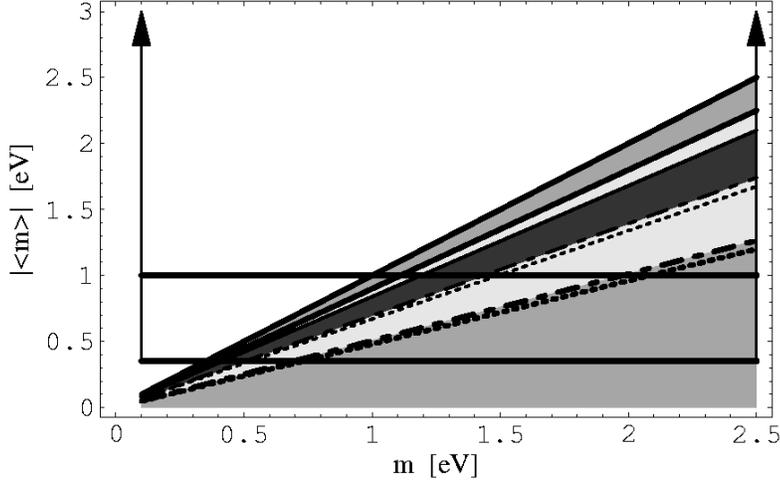, height=6.5cm 
}
\end{center}
\caption[gmassdeg02]{The same as in 
Fig.(\ref{figure:massdeg01}) for the
LOW-QVO solution of the 
$\nu_{\odot}-$problem \cite{Gonza3nu}. 
If CP is conserved and at 
90\% C.L. (99 \% C.L.) \cite{Gonza3nu},  
\meff should lie
in the two medium grey 
(the two light  grey  and medium grey)
triangular regions: 
i) for $\phi_1=\phi_2=  \phi_3$ - 
on the line $\meff = m$, 
ii) for $\phi_1=\phi_2= - \phi_3$ -
in the upper medium 
grey (light grey and  medium grey) 
triangular region,
iii) if $\phi_1= - \phi_2= \pm \phi_3$ - 
in the lower medium 
grey (light grey and  medium grey)
triangular region with dash-dotted 
($\phi_1 = + \phi_3$) and 
dotted ($\phi_1 = - \phi_3$) contours. 
The ``just-CP-violation'' 
region is denoted by dark-grey color. 
} 
\label{figure:massdeg02}
\end{figure}

 If CP-parity is not conserved, one has to take into account 
the two relevant CP-violating phases $\alpha_{21}$ 
and $\alpha_{31}$ in the expression for \meff.
The effective mass \meff can be computed 
using a graphical method \cite{FKNT00,KPS00}. 
Each of the three contributions 
to \meff can be represented by a vector, 
${\bf n}_j,~j=1,2,3$,
in the complex plane, as sketched in 
Fig.~\ref{figure:massdeg00}:
${\bf n}_1 = m_1\ueuno (1,0)$,
${\bf n}_2 = m_2\uedue (\cos\alpha_{21},\sin\alpha_{21})$,
${\bf n}_3 = m_2\uetre (\cos\alpha_{31},\sin\alpha_{31})$.
The two phases $\alpha_{21}$ 
and $\alpha_{31}$ can take values 
from 0 to $2 \pi$ 
and  the effective Majorana mass \meff is given by 
the modulus of the sum of the three vectors:
$\meff = |{\bf n}_1 + {\bf n}_2 + {\bf n}_3|$.

  In order to have a complete cancellation
between the three terms in \meff, eq. (\ref{meffmdeg}), 
the following condition has to be satisfied 
(see also \cite{KPS00}):
\begin{equation}
\left| m_1 U^2_{\mathrm{e} 1} \right| 
< \left| m_2 U^2_{\mathrm{e} 2} \right| +
\left| m_3 U^2_{\mathrm{e} 3} \right|.
\end{equation}

\noindent The relation (\ref{mdegspec}) between the masses 
$m_j$, eq. (\ref{mbig}) and the identification (\ref{degparam})
imply that this inequality reduces to a condition
on the angle $\theta_\odot$ and on $\uetre$:
\begin{equation}
\cos^2 \theta_\odot < \frac{1}{2 (1 - |U_{\mathrm{e}3}|^2)}.
\end{equation}

Using the results of the different data analyzes which are
summarized in Section 2 and in Tables 1 and 2,
we can conclude that:
\begin{enumerate}
\item
a complete cancellation between the three terms in \meff,
eq. (\ref{meffmdeg}), is impossible for the
SMA solution of the solar neutrino problem;
\item
the results on the possible values of $\theta_{\odot}$
in \cite{Gonza3nu} and 
 \cite{Fogli00} (Table 2) 
show that there can be a complete cancellation 
both in the case of the  LMA and LOW-QVO solutions;
\item 
according to the analysis of ref.~\cite{SKYSuz00}, 
there can be a cancellation in the LOW-QVO case but not in the LMA one.  
\end{enumerate}

  The lower and upper limits on \meff can be found 
taking respectively $\alpha_{21} =\alpha_{31} = \pm  \pi$
 (e.g. $\phi_1=- \phi_2=- \phi_3$) and 
$\alpha_{21} =\alpha_{31} =  0$ (e.g. $\phi_1= \phi_2= \phi_3$). 
Recalling the limits already found in the CP-conserving 
cases, the bounds holding in the CP-violating one are 
reported in Table~\ref{table5}.

\begin{table}[H]
\caption{Values of $\meff/m$, 
corresponding to the different 
solutions of the solar neutrino
problem, neutrino oscillation 
interpretation
of the atmospheric neutrino data 
and the CHOOZ limit, in the case of 
quasi-degenerate neutrino mass spectrum, 
eq. (\ref{mdegspec}), 
and CP-nonconservation (see text for details).
}
\label{table5}
\setlength{\extrarowheight}{4pt}
\begin{center}
\begin{tabular}{|c|c|c|} \hline
Data from  &                 & $ \meff / m$  \\ \hline \hline
Ref. \cite{SKYSuz00}        &   LMA           &  $0.10  \div 1.0$ \\
 $(95 \%\,  \mathrm{ C.L.} )$   & LOW-QVO         &  $ 0.18 \div 1.0$ \\ \hline  
Ref. \cite{Fogli00}       &   LMA           & $ 0.12 \ (0) \div 1.0 \  (1.0)$ \\
 $(90 \ (99 \%) \, \mathrm{ C.L.} )$   & SMA             & $ (0.8 \div 1.0) $\\
\mbox{}     & LOW-QVO         & $0 \ (0) \div 1.0 \  (1.0) $ \\ \hline
Ref. \cite{Gonza3nu}  & LMA         & $0 \ (0)  \div 1.0 \ (1.0)$\\
   $(90 \ (99 \%) \, \mathrm{ C.L.} )$       & SMA         & $  0.9 \ (0.8) \div 1.0 \ (1.0)$ \\
\mbox{}         & LOW-QVO     & $0 \ (0)  \div 1.0 \ (1.0) $ \\ \hline
\end{tabular}
\end{center}
\end{table}

 We see that in the case of CP-violation we can have 
$\meff \geq 0.1~$eV as well.
The upper bound on $m$ following from the 
$^{3}$H $\beta-$decay data implies 
$\meff < 2.5~$eV, while the results (\ref{76Ge00}) of the
Heidelberg-Moscow \betabeta-decay experiment 
(including a factor of 3 uncertainty
in the value of the relevant 
nuclear matrix element)
limit further the maximal value of 
\meff to ${\rm max}(\meff) < (0.35 - 1.05)~$eV.
Let us remind the reader that, as our results 
in Sections 4 and 5 showed, 
one has ${\rm max}(\meff) \ltap 0.02~$eV 
if the neutrino mass spectrum is hierarchical,
eq. (\ref{hierarchy}),
while in the case of inverted mass 
hierarchy, eq. (\ref{invmhspec}), 
${\rm max}(\meff) \ltap (0.08 - 0.09)~$eV.
Thus, the observation of the \betabeta-decay 
with a rate which corresponds to 
$\meff \gtap (0.20 - 0.30)~$eV,
would be a very strong 
experimental evidence
in favor of the quasi-degenerate
neutrino mass spectrum:
this would practically rule out 
the neutrino mass spectrum of the
hierarchical type and would strongly
disfavor (if not rule out) that of the 
inverse mass hierarchy type.
The above conclusion would be valid,
of course, only if the \betabeta-decay
is generated by the charged current 
weak interaction with 
left-handed currents and 
three-neutrino mixing,
via the exchange of 
three massive Majorana neutrinos.

\begin{figure}[p]
\begin{center}
\epsfig{file=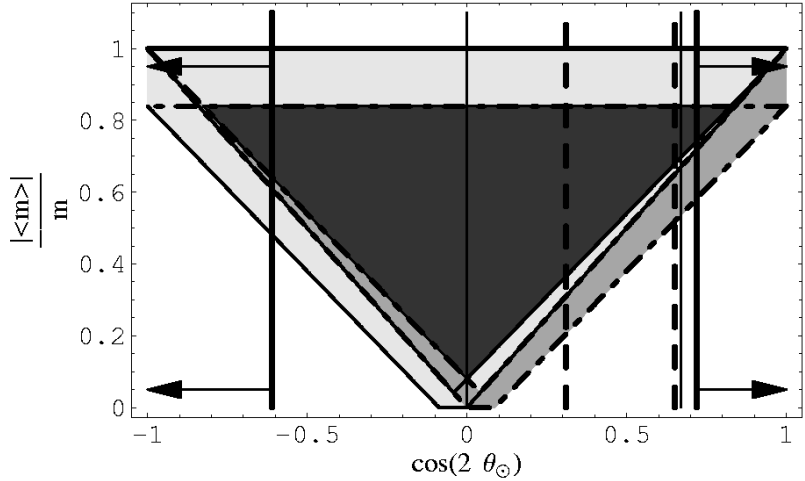, height=6.5cm  
}
\end{center}
\caption[gmassdeg03]{
The dependence of $\meff/{\rm m}$ 
on $\cos 2 \theta_\odot$ 
for the quasi-degenerate 
neutrino mass spectrum, 
eqs. (\ref{mdegspec}) - (\ref{mbig}),
and the LMA solution of the 
$\nu_{\odot}-$problem.
If CP-invariance holds, the values of 
$\meff/{\rm m}$ lie:
i) for $\phi_1 = \phi_2 = \phi_3$ -
on the line $\meff / m = 1$, 
ii) for $\phi_1 = \phi_2 = - \phi_3$ - 
in the region between the  thick horizontal solid and 
dash-dotted lines (in light  grey and 
medium grey colors),
iii) for $\phi_1 =- \phi_2 = + \phi_3$ - in 
the light  grey polygon with solid-line contours
and iv) for $\phi_1 = -  \phi_2 = - \phi_3$ -
in the medium grey 
polygon with the dash-dotted-line contours.
The ``just-CP-violation'' region is denoted 
by dark-grey color. The 
values of $\cos 2 \theta_\odot$ 
between the doubly thick solid,
the normal solid and the 
doubly thick dashed lines
correspond to the 
99\%, 99\%  and 95\% C.L.
LMA solution regions
in refs. \cite{Gonza3nu}, 
\cite{Fogli00}
and \cite{SKYSuz00}, respectively. 
} 
\label{figure:massdeg06}
\end{figure}

\begin{figure}[p]
\begin{center}
\epsfig{file=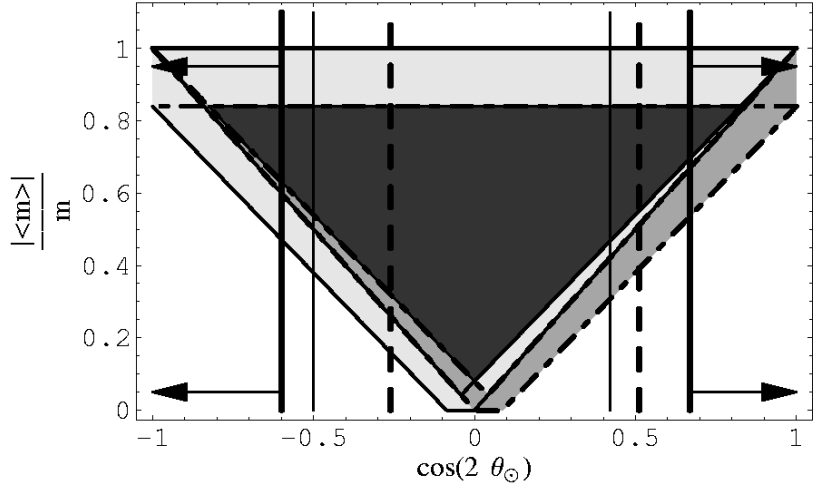, height=6.5cm 
}
\end{center}
\caption[gmassdeg04]{The same as in Fig. (\ref{figure:massdeg06})
for the LOW-QVO solution of the solar neutrino problem.
The ``just-CP-violating'' region, in particular,
is denoted by dark-grey color. 
} 
\label{figure:massdeg07}
\end{figure}

 The improvement of the 
sensitivity of the \hbeta experiments would allow
one to test further the pattern of neutrino masses
(\ref{mdegspec}) - (\ref{mbig}). 
The \betabeta-decay experiments could  
distinguish between the different 
CP-parity configurations 
of the massive neutrinos $\nu_j$
and furthermore either  
establish the existence of CP-violation in the 
leptonic sector or put an upper bound on its possible
magnitude. This is illustrated  
in Fig.~\ref{figure:massdeg06} and in 
Fig.~\ref{figure:massdeg07} where we show $\meff / m$ as a 
function of $\cos 2 \theta_\odot$. The dark grey region 
is the ``just-CP-violation'' region which 
can be spanned by the values of 
$\meff /{\rm m}$ {\it only if the CP-symmetry is violated}.

\begin{figure}[p]
\begin{center}
\epsfig{file=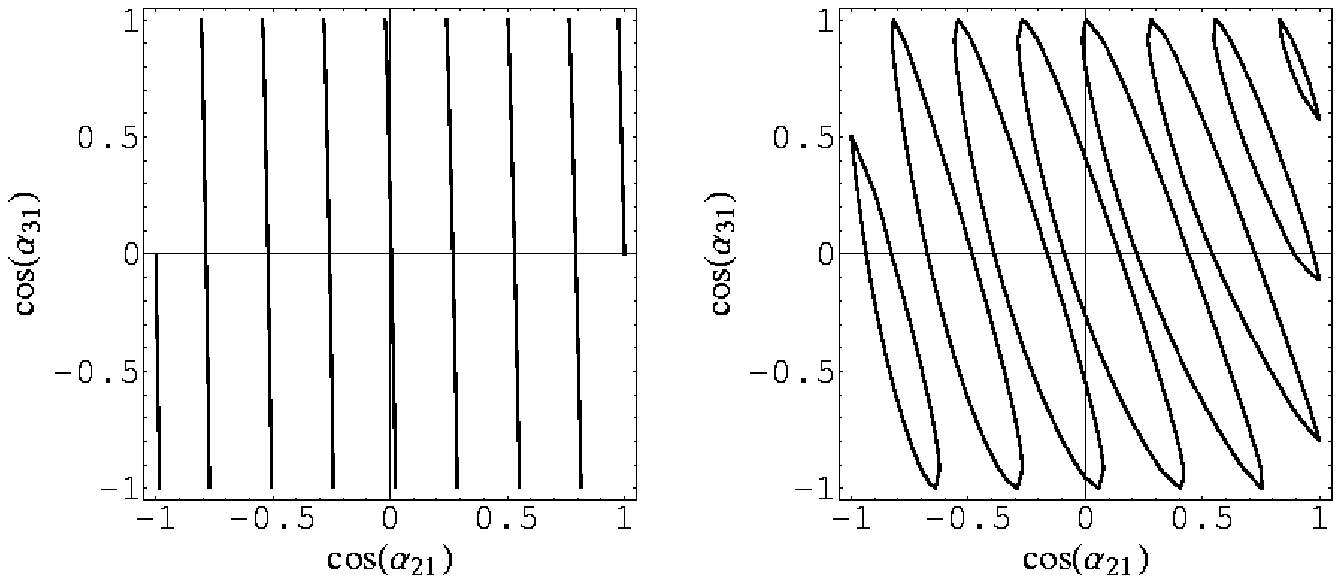, height=6.5cm
}
\end{center} 
\caption[gmassdeg05]{The interdependence 
of the two CP-violating 
phases, $\alpha_{21}$ and $\alpha_{31}$, for a given value of the ratio
$\meff/{\rm m}$ in the case of quasi-degenerate neutrino mass spectrum.
The figures are obtained for 
$ \meff/{\rm m} = \sqrt{0.2 + 0.1 n} \eV$ with 
$n= 0,1 \ldots 8$  (with increasing \meff from left to right) 
and the best fit values of the 
solar and atmospheric neutrino oscillation parameters
from \cite{Gonza3nu}, quoted in Section 2
(left-hand plot), and for 
$ \meff/{\rm m} = \sqrt{0.24 + 0.10 n} \eV $
 with $n= 0,1 \ldots 7$ (with increasing \meff from left to right),
$|U_{\mathrm{e}3}|^2 = 0.08$ and the best fit values of the 
other solar and atmospheric neutrino oscillation parameters
\cite{Gonza3nu} (right-hand plot). 
The values of $\cos \alpha_{21,31} = 0,\pm 1$, 
correspond to CP-invariance.
} 
\label{figure:massdeg03}
\end{figure}

 If $\theta_\odot$, \uetre, $m$ and \meff are known with 
a sufficient accuracy, it would be possible to 
tightly constrain the allowed values of the 
two CP-violating Majorana phases $\alpha_{21}$ 
and $\alpha_{31}$ as they must satisfy the relation:

\begin{equation}
\begin{split}
\frac{\meff^2}{  m^2}-  \left( \cos^4 \theta_\odot  
(1 - |U_{\mathrm{e}3}|^2)^2 +   
\sin^4 \theta_\odot  (1 - |U_{\mathrm{e}3}|^2)^2 + 
|U_{\mathrm{e}3}|^4  \right) = \\
 2  \cos^2 \theta_\odot  |U_{\mathrm{e}3}|^2 
(1 - |U_{\mathrm{e}3}|^2) \cos \alpha_{31} 
+  2  \sin^2 \theta_\odot  \cos^2 \theta_\odot  
(1 - |U_{\mathrm{e}3}|^2)^2         \cos \alpha_{21} \\
 +   2  \sin^2 \theta_\odot  |U_{\mathrm{e}3}|^2 
(1 - |U_{\mathrm{e}3}|^2) \cos ( \alpha_{21} - \alpha_{31}).
\end{split}  
\label{formulacos}
\end{equation}
In Fig.~\ref{figure:massdeg03}  we exhibit the allowed values 
of $\cos \alpha_{21}$ and $\cos \alpha_{31}$ for a 
given value of $\meff/m$: for the parameters entering into 
the expression (\ref{formulacos}) 
we use their best fit values from ref.~\cite{Gonza3nu} 
(which lie in the LMA solution region, etc.),
$|U_{\mathrm{e}3}|^2 = 0.005$ and 
$|U_{\mathrm{e}3}|^2 = 0.08$ and 
we allow \meff to vary respectively in the intervals 
$\meff = 0.40  \ m \div 0.70 \  m$ and 
$\meff = 0.20 \  m \div 0.50 \  m$.


\section{\textbf{The Partial Mass Hierarchy and Partial Inverted Mass
Hierarchy Cases}}
\label{massparhie}

\indent The partial mass hierarchy and the partial inverted mass
hierarchy spectra are characterized respectively 
by the following relations between the three neutrino masses:
\begin{equation}
m_1 \simeq  m_2 < m_3,~~~{\rm or}~~~~ 
\deltadue = \deltasol 
\ll \deltatre = \deltaatm 
\sim \boldsymbol{O}( m_1^2)~, 
\label{pmhspec}
\end{equation}
\noindent and 
\begin{equation}
m_1 <  m_2 \simeq m_3,~~~{\rm or}~~~~ 
\deltatre = \deltasol 
\ll \deltadue = \deltaatm 
\sim \boldsymbol{O}(m_1^2)~. 
\label{pinvmhspec}
\end{equation}
\noindent 
The pattern (\ref{pmhspec}) ((\ref{pinvmhspec})) interpolates 
between the hierarchical (inverted mass hierarchy) 
neutrino mass spectrum and the 
quasi-degenerate one. The two intermediate 
types of neutrino mass spectrum 
(\ref{pmhspec}) and (\ref{pinvmhspec}) 
were discussed also in \cite{KPS00}.
We will consider in the 
subsequent analysis values 
of $m_1$ from the interval:
$0.02~{\rm eV} \leq m_1 \leq 0.2~$eV. 
Theses values of $m_1$ are determined by the
values of \deltaatm, obtained in the analyzes of 
the atmospheric neutrino data, 
by requiring that $1/5 \  \deltaatm|_{\mathrm{MIN}} 
\leq m_1^2 \leq 5  \  \deltaatm|_{\mathrm{MAX}}$. 
For $m_1 > 0.2 \eV$ one recovers 
the quasi-degenerate neutrino mass
spectrum. If  $m_1 \ll 0.02 \eV$, 
we get the hierarchical 
or the inverted mass hierarchy spectrum. 
We will consider values of $m_1$ not smaller than
0.02 eV also because for $m_1 \ll 0.02 \eV$ 
the contributions of the term  $\sim m_1$ in
\meff amounts to a relatively small correction.

\subsection{Spectrum with Partial Mass Hierarchy}

\indent  Using eqs. (\ref{meffective}) and (\ref{pmhspec})
and neglecting corrections $\sim \deltasol/m_1^2$, one can express 
the effective Majorana mass parameter 
\meff in the form: 
\begin{equation}
\meff = m_1 \left| \cos^2 \theta_\odot 
( 1 - |U_{\mathrm{e} 3}|^2) \, e^{i \alpha_1} \, 
+ \, \sin^2 \theta_\odot ( 1 - |U_{\mathrm{e} 3}|^2) \,  e^{i \alpha_2} \,  
+ \, \sqrt{1 + \frac{\deltaatm}{m^2_1}} \; |U_{\mathrm{e} 3}|^2  
\, e^{i \alpha_3} \right|.
\label{meffpmh}
\end{equation}

 If CP is conserved, there are several different 
possibilities for the values of the relative 
CP-parities of the three Majorana neutrinos $\nu_{1,2,3}$.

\vskip 0.3truecm
{\bf Case A.} The three neutrinos $\nu_{1,2,3}$ have the same 
CP-parities,  $\phi_1 = \phi_2 = \phi_3$ (\textit{i.e.,} 
$\alpha_{21} = \alpha_{31} = 0$).
The effective Majorana mass \meff is given by:
\begin{equation}
\meff \cong m_1~( \ 1 - |U_{\mathrm{e} 3}|^2 + 
\sqrt{1 + \frac{\deltaatm}{m^2_1}} \; |U_{\mathrm{e} 3}|^2 ).
\end{equation}
\noindent Note that (in the approximation we are working)
\meff does not depend on the 
value of $\theta_\odot$.
Therefore, for all analyzes of the solar neutrino data
\meff is bounded to lie in the interval:

\begin{equation}
2.0 \times 10^{-2}  \eV \ltap \meff \ltap  2.0 \times 10^{-1} \eV. 
\label{pmhcp1b}
\end{equation}
\noindent These values of \meff can be probed in the future
\betabeta-decay experiments \cite{NEMO3,CUORE,GENIUS,EXO}.

\vskip 0.3truecm
{\bf Case B.} If $\phi_1 = \phi_2 = - \phi_3$ 
(i.e., 
$\alpha_{21} =0,
~\alpha_{31} =  \pm  \pi$), 
\meff has the form:
\begin{equation}
\meff \cong m_1~(1 - |U_{\mathrm{e} 3}|^2 - \sqrt{1 + 
\frac{\deltaatm}{m^2_1}} \; |U_{\mathrm{e} 3}|^2 ).
\label{pmhcp2}
\end{equation}
\noindent The three terms in the brackets in (\ref{pmhcp2}) 
can mutually compensate each other if
$|U_{\mathrm{e} 3}|^2$ has a 
value determined by the condition
\begin{equation}
|U_{\mathrm{e} 3}|^{2}_{0}= 
\frac{1}{1 + \sqrt{\frac{\deltaatm}{m^2_1}}}.
\label{pmhUe3o}
\end{equation}
\noindent Using ${\rm max}(\deltaatm) = 
8\times 10^{-3}~{\rm eV^2}$ (see eq. (5)) 
and ${\rm min}(m_1) = 0.02~{\rm eV}$,
we find that the term in the right-hand side of 
eq. (77)  
$(1 + \sqrt{\deltaatm/m^2_1})^{-1} \gtap 0.18$.
Thus, condition (\ref{pmhUe3o}) 
cannot be satisfied by the experimentally 
allowed values of $|U_{\mathrm{e} 3}|^2 $. 
Correspondingly, cancellation in 
eq. (\ref{pmhcp2}) 
is impossible and 
there exists a non-trivial lower bound on 
\meff: 
\begin{equation}
 1.6 \ (1.5) \times 10^{-2}  \eV  \leq \meff 
\leq 2.0 \ ( 2.0 ) \times 10^{-1} \eV.
\label{pmhcp2b}
\end{equation}
%
\noindent The predictions for \meff
in this case can be tested in the 
next generation of \betabeta-decay experiments
as well.

\begin{figure}[p]
\begin{center}
\epsfig{file=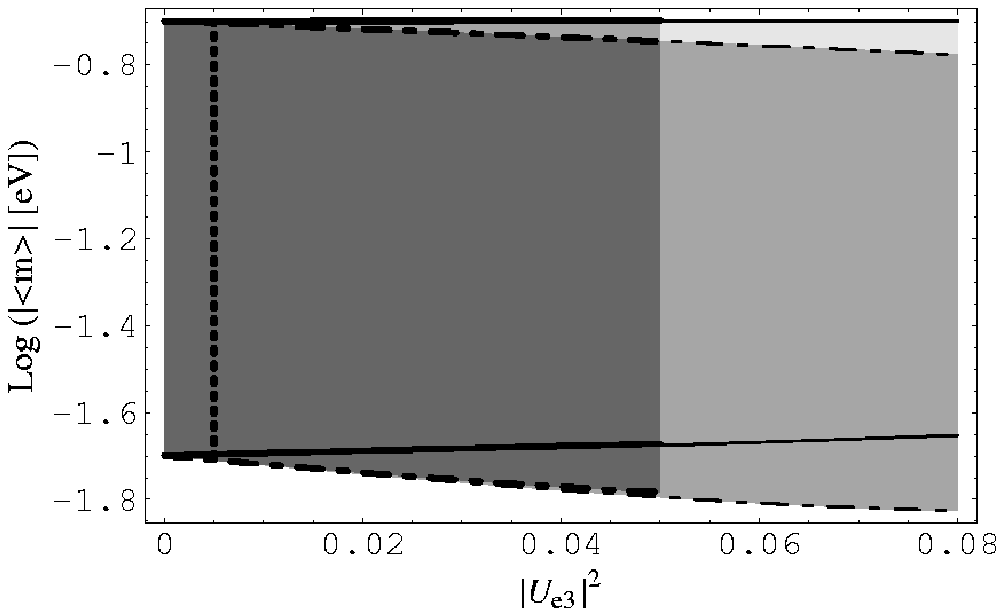, height=8cm
}
\end{center}
\caption[gmasspar01]{The effective Majorana mass 
\meff, allowed by the data from the solar 
and atmospheric neutrino
and CHOOZ experiments, as a function 
of $|U_{\mathrm{e} 3}|^2$ 
in the case of 
partial hierarchy mass spectrum, eq. (\ref{pmhspec}). 
The values of \meff  are obtained 
for \deltasol, $\sin^2 \theta_\odot$ from the 
LMA MSW solution region,
and \deltaatm and $|U_{\mathrm{e} 3}|^2$, 
derived in \cite{Gonza3nu} at  
$90\%$ C.L. (medium grey and dark grey regions at 
$|U_{\mathrm{e} 3}|^2 <0.05$)
and at $99\%$ C.L. (light, medium   
and dark grey regions at $|U_{\mathrm{e} 3}|^2 <0.08$). 
The $90\%$ ($99\%$) C.L. allowed regions 
located \textit{i)} between the two thick 
(thin) solid lines and \textit{ii)} between 
the two  thick (thin) dashed lines, correspond to the two cases of CP conservation: 
\textit{i)} 
$\phi_1 = \phi_2 = \phi_3$, and \textit{ii)} 
$\phi_1 = \phi_2 = - \phi_3$, 
$i\phi_{2,3}$ being the CP-parities of 
$\nu_{2,3}$.
The values of \meff, calculated for the best 
fit values of \deltasol, \deltaatm, $\sin^2 \theta_\odot$, $|U_{\mathrm{e} 3}|^2$ and for \  $0.02 \eV \leq m_1 \leq 0.2 \eV$ \  are 
indicated by the thick dotted line.
}
\label{figure:gmasspar01}
\end{figure}

  The dependence of \meff on  
$|U_{\mathrm{e} 3}|^2$ for 
$\phi_1 = \phi_2 = \phi_3$ and $\phi_1 = 
\phi_2 = - \phi_3$ is shown 
in Fig.~\ref{figure:gmasspar01}.
The figure is obtained for the 
90 (99)\% C.L. results of the analysis 
of ref.~\cite{Gonza3nu} and 
taking into account the 
dependence of the allowed values of 
\deltaatm on the value 
of $|U_{\mathrm{e} 3}|^2$. 
Since in both cases \meff does not 
depend on $\theta_\odot$ and \deltasol,
the results exhibited in 
Fig.~\ref{figure:gmasspar01}
are valid for all the 
different solutions of 
the solar neutrino problem.

\vskip 0.3truecm
{\bf Case C.} For $\phi_1 = - \phi_2 =   \phi_3$ 
(\textit{i.e.,}
$\alpha_{21} = \pm \pi, \ \alpha_{31}=0$), 
the effective mass \meff is given by:
\begin{equation}
\meff \cong  m_1 \left| \cos 2 \theta_{\odot}\  (1 - |U_{\mathrm{e} 3}|^2)  + \sqrt{1 +\frac{ \deltaatm}{m_1^2}}\ |U_{\mathrm{e} 3}|^2 \right|.
\label{meffpmhcp3}
\end{equation}

\noindent Now \meff depends on
$\cos 2\theta_\odot$ and we will 
get different predictions for
\meff for the different solutions 
of the solar neutrino problem.
Since for the SMA solution one has  
$\cos 2 \theta_\odot \cong 1$, 
the results for \meff corresponding to the
SMA solution practically coincide with those
in the Case~1 considered above and given in
eq. (\ref{pmhcp1b}). 
According 
to the 99\% C.L. results of ref.~\cite{Gonza3nu},  
for the LMA and LOW-QVO solutions  
$\cos 2 \theta_\odot$  can be either positive or negative 
and therefore  there can be a cancellation for any 
allowed value of $|U_{\mathrm{e} 3}|^2$. 
The intervals of allowed values of \meff read:
\begin{align}
 2.2 \times 10^{-3}~(0.0) \eV \quad & \leq &  \meff \quad  & 
\leq & 1.3 \  ( 1.4 ) \times 10^{-1}  \eV, & 
\quad\quad\quad \quad \quad \quad \quad   \text{LMA,} \\
 0 \ (0) \quad & \leq  & \meff \quad &  
\leq & 7.0 \  ( 11) \times 10^{-2} \eV, & 
\quad \quad\quad \quad \quad \quad \quad \text{LOW-QVO,} \\
  2.0 \ (2.0) \times 10^{-2} \eV \quad & \leq  & \meff \quad  & 
\leq & 2.0 \ ( 2.0) \times 10^{-1} \eV, & 
\quad  \quad \quad \quad \quad \quad  \quad \text{SMA.} 
\end{align}

\noindent Since for the LMA solution the corrections
$\sim \Delta m^2_{21}/m^2_{1} = \deltasol/m^2_1$
can be as large as $\sim 2.3\times 10^{-3}~$eV,
the (99\% C.L.) lower limit,
${\rm min}(\meff) = 0~$eV, quoted above,
can actually be any value between
$\sim 2.3\times 10^{-3}~$eV and 0 eV.
In contrast, from the 
analysis of ref.~\cite{SKYSuz00} it follows 
that $\cos 2 \theta_\odot > 0$ for the
LMA and LOW-QVO 
solutions. Thus, according to the results
of \cite{SKYSuz00},
cancellation of the two-contributions 
in eq. (\ref{meffpmhcp3})
is excluded for both the LMA and the LOW-QVO 
solutions, leading to the following intervals 
of possible values of \meff:
\begin{align}
 4.0 \times 10^{-3} \eV \quad & \leq & \meff \quad & 
\leq  &  1.4  \times 10^{-1}  \eV & 
\quad \quad\quad \quad \quad \quad \quad \quad  \quad   \text{LMA,} \\
 6.6 \times 10^{-3} \eV \quad & \leq &  \meff\quad &  \leq & 1.2 \times 10^{-1} \eV & 
\quad \quad\quad \quad \quad \quad \quad \quad  \quad  \text{LOW-QVO.} 
\end{align}

\vskip 0.3truecm
{\bf Case D.} For $\phi_1 = - \phi_2 = -  \phi_3$ 
(\textit{i.e.,} 
$\alpha_{21} = \alpha_{31}= \pm \pi$), 
the expression for \meff can be written as:
\begin{equation}
\meff \simeq m_1 \left| \cos 2 \theta_{\odot} (1 - |U_{\mathrm{e} 3}|^2)  
- \sqrt{1 +\frac{ \deltaatm}{m_1^2}}\ |U_{\mathrm{e} 3}|^2 \right|.
\label{meffpmhcp4}
\end{equation}
\noindent Most of the observations made for the previous case 
hold also for this one. The results obtained in 
ref.~\cite{Gonza3nu} imply the following allowed 
intervals of values of \meff:
\begin{align}
 8.0 \times 10^{-4}~(0.0)\eV\quad  & \leq & \meff\quad  & \leq & 1.2 \  ( 1.3 ) \times 10^{-1}  \eV, & \quad  \quad \quad \quad \quad \quad  \quad  \text{LMA,} \\
 0 \ (0)\quad  & \leq & \meff\quad  & \leq & 5.0  \  ( 14) \times 10^{-2} \eV, & 
\quad  \quad \quad \quad \quad \quad  \quad  \text{LOW-QVO,} \\
 1.6  \ (1.5) \times 10^{-2} \eV  \quad & \leq & \meff\quad  & 
\leq & 2.0  \ ( 2.0) \times 10^{-1}   \eV, 
& \quad  \quad \quad \quad \quad \quad  \quad  \text{SMA.} 
\end{align}
 
\begin{figure}[p]
\begin{center}
\epsfig{file=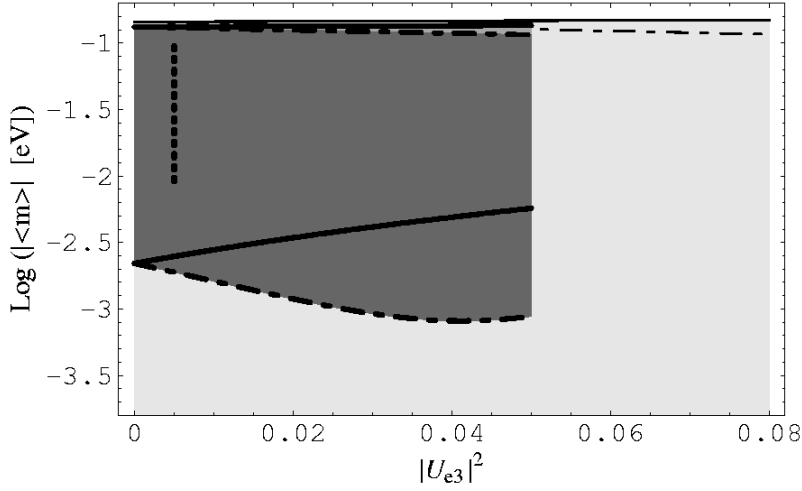, height=6.5cm
}
\end{center}
\caption[gmasspar02]{The same as Fig. \ref{figure:gmasspar01}
but for two different sets of values 
of the relative CP-parities of the massive Majorana 
neutrinos. The allowed regions are derived using 
the results of \cite{Gonza3nu} at  
$90\%$ C.L. (medium grey and dark grey regions at 
$|U_{\mathrm{e} 3}|^2 <0.05$)
and at $99\%$ C.L. (light, medium   
and dark grey regions at $|U_{\mathrm{e} 3}|^2 <0.08$). 
The $90\%$ ($99\%$) C.L. allowed regions 
located \textit{i)} between the two thick 
 solid lines (between the thin solid  line 
and the horizontal axis) and \textit{ii)} between 
the two  thick  dashed lines (between the thin dashed line and the horizontal axis), correspond to the two cases  of CP conservation: 
\textit{i)} 
$\phi_1 = - \phi_2 = \phi_3$, and \textit{ii)} 
$\phi_1 = - \phi_2 = - \phi_3$, 
$i\phi_{2,3}$ being the CP-parities of 
$\nu_{2,3}$.
The values of \meff, calculated for the best 
fit values of \deltasol, \deltaatm, $\sin^2 \theta_\odot$, $|U_{\mathrm{e} 3}|^2$ and for $0.02 \eV \leq m_1 \leq 0.2 \eV$  are 
indicated by the thick dotted line.
}
\label{figure:gmasspar02}
\end{figure}

\begin{figure}[p]
\begin{center}
\epsfig{file=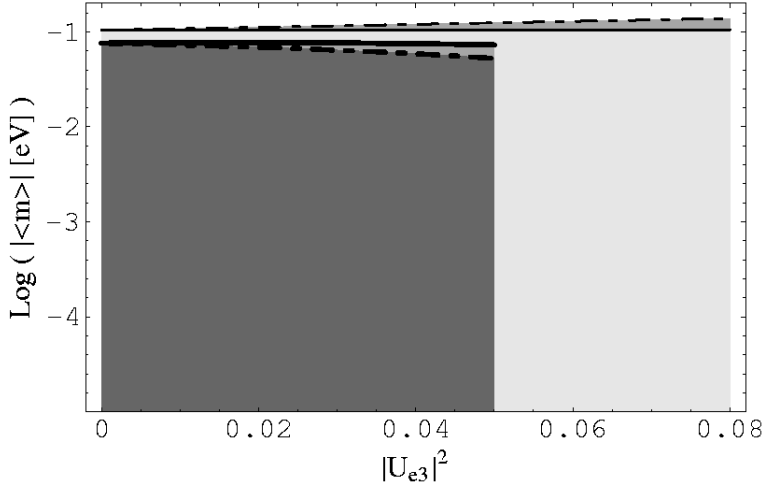, height=6.5cm
}
\end{center}
\caption[gmasspar03]{The same as in 
Fig. \ref{figure:gmasspar02} but 
for the LOW-QVO solution of the 
solar neutrino problem. 
The $90\%$ ($99\%$) C.L. allowed regions 
located \textit{i)} between the  thick (thin) 
solid line 
and the horizontal axis and \textit{ii)} between 
the   thick (thin) dashed line 
and the horizontal axis , 
correspond to the two cases of CP conservation: 
\textit{i)} 
$\phi_1 = - \phi_2 = \phi_3$, and \textit{ii)} 
$\phi_1 = - \phi_2 = - \phi_3$, 
$i\phi_{2,3}$ being the CP-parities of 
$\nu_{2,3}$.
}
\label{figure:gmasspar03}
\end{figure}

 The dependence of \meff on  
$|U_{\mathrm{e} 3}|^2$
for the two cases $\phi_1 = - \phi_2 = \pm  \phi_3$
is shown in
Fig.~\ref{figure:gmasspar02} and in 
Fig.~\ref{figure:gmasspar03}, respectively.
The 90 and 99\% C.L. results of 
ref.~\cite{Gonza3nu} were used for the figures
(in particular, the dependence of \deltaatm and 
$\theta_\odot$ on $|U_{\mathrm{e} 3}|^2$
was taken into account).

\begin{figure}[p]
\begin{center}
\epsfig{file=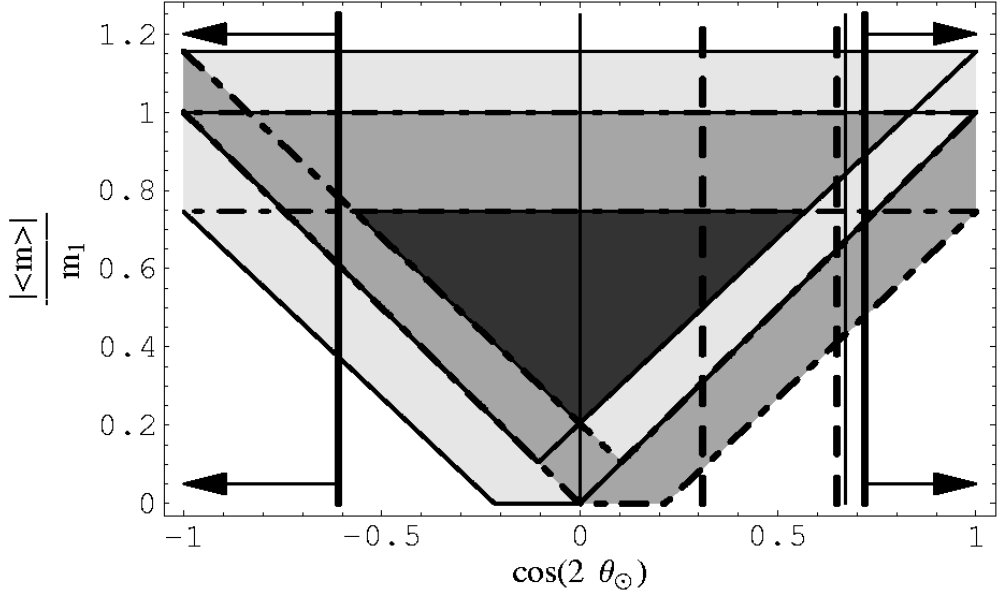, height=8cm
}
\end{center}
\caption[gmasspar04]{The dependence of $\meff/m_1$ 
on $\cos 2 \theta_\odot$ 
for the partial hierarchy  mass spectrum, 
eq. (\ref{pmhspec}),
and the LMA solution of the 
$\nu_{\odot}-$problem, 
for the 99\%~C.L. allowed values of 
the solar and atmospheric neutrino oscillation parameters 
and  $0.02 \eV \leq m_1 \leq 0.2 \eV$.
If CP-invariance holds, the values of 
$\meff/m_1$ lie:
i) for $\phi_1 = \phi_2 = \phi_3$ -
in the region between the two thin solid horizontal lines (in light and medium grey), 
ii) for $\phi_1 = \phi_2 = - \phi_3$ - 
in the region between the two thick horizontal 
dash-dotted lines (in light  grey and 
medium grey colors),
iii) for $\phi_1 =- \phi_2 = + \phi_3$ - in 
the light grey polygon with solid-line contours
and iv) for $\phi_1 = -  \phi_2 = - \phi_3$ -
in the medium grey 
polygon with the dash-dotted-line contours.
The ``just-CP-violation'' region is denoted 
by dark-grey color. The 
values of $\cos 2 \theta_\odot$ 
between the doubly thick solid,
the normal solid and the 
doubly thick dashed lines
correspond to the 
99\%, 99\% and 95\% C.L.
LMA solution regions in refs. \cite{Gonza3nu}, 
\cite{Fogli00} and \cite{SKYSuz00}, respectively. 
 }
\label{figure:masspar04}
\end{figure}

 In order to illustrate the dependence of \meff on 
$\theta_\odot$,  
we plot in Fig.~\ref{figure:masspar04} 
the ratio $\meff / m_1$  versus $\cos 2 \theta_\odot$ 
for the four CP-conserving cases discussed above. 
The region marked with dark-grey scale
corresponds to ``just-CP-violation''. 
A value of $\meff/m_1$ in this region
would signal the 
existence of CP-violation 
in the leptonic sector,
induced by the Majorana CP-violating phases:
$\meff/m_1$ cannot take a value 
in the indicated region
if CP-parity is conserved.

Note that, as we have shown in Section 5,
if the neutrino mass spectrum is 
of the inverted hierarchy type,
one typically has 
${\rm max}(\meff) \ltap 0.06~$eV,
with the largest allowed value
of $\meff \ltap (0.08 - 0.09)~$eV
possible only in the case of the
SMA solution of the 
$\nu_{\odot}-$ problem. 
This has to be compared with 
${\rm max}(\meff) = 
{\rm max}(m_1) \cong 0.2~$eV
one finds for the partial mass
hierarchy spectrum 
under study.

  If the CP-symmetry is violated, the two CP-violating phases
entering into the expression (\ref{meffpmh}) for \meff,
$\alpha_{21}$ and $\alpha_{31}$, 
must obey the following relation: 

\begin{equation}
\begin{split}
 \meff^2 - m_1^2 \cos^4 \! \theta_\odot \, 
(1- |U_{\mathrm{e} 3}|^2)^2 - m_1^2 \sin^4 \! \theta_\odot \, 
(1- |U_{\mathrm{e} 3}|^2)^2 - (m_1^2 + \deltaatm) |U_{\mathrm{e} 3}|^4 = \\
  2 \,  m_1^2 \cos^2 \! \theta_\odot \, \sin^2 \! 
\theta_\odot \, (1- |U_{\mathrm{e} 3}|^2)^2 \cos \alpha_{21} + 
2 m_1 \sqrt{m_1^2 + \deltaatm} \cos^2 \! \theta_\odot \, 
|U_{\mathrm{e} 3}|^2 \, (1- |U_{\mathrm{e} 3}|^2) \cos \alpha_{31} \\ + 
2  m_1 \sqrt{m_1^2 + \deltaatm} \sin^2 \!  \theta_\odot \, 
|U_{\mathrm{e} 3}|^2 \, (1- |U_{\mathrm{e} 3}|^2)
(\cos \alpha_{21} \cos \alpha_{31} + \sin \alpha_{21} \sin \alpha_{31}) .
\end{split}  
\label{formulacospar}
\end{equation}

\begin{figure}[p]
\begin{center}
\epsfig{file=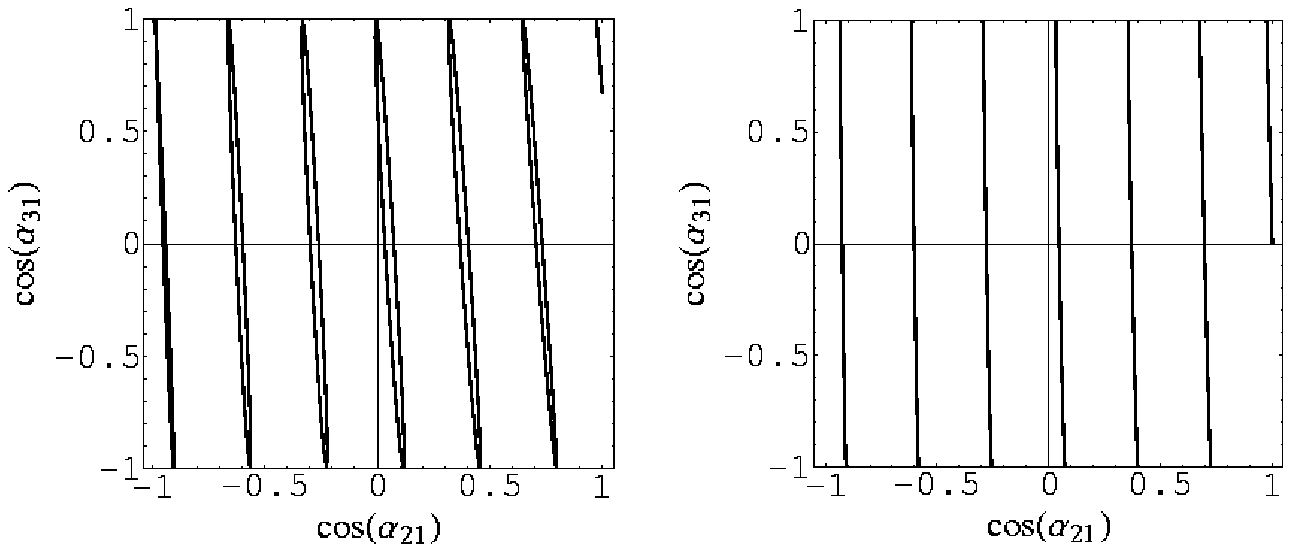, height=6.5cm 
}
\end{center} 
\caption[gmasspar05]{The interdependence 
of the two CP-violating 
phases, $\alpha_{21}$ and $\alpha_{31}$, for a given value of the ratio
$\meff/m_1$ in the case of partially  hierarchical  neutrino mass spectrum.
The figures are obtained for 
$ \meff = \sqrt{1 + 0.5 n} \times 10^{-2} \eV$ with 
$n= 0,1 \ldots 6$ (with increasing \meff from left to right), the best fit values of the 
solar and atmospheric neutrino oscillation parameters
from \cite{Gonza3nu}, quoted in Section 2, and $m_1 = 0.02 \eV$
(left-hand plot), and for 
$ \meff = \sqrt{ 0.5 n} \times 10^{-1} \eV $
 with $n= 0,1 \ldots 6 $ (with increasing \meff from left to right), the best fit values of the 
solar and atmospheric neutrino oscillation parameters
from \cite{Gonza3nu} and $m_1 = 0.2 \eV$
(right-hand plot).
The values of $\cos \alpha_{21,31} =0,\pm 1$, 
correspond to CP-invariance.
} 
\label{figure:gmasspar05}
\end{figure}

\begin{figure}[p]
\begin{center}
\epsfig{file=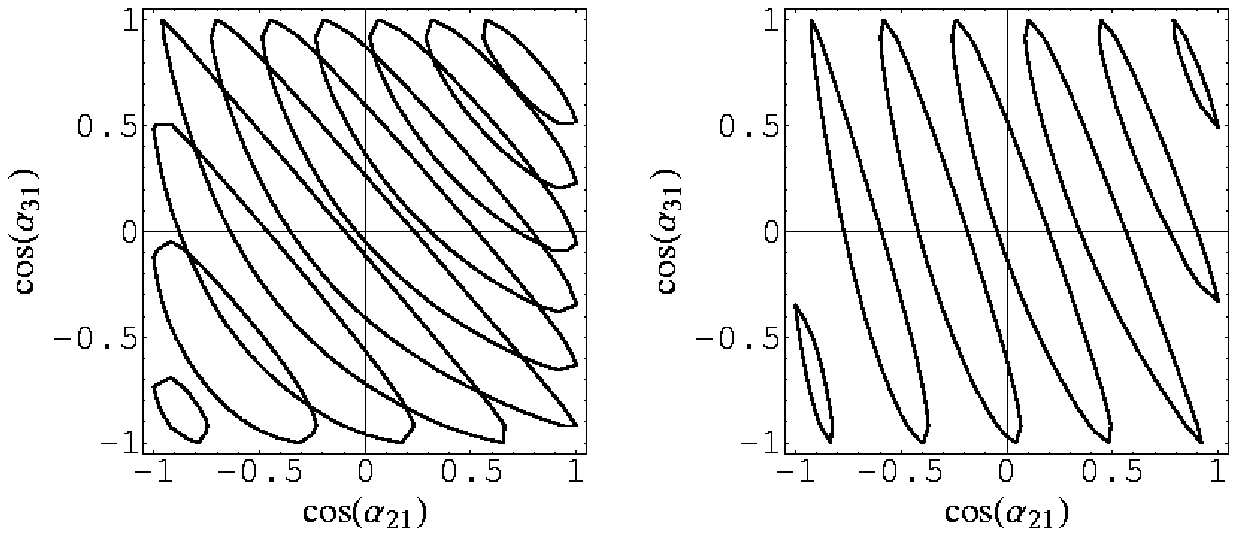, height=6.5cm
}
\end{center} 
\caption[gmasspar06]{The same as in Fig.(\ref{figure:gmasspar05}) but for 
$ \meff = \sqrt{ 0.5 n} \times 10^{-2} \eV$ with 
$n= 1 \ldots 10$ (with increasing \meff from left to right), $|U_{\mathrm{e} 3}|^2 = 0.08$,  the best fit values of the other  
solar and atmospheric neutrino oscillation parameters
from \cite{Gonza3nu}, quoted in Section 2, and $m_1 = 0.02 \eV$
(left-hand plot), and for 
$ \meff = \sqrt{1 +  0.5 n} \times 10^{-1} \eV $
 with $n= 0,1 \ldots 6 $ 
(with increasing \meff from left to right), $|U_{\mathrm{e} 3}|^2 = 0.08$,
 the best fit values of the other 
solar and atmospheric neutrino oscillation parameters
from \cite{Gonza3nu} and $m_1 = 0.2 \eV$
(right-hand plot).
The values of $\cos \alpha_{21,31} = 0,\pm 1$, 
correspond to CP-invariance.
}
\label{figure:gmasspar06}
\end{figure}

\noindent One could constrain  
$\alpha_{21}$ and $\alpha_{31}$
once the values of all other observables
in eq. (\ref{formulacospar})
are measured with a sufficiently good precision.
In Fig.~\ref{figure:gmasspar05} 
and  Fig.~\ref{figure:gmasspar06}  we 
illustrate this possibility
by plotting $\cos \alpha_{21}$ versus 
$\cos \alpha_{31}$ for 
the best fit values of \deltasol, $\theta_{\odot}$, 
\deltaatm and $|U_{\mathrm{e} 3}|^2$ found
in \cite{Gonza3nu} (see Section 2),
$m_1 = 0.02;~0.2 \eV$, and 
for  $|U_{\mathrm{e} 3}|^2 = 0.08$, 
best fit values of the 
remaining three parameters and 
$m_1 = 0.02;~0.2 \eV$, respectively.
Let us note, however, that
obtaining experimental information
about the mass $m_1~(\simeq m_2)$
which in the case under study
is supposed to have a value
$\sim (0.02 - 0.20)~$eV, seems 
at present to be extremely difficult.

\subsection{Spectrum with Partial Inverted Hierarchy}

\indent  In this case $|U_{\mathrm{e} 1}|^2 = \sin^2\theta$ 
and is constrained by the CHOOZ limit. The general expression for \meff
has the form
\begin{equation}
\meff = \left| m_1|U_{\mathrm{e} 1}|^2~e^{i\alpha_1}
+ (1 - |U_{\mathrm{e} 1}|^2)\sqrt{m_1^2 + \deltaatm}
(\cos^2\theta_{\odot}  + \sin^2\theta_{\odot}~  
e^{i(\alpha_3 - \alpha_2)})\, e^{i\alpha_2} \right|.
\label{meffpimh}
\end{equation}

\noindent For relatively small values of 
$|U_{\mathrm{e} 1}|^2$, e.g., $|U_{\mathrm{e} 1}|^2 \ltap 0.01$,
the term $m_1|U_{\mathrm{e} 1}|^2 \ltap 2\times 10^{-3}~$eV
and neglecting it we get 
the following simple expression for \meff :
\begin{equation}
\meff \cong \left| \cos^2\theta_\odot  + \sin^2\theta_\odot~ 
e^{i (\alpha_3- \alpha_2)} \right| \sqrt{m_1^2 + \deltaatm}.
\label{meffpimh1}
\end{equation}

  In the CP-conserving case, there are two major possibilities 
for the values of the neutrino CP-parities.

\vskip 0.3truecm
{\bf Case A.} If $ \phi_2 =  \phi_3 = \pm \phi_1$ (\textit{i.e.,} 
$\alpha_{21} = \alpha_{31} = 0, \pm \pi$), 
the effective Majorana mass parameter \meff is given by:
\begin{equation}
\meff  = \sqrt{m_1^2 + \deltaatm}(1 - |U_{\mathrm{e} 1}|^2) 
\pm m_1|U_{\mathrm{e} 1}|^2,
\label{pimhcp1}
\end{equation}

\noindent where $m_1|U_{\mathrm{e} 1}|^2 \leq 1.8\times 10^{-2}~$eV.
A cancellation between the two terms in (\ref{pimhcp1}) 
is impossible and one has a non-trivial lower bound on 
\meff. The allowed values of \meff do not depend 
on $\theta_\odot$ and therefore are the same for 
all solutions of the $\nu_{\odot}-$problem
and the different analyzes of the solar neutrino data.
They depend very weakly on $|U_{\mathrm{e} 1}|^2$. 
We have: 
\begin{equation*}
 2.8 \ (1.8) \times 10^{-2}   \eV \leq \ltap 
\meff \leq  \ltap  2.1 \ ( 2.2 ) \times 10^{-1} \eV.
\label{pimhcp1meff}
\end{equation*}

\begin{figure}[p]
\begin{center}
\epsfig{file=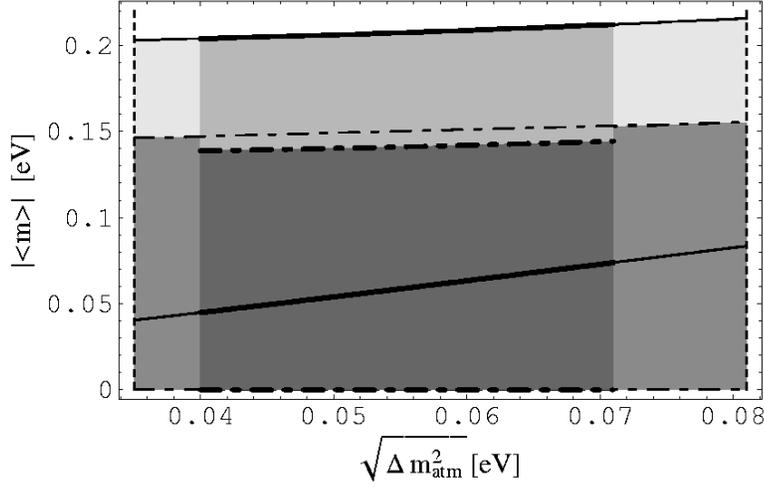, height=6.5cm
}
\end{center}
\caption[gmassparinv01]{The effective Majorana mass 
\meff as a function of $\sqrt{\deltaatm}$ 
for the neutrino mass spectrum of the partial  inverted hierarchy
type, eq. (\ref{pinvmhspec}). 
The allowed regions (in grey) 
correspond to the LMA solution 
of ref.~\cite{Gonza3nu}.  
In the case of CP-invariance and for the
$90\%~ (99\%)$ C.L. 
solution regions, \meff can have values 
{\it i}) for $\phi_2 = \phi_3$ - in the  medium-light  and dark grey  
(light grey, medium-light, medium-dark grey   and dark grey)
 upper  region, limited by the
doubly thick (thick and doubly thick) solid lines, and 
ii) for $\phi_2 = - \phi_3$ - in the dark grey  
(medium-light grey, medium-dark grey and dark grey) 
region limited
by the doubly thick (thick and doubly thick) 
dash-dotted lines.   
If CP is not conserved, \meff can lie in any of the 
regions marked by different grey scales.      
}
\label{figure:massparinv01}
\end{figure}

\begin{figure}[p]
\begin{center}
\epsfig{file=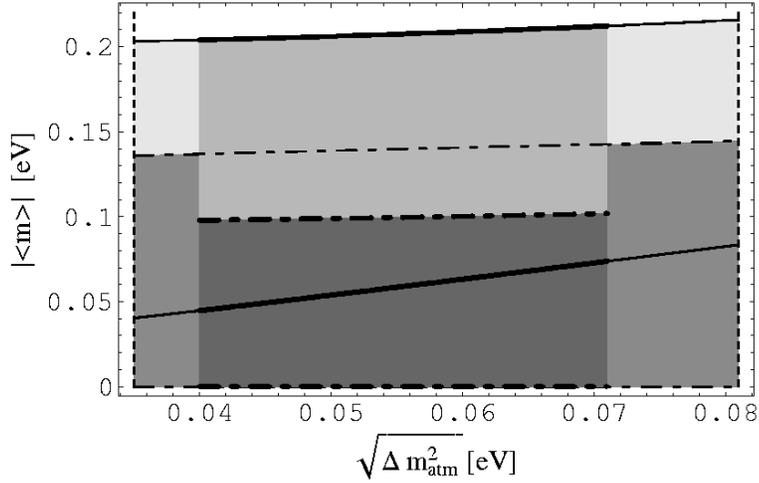, height=6.5cm
}
\end{center}
\caption[gmassparinv02]{
The same as Fig.(\ref{figure:massparinv01}) 
for the $90\%$ ($99\%$) C.L. LOW-QVO solution
of ref. \cite{Gonza3nu}. 
}
\label{figure:massparinv02}
\end{figure}

\vskip 0.3truecm 
{\bf Case B.} For  $\phi_2 = - \phi_3 \pm \phi_1$ (\textit{i.e.,} 
$(\alpha_3 - \alpha_2) = \pm \pi,~\alpha_{21}, \alpha_{31} = 0,\pm \pi$), 
\meff takes the form:
\begin{equation}
\meff_{\pm} \simeq \left| \cos 2 \theta_\odot~\sqrt{m_1^2 + \deltaatm}
(1 - |U_{\mathrm{e} 1}|^2) 
\pm m_1|U_{\mathrm{e} 1}|^2 \right|.
\end{equation}
\noindent For the SMA solution we have 
$\cos 2 \theta_\odot \cong 1$ and 
the results for \meff are equivalent to those
in the Case 1. For the LMA and 
LOW-QVO solutions the values \meff can have
depend on the solution
and on the data analysis. 
In Fig.~\ref{figure:massparinv01} and in 
Fig.~\ref{figure:massparinv02} we plot the allowed 
regions of values of \meff versus \deltaatm 
for the LMA and 
LOW-QVO solutions, respectively, of
ref.~\cite{Gonza3nu}. 
Since  for both solutions values of
$|\cos 2 \theta_\odot| < 0.09$  
are possible (at 90\% and 99\% C.L.) \cite{Gonza3nu}, 
and we can have also $|U_{\mathrm{e} 1}|^2 \cong 0$,
one does not get a significant lower bound on \meff.
The values of \meff,
corresponding to the three solutions of the solar neutrino problem
and the different data analyzes in \cite{SKYSuz00}, 
\cite{Fogli00} and \cite{Gonza3nu}, 
are reported in Table~\ref{table6}.
According to the analysis of ref.~\cite{SKYSuz00}, 
$\cos 2 \theta_\odot > 0.2~(0.4)$ for the LMA (LOW-QVO)
solution. This leads to the non-trivial lower
bounds on the possible values of \meff for the 
indicated solutions, cited in Table 6.
 
\begin{figure}[p]
\begin{center}
\epsfig{file=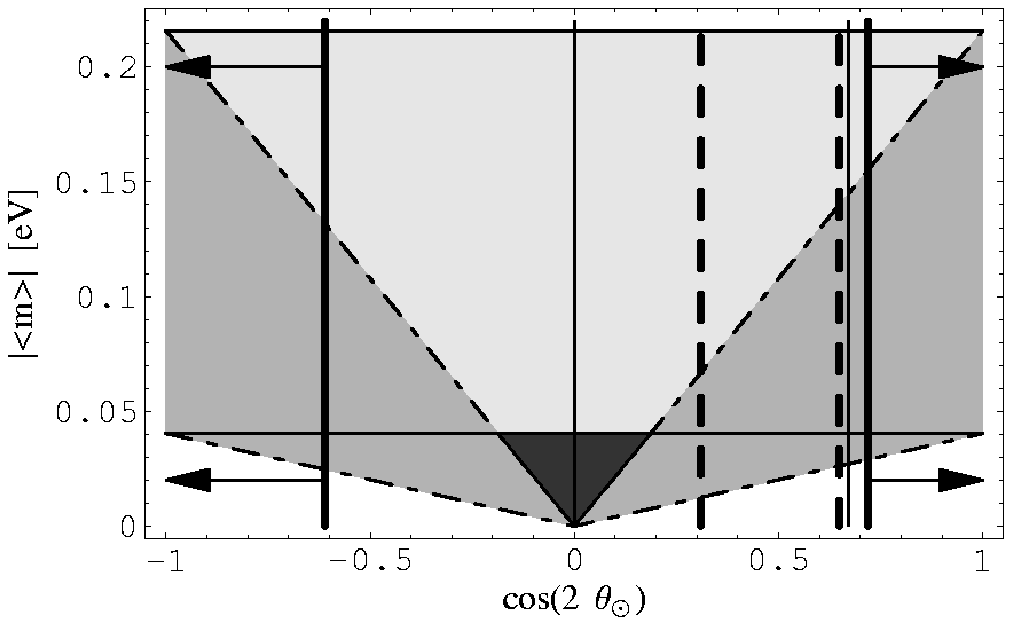, height=8cm
}
\end{center}
\caption[gmassparinv03]{The dependence of 
\meff on $\cos 2 \theta_\odot$ 
for the neutrino mass spectrum
with partial inverted hierarchy, 
eq. (\ref{pinvmhspec}), 
and the LMA solution of the 
$\nu_{\odot}-$problem.
The region between the two thick horizontal 
solid lines (in light grey and medium grey colors)
and the two triangular regions between the thick 
dash-dotted lines (in medium grey color),
correspond to the two CP-
conserving cases, $\phi_2 = \phi_3$ 
and $\phi_2 = - \phi_3$, respectively.
The ``just-CP-violation'' region is denoted 
by dark-grey color. 
The regions between each of the three pairs 
of vertical lines of a given type
- solid, doubly thick solid and doubly thick dashed,
correspond to the intervals of values of 
$\cos 2 \theta_\odot$ 
for the LMA solution derived (at 99\% C.L.) 
in ref.~\cite{SKYSuz00}
(region between the two doubly thick dashed lines),
ref.~\cite{Fogli00} (region between the solid lines)
and ref.~\cite{Gonza3nu}. 
 } 
\label{figure:massparinv03}
\end{figure}

 The dependence of 
\meff on $\cos 2\theta_\odot$ is 
exhibited in Fig.~\ref{figure:massparinv03}. 
The uncertainty in the allowed ranges of values of
\meff is mainly 
due to the relatively large interval 
of possible values of $m_1$ we consider. 

\begin{figure}[p]
\begin{center}
\epsfig{file=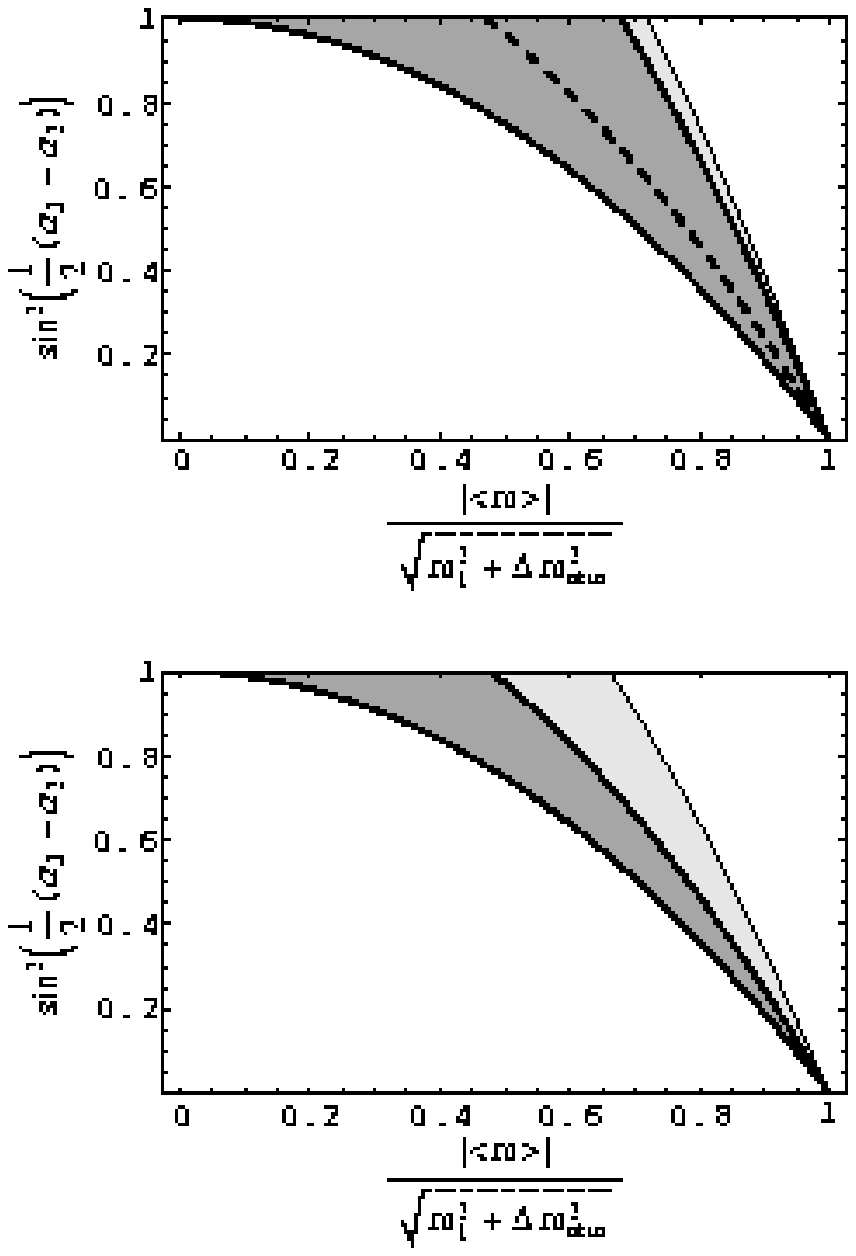, height=15cm
}
\end{center}
\caption[gmassparinv06]{ The CP-violation factor 
$\sin^2 (\alpha_2 - \alpha_3)/2$ 
as a function of \meff
in the case of partially inverted mass hierarchy spectrum, 
eq. (\ref{pinvmhspec}), and the LMA  (upper panel) and 
LOW-QVO (lower panel) solutions of ref.~\cite{Gonza3nu},
obtained at $90\%$ C.L. (medium-grey region 
with thick contours) and $99\%$ C.L.
(light grey + medium grey region limited by 
ordinary solid lines). 
The values of \meff, corresponding to the best fit 
values of the input parameters, found in \cite{Gonza3nu},
are indicated by the dashed line.
A value of $\sin^2 (\alpha_2 - \alpha_3)/2 \neq 0,1$, 
would signal CP-violation. 
}
\label{figure:massparinv04}
\end{figure}

 If CP-is not conserved and 
$|U_{\mathrm{e} 1}|^2$ is shown 
to be sufficiently small,
so that for $m_1 \leq 0.2~$eV 
the term $m_1 |U_{\mathrm{e} 1}|^2$
in eq. (\ref{meffpimh}) can be neglected,
one has the following simple relation between 
$\sin^2(\alpha_3 - \alpha_2)/2$ and the observables
\meff, $\sin^22\theta_{\odot}$ and $m_1^2 + \deltaatm$:
\begin{equation}
\sin^2 \frac{\alpha_3 - \alpha_2}{2} = 
\frac{1}{\sin^2 2 \theta_\odot} \, 
\left( 1- \frac{\meff^2}{m_1^2 + \deltaatm} \right).
\end{equation}
\noindent The allowed ranges of $\sin^2 (\alpha_3 - \alpha_2)/2$ 
as a function of $\meff / \sqrt{m_1^2 + \deltaatm}$   
are reported in Fig.~\ref{figure:massparinv04}  
for the LMA and LOW-QVO solutions \cite{Gonza3nu}, 
respectively. Obviously, knowing \meff, 
$\sin^22\theta_{\odot}$ and $(m_1^2 + \deltaatm)$
experimentally would allow to get information about the
CP-violation caused by the Majorana CP-violating phases.
We note that this conclusion is not valid 
for the SMA solution of the solar neutrino problem
since for this solution $\sin^2 \theta_\odot \ll 1$
and (see eq. (\ref{meffpimh1}))
\begin{equation}
\meff \cong \sqrt{m_1^2 + \deltaatm},~~~~~~~~~~~~~{\rm SMA~MSW}.
\end{equation}
%
\begin{table}[H]
\caption{Values of \meff, 
corresponding to the different 
solutions of the solar neutrino
problem, neutrino oscillation 
interpretation
of the atmospheric neutrino data 
and the CHOOZ limit. The results are obtained 
for the mass spectrum with partial 
inverted hierarchy, 
eq. (\ref{pinvmhspec}), 
CP-conservation, $\phi_2 = - \phi_3 \pm \phi_1$, 
negligible $m_1|U_{\mathrm{e} 1}|^2$, and 
$0.02~{\rm eV}\leq m_1 \leq 0.20~{\rm eV}$
(see text for further details).   
}
\label{table6}
\setlength{\extrarowheight}{4pt}
\begin{center}
\begin{tabular}{|c|c|c|} \hline
Data from    &                 & $ \meff \ [\eV]$  \\ \hline \hline
Ref. \cite{SKYSuz00}        &   LMA           &  $0   \div 1.4\times 10^{-1}$ \\
$(95 \%\,  \mathrm{ C.L.} )$    & LOW-QVO         &  $1.0 \times 10^{-2}  \div 1.1 \times 10^{-1} $ \\ \hline  
Ref. \cite{Fogli00}       &   LMA           & $0 \ (0) \times 10^{-2}  \div 1.4 \ (1.5)  \times 10^{-1} $ \\
    $(90 \ (99 \%) \, \mathrm{ C.L.} )$ & SMA             & $ (1.8  \times 10^{-2}  \div 2.2  \times 10^{-1}) $\\
\mbox{}     & LOW-QVO         & $ 0  \ (0 ) \times 10^{-3}  \div 0.5 \ (1.1 ) \times 10^{-1} $ \\ \hline
Ref. \cite{Gonza3nu}  & LMA         & $ 0  \ ( 0)   \div 1.4 \ (1.5 ) \times 10^{-1} $ \\
     $(90 \ (99 \%) \, \mathrm{ C.L.} )$     & SMA         & $ 2.8 \ (1.8 ) \times 10^{-2}  \div 2.1 \ (2.2 ) \times 10^{-1}$ \\
\mbox{}         & LOW-QVO     & $  0  \ ( 0)   \div 1.0  \ (1.4 ) \times 10^{-1} $ \\ \hline
\end{tabular}
\end{center}
\end{table}
%
\noindent However, the measurement of \meff
in this case would allow to determine $m_1$
and thus the whole neutrino mass spectrum.

\vspace{0.4cm}
\section{Conclusions}

  Assuming three-neutrino mixing 
and massive Majorana neutrinos,
we have studied in detail 
the implications of the neutrino 
oscillation solutions of the solar 
and atmospheric neutrino problems
and of the data of the 
reactor long baseline CHOOZ and Palo Verde 
experiments \cite{CHOOZ,PaloV},
for the predictions of the
effective Majorana mass 
\meff, which determines
the \betabeta-decay rate. The results 
of the neutrino oscillation fits of the
latest solar and atmospheric neutrino data
\cite{SKYSuz00,SKatm00,Fogli00,Gonza3nu}
have been used in our analysis.
We have considered essentially all 
possible types of neutrino
mass spectrum 
compatible with the existing data: 
hierarchical, 
neutrino mass spectrum with 
inverted hierarchy, 
with partial hierarchy, 
with partial inverted hierarchy,
and with three quasi-degenerate neutrinos.
In the present study we have investigated  
the general case of CP-nonconservation.
The \betabeta-decay effective Majorana mass
\meff depends, in general, 
on two physical (Majorana) CP-violating 
phases. For each of the five types of 
neutrino mass spectrum indicated above we 
have derived detailed predictions for
the values of \meff for the three solutions 
of the solar neutrino problem,
favored by the current solar neutrino data:
the LMA MSW, the SMA MSW and the LOW-QVO one.
In each of these cases we have identified 
the ``just-CP-violation'' region
of values of \meff whenever it existed:
a value of \meff in this
region would unambiguously signal the presence of
CP-violation in the lepton sector.
Analyzing the case of CP-conservation, we 
have derived predictions for \meff corresponding 
to all possible sets of values of the
relative CP-parities of the three massive Majorana
neutrinos. This was done for all  
different types of neutrino mass spectrum
we have considered. We have investigated
the possibility of cancellation 
between the different terms contributing to
\meff. The cases when such a cancellation
is impossible and there exists
a non-trivial lower bound
on \meff  were identified and the
corresponding lower bounds
were given.

   More specifically, if the neutrino mass 
spectrum is hierarchical, only the contributions
to \meff due to the exchange of the two heavier 
Majorana neutrinos $\nu_{2,3}$ can be relevant
and \meff depends only on 
one CP-violating phase, $\alpha_{32}$. 
For the SMA and LOW-QVO solutions of 
the solar neutrino problem
one has $\meff \ltap 4.0\times 10^{-3}~{\rm eV}$.
For the LMA solution we get
$\meff \ltap 8.5\times 10^{-3}~{\rm eV}$
if one uses the results of the analyzes in
\cite{SKYSuz00,Fogli00}, while the results of
the 3-neutrino mixing analysis 
of \cite{Gonza3nu} allow a somewhat larger value of
$\meff \ltap (1.0 - 2.0)\times 10^{-2}~{\rm eV}$.
The maximal values of \meff correspond to
CP-conservation and 
$\nu_{2}$ and $\nu_{3}$ having 
identical CP-parities, $\phi_2 = \phi_3$.
If $\phi_2 = - \phi_3$, the
allowed maximal values of \meff
are somewhat smaller. 
For all three solutions there are no 
significant lower bounds on \meff:
the lower bound in any case does not exceed 
$8.0\times 10^{-4}~{\rm eV}$.
Deep mutual compensations
between the terms contributing
to \meff, corresponding to the exchange of different 
virtual massive Majorana neutrinos, are possible.
Such cancellation can take place
both if CP is not conserved 
and in the case of CP-conservation
if $\phi_2 = - \phi_3$.
The degree of compensation depends for given
values of the solar and 
atmospheric neutrino oscillation
parameters \deltasol, $\theta_{\odot}$ and 
\deltaatm, on the value of $|U_{e3}|^2$,
which is constrained by the CHOOZ data.
The problem of compensations is most relevant  
in the case of the LMA MSW solution
of the $\nu_{\odot}-$problem,
since in this case \meff can have 
the largest possible values 
for the hierarchical neutrino mass spectrum.
Due to the relatively large
allowed ranges of values of
\deltasol, $\theta_{\odot}$ and 
\deltaatm, there are no
``just-CP-violation'' regions of \meff .
A better determination of 
\deltasol, $\theta_{\odot}$ and 
\deltaatm together with
a sufficiently accurate 
measurement of \meff  could
allow to get a direct information
on the CP-violating phase
$\alpha_{32}$. These results are 
summarized in Figs. 2 - 5.

   In the case of the neutrino mass 
spectrum with inverted hierarchy, 
eqs. (\ref{invmhspec}) - (\ref{invm23}),
the dominant contribution to \meff
are again due to 
the two heavier Majorana neutrinos 
$\nu_{2,3}$ and \meff is determined 
by \deltaatm, $\sin^22\theta_{\odot}$,
the CP-violating phase $\alpha_{32}$ 
and by $|U_{e1}|^2$,
which is constrained by the CHOOZ data.
The effective Majorana mass 
can be considerably larger
than in the case of a 
hierarchical neutrino mass spectrum:
$\meff \ltap (6.8 - 8.9)\times 10^{-2}~{\rm eV}$. 
The maximal \meff corresponds to
CP-conservation and 
$\phi_2 = \phi_3$ ($\alpha_{32} =0$); 
it is possible for all three
solutions of the $\nu_{\odot}-$problem.
It varies weakly with the 
results of the different analyzes being used.
For $\phi_2 = - \phi_3$ ($\alpha_{32} = \pm \pi$)
the maximal allowed values of
\meff are the same for the SMA MSW solution,
while we have  
$\meff \ltap (4.8 - 5.9) \times 10^{-2}~{\rm eV}$
~~($\meff \ltap (3.7 - 5.4) \times 10^{-2}~{\rm eV}$)
for the LMA MSW (LOW-QVO) solution.
For all three solutions there 
exists a significant lower bound of 
$\meff \gtap (3 - 4) \times 10^{-2}~{\rm eV}$
if $\phi_2 = \phi_3$.
For $\phi_2 = - \phi_3$ one gets an
analogous lower bound 
only in the case of the SMA solution,
while for the LMA and LOW-QVO 
solutions values of 
$\meff \ll  10^{-2}~{\rm eV}$
are possible. Both for the
LMA and LOW-QVO 
solutions there exist 
relatively large ``just-CP-violation'' 
regions of \meff, in which \meff 
has a relatively large value:
$\meff \cong (2 - 8) \times 10^{-2}~{\rm eV}$.
There is no such region
in the case of the SMA MSW solution.
We have also shown that a sufficiently
accurate
measurement of \meff,
\deltaatm, $\sin^22\theta_{\odot}$,
$(1 - |U_{e1}|^2)$ can allow to get 
direct information on the value of
the CP-violating phase $\alpha_{32}$.
Our results for the case of 
neutrino mass spectrum
with inverted hierarchy
are illustrated in Figs. 6 - 11.

   For the quasi-degenerate neutrino mass spectrum,
eqs. (70) - (72), we have $\meff \sim m$, 
where $m$ is the common neutrino mass, 
information about which can be obtained from the
$^{3}$H $\beta-$decay experiments \cite{MoscowH3,Mainz}:
$m < 2.5~{\rm eV}$. Future astrophysical 
and cosmological measurements (see, e.g., \cite{Cosmo,MAP}) 
might further constrain $m$. 
The effective Majorana mass \meff
depends also on $\theta_{\odot}$,
$|U_{e3}|^2$ which is constrained by the CHOOZ data, 
and on two physical CP-violating phases,
$\alpha_{21}$ and $\alpha_{31}$.
The maximal value of \meff is determined by
the value of $m$, $\meff \leq m$. 
It is possible for all three solutions of the solar
neutrino problem. Obviously,
\meff is limited by the upper bounds 
obtained in the $^{3}$H $\beta-$decay
\cite{MoscowH3,Mainz}
and in the \betabeta-decay 
\cite{76Ge00,IGEX00} (see eqs. (3) and (4)) 
experiments: 
$\meff < (0.33 - 1.05)~{\rm eV}$.
If CP-invariance holds and the CP-parities of the
three massive Majorana neutrinos
satisfy the relation 
$\phi_1 = \phi_2 = \pm \phi_3$,
we have $0.8m \ltap \meff \leq m$.
We get a similar result 
for $\phi_1 = - \phi_2 = \pm \phi_3$
in the case of the SMA solution.
The existence of a significant lower bound 
on \meff for the LMA and the LOW-QVO solutions
if $\phi_1 = - \phi_2 = \pm \phi_3$
depends on the ${\rm min}|\cos2\theta_{\odot}|$.
The latter varies with the analysis:
using the results of \cite{SKYSuz00} 
one finds $\meff \gtap (0.1 - 0.2)~m$
~~($\meff \gtap (0.2 - 0.3)~m$) for the
LMA (LOW-QVO) solution.
According to the (99\% C.L.) results 
of the analysis \cite{Fogli00,Gonza3nu}
one can have $\cos2\theta_{\odot} = 0$
and therefore there is no 
significant lower bound
on \meff for both solutions. 
There exist  
``just-CP-violation'' regions for the
LMA and LOW-QVO solutions, in which
\meff can be in the range of sensitivity of
the future \betabeta-decay experiments, 
while in the case of the SMA MSW solution
there is no such region.
The knowledge  of \meff, $m$,
$\theta_{\odot}$ and $|U_{e3}|^2$ 
would imply a non-trivial
constraint on the two CP-violating phases
$\alpha_{21}$ and $\alpha_{31}$.
Some of our results for the quasi-degenerate
neutrino mass spectrum
are presented graphically in Figs. 12 - 16.

  We have analyzed also in detail the predictions for \meff
in the case neutrino mass spectrum with partial mass hierarchy
eq. (83), and with partial inverted mass hierarchy,
eq. (84). These spectra 
interpolate respectively between the hierarchical    
and the inverted mass hierarchy spectra and the
quasi-degenerate one. Accordingly, the results for 
\meff we have found in the case of 
neutrino mass spectrum with partial hierarchy
(partial inverted hierarchy)
``interpolate'' between the results
for the hierarchical (inverted mass hierarchy)
and the quasi-degenerate neutrino mass spectra.
Most of these results are presented 
graphically in Figs. 17 - 26.

     The quasi-degenerate 3-neutrino mass spectrum
can be critically tested  in the future 
$^3$H $\beta-$decay experiment 
KATRIN \cite{KATRIN}, which is planned to 
have a sensitivity to $m_{\nu_e} \geq 0.35$ eV.
For the quasi-degenerate spectrum
one finds $m_{\nu_e} = m \gtap 0.20$ eV,
while for all the other 3-neutrino mass spectra
compatible with the neutrino mass 
and neutrino oscillations data,
we have $m_{\nu_e} \ltap 0.20$ eV.

     Actually, the five different types of
possible neutrino mass spectrum we have considered
correspond, once \deltasol and \deltaatm
are known and the choice 
$\deltasol = \Delta m^2_{21}$ or
$\deltasol = \Delta m^2_{32}$ is made,
to different ranges of values of the 
smallest neutrino mass $m_1$. At the same time
\meff is a continuous function of $m_1$.
In Figs. 27 and 28 we show the
possible magnitude of
\meff as a function of $m_1$
for $m_1$ varying continuously
\footnote{Similar plots were
proposed also in refs. \cite{Pols00,KPS00}.}
from $10^{-5}$ eV to 2.5 eV.
The results presented in Figs.
27 and 28 are derived for 
the allowed regions of 
values of the relevant input 
oscillation parameters,
obtained in \cite{SKYSuz00,SKatm00,CHOOZ}
and in \cite{Gonza3nu}, 
and for the two possible choices of
\deltasol in the case of 
3-neutrino mixing,
$\deltasol = \Delta m^2_{21}$ 
and $\deltasol = \Delta m^2_{32}$.
With the notations we use, 
we can have $\deltasol = \Delta m^2_{21}$
($\deltasol = \Delta m^2_{32}$)
in the cases of hierarchical
(inverted hierarchy),
partial hierarchy 
(partial inverted hierarchy), 
and quasi-degenerate
neutrino mass spectra. 

 To conclude, we have found that  
the observation of the
\betabeta-decay with a rate
corresponding to 
$\meff \gtap 0.02~$eV,
which is in the range of sensitivity of the 
future \betabeta-decay experiments, 
can provide unique information 
on the neutrino mass spectrum.
Combined with information on the lightest
neutrino mass or the type of neutrino mass spectrum, 
it can give also information 
on the CP-violation in the lepton sector,
and if CP-invariance holds - on the 
relative CP-parities
of the massive Majorana neutrinos.
A measured value of 
$\meff \gtap (2 - 3)\times 10^{-2}~$eV
would strongly disfavor (if not rule out), 
under the general assumptions of the present study
(3-neutrino mixing, \betabeta-decay generated 
only by the charged (V-A) current weak interaction 
via the exchange of the three Majorana neutrinos,
neutrino oscillation solutions of the solar 
neutrino problem and atmospheric
neutrino anomaly) the possibility of a 
hierarchical neutrino mass spectrum,
while a value of 
$\meff \gtap (2 - 3)\times 10^{-1}~$eV
would rule out the 
hierarchical neutrino mass spectrum,
strongly disfavor the spectrum with
inverted mass hierarchy and favor
the quasi-degenerate spectrum.

  In the present article we have considered
only the mixing of three
massive Majorana neutrinos.
We have assumed that the \betabeta-decay is 
induced by the (V-A) charged current weak
interaction via the
exchange of the massive Majorana neutrinos
between two neutrons in the initial nucleus.
Results for the case of four-neutrino 
mixing will be presented
elsewhere \cite{BPPbb4nu}.

\vspace{0.5cm}
{\it Note added.} Our attention was brought to 
the paper \cite{Napoli01} which discusses some of the
questions studied in the present article and 
which appeared while the 
present article was being 
prepared for submission
for publication.

\vspace{0.5cm}
\leftline{\bf Acknowledgments.} 
We would like to acknowledge
discussions with L. Wolfenstein,
B. Kayser, J. Nieves and P. Pal
concerning the Majorana 
CP-violating phases.
S.M.B. would like to thank the Elementary Particle
Theory Sector at SISSA for kind hospitality and support. 
S.T.P.  would like to acknowledge the 
hospitality of the Aspen Center for
Physics where part of this work was done.
The work of S.T.P. was supported in part by the EEC grant 
ERBFMRXCT960090
and by the  Italian MURST under the program ``Fisica
Teorica delle Interazioni Fondamentali''.


\newpage


\clearpage

%

\clearpage

\clearpage

\clearpage


\begin{figure}[p]
\begin{center}
\epsfig{file=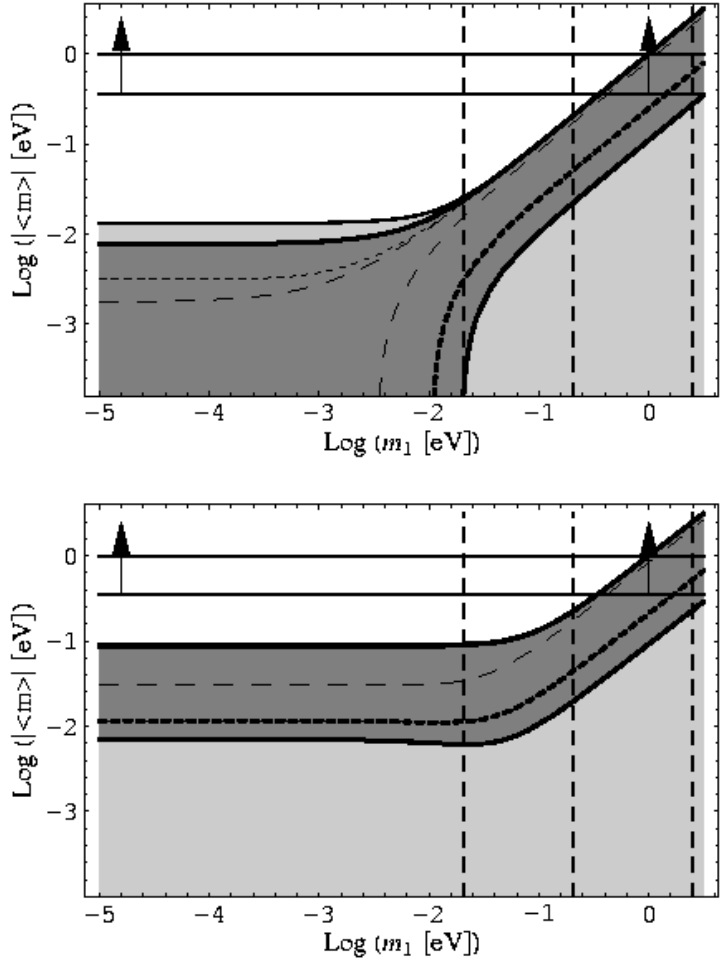, height=12.9cm
}
\end{center}
\caption[gmassgl01]{The dependence of \meff on $m_1$ 
in the case of $\deltasol = \Delta m_{21}^2$
(upper panel) and of 
$\deltasol = \Delta m_{32}^2$ (lower panel).
For  $\deltasol = \Delta m_{21}^2$ (upper panel), 
the allowed values of \meff are obtained {\it i})
using the 99\%~C.L. results of 
ref. \cite{Gonza3nu}
for the LMA solution
of the $\nu_{\odot}-$problem  
(light grey and dark grey region 
between the thick solid line and the axes),
for the LOW-QVO solution (light grey and dark grey
region between the thin
dotted line and the axes), and for 
the SMA one
(the dark grey region between the thin  dashed lines)
 and {\it ii}) using the 
results of ref. \cite{SKYSuz00}
 for the LMA  
(dark grey region between 
the two doubly thick solid lines)
 and for the LOW-QVO  
(dark grey region between the upper 
doubly thick solid line 
and the doubly thick dotted line)
solutions.
For  $\deltasol = \Delta m_{32}^2$ (lower panel), 
the allowed values are obtained 
{\it i}) using the 
99\%~C.L. results of the analysis of 
ref. \cite{Gonza3nu}  
 for the LMA and LOW-QVO solutions 
(light grey and dark grey 
region between the doubly thick line and the axes)
and the SMA one
 (the dark grey region between 
the upper doubly  thick line and the thin  dashed one) 
and {\it ii}) using the  
results of ref. \cite{SKYSuz00}
 for the LMA  
(dark grey region between the 
two doubly thick solid lines) 
and the LOW-QVO  
(dark grey region between 
the upper doubly thick solid line 
and the doubly thick dotted line) solutions. 
 The regions divided by the vertical dashed 
lines on the upper (lower) panel correspond to 
{\it i}) $m_1 \ll 0.02 \eV $, and if
$m_1 \ll \sqrt{\Delta m^2_{\odot}}$
~($m_1 < \sqrt{\Delta m^2_{\odot}} 
\ll \sqrt{\Delta m^2_{atm}}$) - 
to a hierarchical (inverted hierarchy) 
neutrino mass spectrum,
 {\it ii}) $0.02 \eV \leq m_1 \leq 0.2 \eV$, 
i.e.,  spectrum with partial hierarchy 
(partial inverted hierarchy), and
to {\it iii}) $m_1 \geq 0.2 \eV$,
i.e., quasi-degenerate neutrinos. 
For both cases the upper bounds 
from ref. \cite{76Ge00}, eq. (\ref{76Ge00}),
are shown by the horizontal 
upper doubly thick solid lines.
}
\label{figure:massgl01}
\end{figure}

\begin{figure}[p]
\begin{center}
\epsfig{file=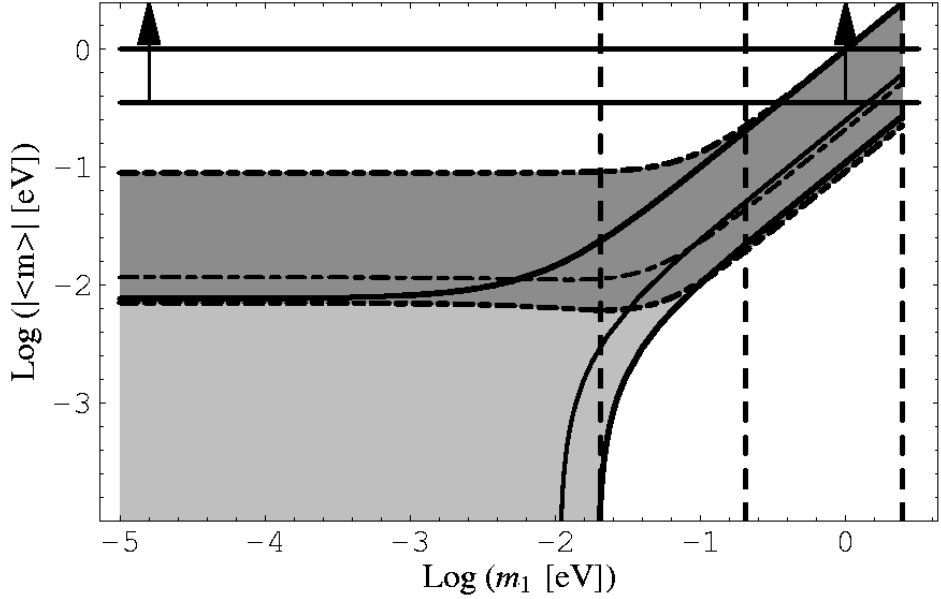, height=8cm
}
\end{center}
\caption[gmassgl02]{The dependence of \meff 
on $m_1$ 
{\it i}) for $\deltasol = \Delta m_{21}^2$
and using the  results of the analysis of  
ref. \cite{SKYSuz00} 
for the LMA solution
 (light grey and dark grey region 
between the two doubly thick solid lines)
 and for the LOW-QVO solution 
(light grey and dark grey region 
between the upper doubly thick solid line
 and the  thick solid line), and
 {\it ii}) for $\deltasol = \Delta m_{32}^2$
and in the case of the LMA solution 
(dark grey region between 
the two doubly thick dash-dotted  lines) 
and the LOW-QVO solution
 (dark grey region between 
the upper doubly thick dashed-dotted line 
and the  thick dashed-dotted line). The 
upper bounds on \meff 
from ref. \cite{76Ge00}, eq. (\ref{76Ge00}),
are shown by the horizontal upper doubly thick solid lines.
The regions separated by the vertical dashed lines 
correspond to {\it i}) $m_1 \ll 0.02 \eV $,
and if $m_1 \ll \sqrt{\Delta m^2_{\odot}}$
~($m_1 < \sqrt{\Delta m^2_{\odot}} 
\ll \sqrt{\Delta m^2_{atm}}$) -
to a hierarchical (inverted hierarchy) 
neutrino mass spectrum,
 {\it ii}) $0.02 \eV \leq m_1 \leq 0.2 \eV$,
i.e., partial hierarchy 
and partial inverted hierarchy spectrum and 
to {\it iii}) the $m_1 \geq 0.2 \eV$, 
i.e., quasi-degenerate spectrum.
}
\label{figure:massgl02}
\end{figure}

\end{document}